\documentclass[a4paper,10pt]{article}
\usepackage{graphicx}
\usepackage[curve,matrix,arrow,color]{xy}

\usepackage{subcaption}
\usepackage{pstricks}
\usepackage{pstricks-add}
\usepackage{tikz}
\usepackage{verbatim}
\usetikzlibrary{calc}
\usetikzlibrary{decorations.pathreplacing}
\usetikzlibrary{decorations.markings}

\usepackage{mathtools}
\usepackage{upgreek}
\usepackage{amsmath}
\usepackage{amsfonts}
\usepackage[title]{appendix}
\usepackage[margin=2cm]{geometry}

\usepackage{amsmath, amssymb, amsthm}
\usepackage{color,soul}
\usepackage{float}
\usepackage{placeins}
\usepackage{hyperref}
\usetikzlibrary{intersections} 
\usetikzlibrary{er,positioning}
\usetikzlibrary{arrows,shapes}
\usetikzlibrary{decorations.markings}
\usepackage{blkarray}
\usepackage{cite} % To regroup citations: [1-3] instead of [1,2,3]
\usepackage{boondox-cal} % Lower-case calligraphic font
\newcommand{\beq}{\begin{equation}}

\newcommand{\eeq}{\end{equation}}

\newcommand{\q}{\mathfrak{q}}

\newcommand{\aSt}{\mathcal{W}}
\newcommand{\coaSt}{\widetilde{\mathcal{W}}}

\newcommand{\AStTL}[2]{\aSt_{#1,#2}}
\newcommand{\coAStTL}[2]{\coaSt_{#1,#2}}

\newcommand{\bAStTL}[2]{\overline{\mathcal{W}}_{#1,#2}}
\newcommand{\ATL}[1]{\mathsf{T}^a_{#1}}

\newcommand{\Verma}[1]{\mathsf{V}_{#1}}
\newcommand{\Vermab}[1] {\overline{\mathsf{V}}_{#1}}
\newcommand{\IrrV}[1]{\mathsf{X}_{#1}}
\newcommand{\IrrVb}[1]{\overline{\mathsf{X}}_{#1}}
\newcommand{\VirN}{\hbox{Vir}\otimes\overline{\hbox{Vir}}}

\newcommand{\HN}{\mathcal{H}}
\newcommand{\FN}{F}
\newcommand{\PN}{\mathcal{P}}

\newcommand{\m}{\mathfrak{m}}
\newcommand{\Pscaled}{{\mathcal p}}

\newcommand{\KSgen}{\mathcal{L}}

\newcommand{\spindagger}{\dag}
\newcommand{\loopdagger}{\ddag}
\newcommand{\spininner}[2]{\scalebox{1.5}[1]{$\langle$}  #1, #2 \scalebox{1.5}[1]{$\rangle$} }
\newcommand{\loopinner}[2]{( #1, #2 )}

\newcommand{\tAStTL}[2]{\widetilde{\mathcal{W}}_{#1,#2}}

\newcommand{\arcsize}{0.2}
\newcommand{\loopU}{\vcenter{\hbox{
	\begin{tikzpicture}
 	\draw[thick, dotted] (-\arcsize -0.05,-\arcsize - 0.05) -- (-\arcsize-0.05,  0.05);
  	\draw[thick, dotted] (3*\arcsize + 0.05,-\arcsize - 0.05) -- (3*\arcsize+ 0.05,  0.05);
  	\draw[thick, dotted] (-\arcsize - 0.05,  0.05) -- (3*\arcsize+ 0.05,  0.05);
  	\draw[thick, dotted] (-\arcsize - 0.05, -\arcsize - 0.05) -- (3*\arcsize+ 0.05, -\arcsize - 0.05);
	\draw[thick] (0,0) arc (-180:0:\arcsize);
	\end{tikzpicture} }}}
\newcommand{\loopJL}{\vcenter{\hbox{
\begin{tikzpicture}
 	\draw[thick, dotted] (-\arcsize -0.05,-\arcsize - 0.05) -- (-\arcsize-0.05,  0.05);
  	\draw[thick, dotted] (3*\arcsize + 0.05,-\arcsize - 0.05) -- (3*\arcsize+ 0.05,  0.05);
  	\draw[thick, dotted] (-\arcsize - 0.05,  0.05) -- (3*\arcsize+ 0.05,  0.05);
  	\draw[thick, dotted] (-\arcsize - 0.05, -\arcsize - 0.05) -- (3*\arcsize+ 0.05, -\arcsize - 0.05);
	\draw[thick] (0,0) arc (0:-90:\arcsize);
	\draw[thick] (\arcsize*2,0) arc (-180:-90:\arcsize);
	\end{tikzpicture} }}  }
\newcommand{\loopII}{\vcenter{\hbox{
\begin{tikzpicture}
 	\draw[thick, dotted] (-\arcsize -0.05,-\arcsize - 0.05) -- (-\arcsize-0.05,  0.05);
  	\draw[thick, dotted] (3*\arcsize + 0.05,-\arcsize - 0.05) -- (3*\arcsize+ 0.05,  0.05);
  	\draw[thick, dotted] (-\arcsize - 0.05,  0.05) -- (3*\arcsize+ 0.05,  0.05);
  	\draw[thick, dotted] (-\arcsize - 0.05, -\arcsize - 0.05) -- (3*\arcsize+ 0.05, -\arcsize - 0.05);
	\draw[thick] (0,0) -- (0,-\arcsize);
	\draw[thick] (\arcsize*2,0) -- (\arcsize*2,-\arcsize) ;
	\end{tikzpicture} }}  }

\newcommand{\e}{\mathrm e}
\renewcommand{\i}{\mathrm i}

\usepackage{listings}
\usepackage{stackengine}

\makeatletter
\newcommand*\dashline{\rotatebox[origin=c]{90}{$\dabar@\dabar@\dabar@$}}
\makeatother

\usepackage{physics}
\usepackage{wasysym}
\usepackage{array}% for extended column definitions    

\usepackage{pdflscape}
\usepackage{rotating}

\begin{document}

\title{The action of the Virasoro algebra in the \\ two-dimensional Potts and loop models at generic $Q$}

\author{Linnea Grans-Samuelsson$^1$, Lawrence Liu$^4$, Yifei He$^1$, Jesper Lykke Jacobsen$^{1,2,3}$, Hubert Saleur$^{1,4}$ \\
[2.0mm]
${}^1$ \small Universit\'e Paris-Saclay, CNRS, CEA, Institut de Physique Th\'eorique, 91191, Gif-sur-Yvette, France \\
${}^2$ \small Laboratoire de Physique de l'\'Ecole Normale Sup\'erieure, ENS, Universit\'e PSL, CNRS,\\ 
      \small Sorbonne Universit\'e, Universit\'e de Paris, F-75005 Paris, France \\
${}^3$ \small Sorbonne Universit\'e, \'Ecole Normale Sup\'erieure, CNRS, Laboratoire de Physique (LPENS), 75005 Paris, France \\
${}^4$ \small Department of Physics, University of Southern California, Los Angeles, CA 90089-0484, USA}

%\date{\today}

\maketitle

\begin{abstract}

The spectrum of conformal weights for the CFT describing the two-dimensional critical $Q$-state Potts model (or its close cousin, the dense loop model) has been known for more than 30 years \cite{dFSZ}. However, the exact nature of the corresponding  $\VirN$ representations has remained unknown up to now. Here, we solve the problem for generic values of $Q$. This is achieved by a mixture of different techniques: a careful study of  ``Koo--Saleur generators'' \cite{KooSaleur}, combined with measurements of four-point amplitudes, on the numerical side, and OPEs and the  four-point amplitudes recently determined using the ``interchiral conformal bootstrap'' in \cite{HJS} on the analytical side. We find that null-descendants of diagonal fields having weights $(h_{r,1},h_{r,1})$ (with $r\in \mathbb{N}^*$) are truly zero, so these fields come with simple $\VirN$ (``Kac'') modules. Meanwhile, fields with weights $(h_{r,s},h_{r,-s})$ and $(h_{r,-s},h_{r,s})$ (with $r,s\in\mathbb{N}^*$) come in indecomposable but not fully reducible representations  mixing four simple $\VirN$ modules with a familiar ``diamond'' shape. The ``top'' and ``bottom'' fields in these diamonds have weights $(h_{r,-s},h_{r,-s})$, and form a two-dimensional Jordan cell for $L_0$ and $\bar{L}_0$. This establishes, among other things, that the Potts-model CFT is logarithmic for $Q$ generic. Unlike the case of non-generic (root of unity) values of $Q$, these indecomposable structures are not present in finite size, but we can nevertheless show from the numerical study of the lattice model how the rank-two Jordan cells
build up in the infinite-size limit.

\end{abstract}

%\newpage
\tableofcontents
%\newpage

\section{Introduction}

The full solution of the conformal field theory (CFT) describing the critical $Q$-state Potts model for $Q$ generic (or its cousins, the critical  and dense $O(n)$ models) in two dimensions still eludes us, more than 30 years after the pioneering work \cite{DotsenkoFateev}. While most critical exponents of interest were quickly determined (for some, even before the advent of CFT, using Coulomb-gas techniques) \cite{DenNijs,Nienhuis,SaleurPolyCFT}, the non-rationality of the theory (for $Q$ generic) as well as its non-unitarity (inherited from the geometrical nature of the lattice model) made further progress using ``top-down" approaches (such as the one used for minimal unitary models \cite{FQS}) considerably more difficult. Several breakthroughs took place, however, in the last decade. First, many three-point functions were determined using connections with Liouville theory at $c<1$ \cite{DelfinoViti,PiccoViti,IJS}. Second, a series of attempts using conformal bootstrap ideas \cite{GoriViti,Piccoetal1, JS,Piccoetal2,HGJS,HJS} led to the determination of some of the most fundamental four-point functions in the problem (namely, those defined geometrically, and hence for generic $Q$), also  shedding light on the operator product expansion (OPE) algebra and the relevance of the partition functions determined in \cite{dFSZ}. In particular, the set of operators---the so-called spectrum---required to describe the partition function \cite{dFSZ} and correlation functions \cite{JS} in the Potts-model CFT was settled. While the picture remains incomplete, a complete solution of the problem now appears within reach.

An intriguing aspect of the spectrum proposed in \cite{dFSZ,JS} is the appearance of fields with conformal weights given by the Kac formula $\Delta=h_{r,s}$, with $r,s\in\mathbb{N}^*$ (we call these ``degenerate'' weights).  It is known that for some of these fields---such as the energy operator with weights $(h_{2,1},h_{2,1})$---the null-state descendants are truly zero, and the corresponding four-point functions obey the Belavin--Polyakov--Zamolodchikov (BPZ) differential equations \cite{BPZ}. It is also expected that this does not hold for {\sl all} fields with degenerate weights. In fact, it was suggested  in \cite{JS,HJS} that, in the Potts-model case, {\sl only} fields with weights $(h_{r,1},h_{r,1})$ give rise to null descendants.  Since the spectrum of the model is expected to contain non-diagonal fields with weights $(h_{r,s},h_{r,-s})$ and $(h_{r,-s},h_{r,s})$ for $r,s\in\mathbb{N}^*$, this means that the theory should contain fields with degenerate (left or right) weights whose null descendants are nonzero, even though their two-point function vanishes. It is well understood since the work of Gurarie \cite{Gurarie} that in this case, ``logarithmic partners'' must be invoked  to compensate for the corresponding divergences occurring in the OPEs. Such partners give rise to Jordan cells for $L_0$ or $\bar{L}_0$, and make the theory a logarithmic CFT---i.e., a theory where the action of the product of left and right Virasoro algebras $\VirN$ is not fully reducible. This, in turn,  is made possible by the theory not being unitary in the first place \cite{JPhysAreviews}.

A great deal of our understanding of the fields with degenerate weights in the Potts model comes from indirect arguments, such as the solution of the bootstrap equations for correlation functions and the presence of an underlying  ``interchiral'' 
algebra, responsible for relations between some of the conformal-block amplitudes \cite{HJS}. The purpose of this paper is to explore this issue much more directly using the  lattice regularization of $\VirN$ first introduced in \cite{KooSaleur}, and explored in further detail, in particular, in a companion paper on XXZ spin chains \cite{GSJS} (see also \cite{Vidal,Vidal2a,Vidal2b} for other applications). 

The paper is organized as follows. In section \ref{sec:PottsGen} we start by reminding the reader of basic facts about the two-dimensional Potts model and its CFT.  In section \ref{sec:TL-alg-def} we discuss the algebra of local energy and momentum densities---the Temperley--Lieb algebra---together with its representations in the periodic case. Albeit a bit technical, this section is crucial, since it will be used as a starting point to understand the corresponding representations of $\VirN$ in the continuum limit.  In section \ref{sec:disVir} we remind the reader of the general strategy to study the action of $\VirN$ starting from the lattice model. New results then appear in section \ref{sec:degenerate} where we argue, based on several lattice arguments, for the existence of indecomposable modules of $\VirN$ in the continuum limit of the Potts model for $Q$ generic. Our main results are given in equations \eqref{result1}, \eqref{MainRes3}, and \eqref{moduleL}. In section~\ref{OPEsection} we present a CFT argument in which we analyze the OPE of two copies of a generic field $\Phi_{\Delta}$, which we suppose to produce a field $X_\epsilon$ that tends to $\phi_{1,2}$ when $\epsilon \to 0$. Regularizing the divergences of this OPE leads to the same indecomposable structure \eqref{moduleL} as before and allows us to compute the corresponding indecomposability parameters.  We note that some of these our results overlap with the recent work \cite{SylvainNextPaper}. In section \ref{orderpara} we consider the particular case where $\Phi_{\Delta}$ is the Potts-model order parameter $\phi_{1/2,0}$. We first give two different CFT derivations of the corresponding logarithmic conformal block. Then we go back to the lattice Potts model and provide numerical evidence that the indecomposable structure \eqref{moduleL} builds up when the continuum limit is approached, although in this case there is no indecomposability in finite size. To round off the paper, we briefly discuss in section \ref{currents} the cognate ``ordinary'' loop model with $U(\m)$ symmetry and comment on the relation with recent results by Gorbenko and Zan \cite{GorbenkoZan} on the dilute $O(n)$ model. Our conclusions are given in section~\ref{sec:conclusion}. Two appendices provide details on our numerical work which is referred to throughout the article. 

\subsection*{Notations and definitions}

We gather here some general notations and definitions that are used throughout the paper:

\begin{itemize}

\item$\ATL{N}(\m)$ --- the affine Temperley--Lieb algebra on $N=2L$ sites with parameter $\m$. We shall later parametrize the loop weight as $\m=\q+\q^{-1}$, with $\q=\e^{\i\gamma}$ and the parametrization ($x \in \mathbb{R}^+ \setminus \mathbb{Q}$) %\lawrence{$x$ positive/$\ge 1$?}

\begin{equation}
\label{gammaparam}
 \gamma=\frac{\pi}{x+1} \,.
\end{equation}

\item$\AStTL{j}{\e^{\i\phi}}$ --- standard module of the affine Temperley--Lieb algebra with $2j$ through-lines and pseudomomentum $\phi/2$. 
We define a corresponding electric charge as
\begin{equation} \label{e_phi}
 e_\phi\equiv \frac{\phi}{2\pi} \,.
\end{equation}

\item$\Verma{r,s}$ --- Verma module for the conformal weight $h_{r,s}$, when either $r \notin \mathbb{N}^*$ or $s \notin \mathbb{N}^*$.

\item$\Verma{r,s}^{\rm d}$ --- the (degenerate) Verma module for the conformal weight $h_{r,s}$, when $r,s\in\mathbb{N}^*$.

%\item$\FF{r,s}^{\rm d}$ --- the (degenerate) co-Verma module for the conformal weight $h_{r,s}$, when $r,s\in\mathbb{N}^*$.

\item$\IrrV{r,s}$ --- irreducible Virasoro module for the conformal weight $h_{r,s}$.

\item{} A conformal weight $h_{r,s}$ with $r,s\in\mathbb{N}^*$ will be called degenerate. For such a weight, 
there exists a  descendant state that is also primary: this descendant is often  called a null  (or singular) vector (or state).  We will denote by $A_{r,s}$  the combination of Virasoro generators producing the null state at level $r s$ corresponding to the degenerate weight $h_{r,s}$. $A_{r,s}$ is normalized so that the coefficient of $L_{-rs}$ is equal to unity. Some examples are
\begin{subequations}
\label{A_operators}
\begin{eqnarray}
A_{1,1}&=&L_{-1} \,, \\
A_{1,2}&=&L_{-2}-{3\over 2(2h_{1,2}+1)}L_{-1}^2 \,, \\
A_{2,1}&=&L_{-2}-{3\over 2(2h_{2,1}+1)}L_{-1}^2 \,.
\end{eqnarray}
\end{subequations}

\item{} We will in this paper restrict to generic values of the parameter $\q$ (i.e., $\q$ not a root of unity), and thus to generic values of $x$ (i.e., $x$ irrational). Even in this case, we will encounter situations where some of the modules of interest are no longer irreducible. We will refer to these situations as ``non-generic'' when applied to modules of the affine Temperley--Lieb algebra, and ``degenerate'' when applied to modules of the Virasoro algebra. In earlier papers (see e.g. \cite{GJS18}), we have referred to such cases as ``partly non-generic'' and ``partly degenerate,'' respectively, since having $\q$ a root of unity adds considerably more structure to the modules. We will not do so here, the context clearly excluding $\q$ a root of unity. 

\item{} Finally, we shall discuss two scalar products, denoted by $\spininner{-}{-}$ and $\loopinner{-}{-}$, which are defined such that for any two primary states $V_1,V_2$ we have $\spininner{V_1}{L_n V_2} = \spininner{L_n^\spindagger V_1}{V_2}$ and $\loopinner{V_1}{L_n V_2} = \loopinner{L_n^\loopdagger V_1}{V_2}$, where $L_n^\spindagger$ is discussed below and $L_n^\loopdagger=L_{-n}$ is the usual conformal conjugate \cite{BPZ}. The scalar product $\spininner{-}{-}$ is positive definite and will be used for most parts of the paper. When using this scalar product we shall also use the bra-ket notation: $|V\rangle$ denotes a state $V$ (primary or not) and $\langle V|$ its dual, $\langle V_1 | V_2\rangle \equiv \spininner{V_1}{V_2} $ and $\langle V_1| \mathcal{O} |V_2 \rangle \equiv \spininner{V_1}{\mathcal{O}V_2}$ for an operator $\mathcal{O}$ acting on $|V_2\rangle$ (with $V_1,V_2$ being primary or not).

\item{} We denote by $\phi_{r,s}$ a chiral primary field with conformal weight $h_{r,s}$ and $r,s\in\mathbb{R}$: the structure of the underlying Virasoro module when $r,s\in \mathbb{N}^*$ will be made clear from the context, but will not appear in the notation. We will also freely make use of the symmetries $h_{r,s}=h_{-r,-s}$.

\end{itemize}

\section{The $Q$-state Potts model and its CFT}\label{sec:PottsGen}

We shall assume in this paper that the reader is familiar with the $Q$-state Potts model and its definition for $Q$ non-integer using the Fortuin--Kasteleyn (FK) expansion (we shall sometimes refer to this as the ``FK formulation'' of the Potts model).  More details can be found in  our papers \cite{JS,HGJS}, and in subsection \ref{sec:loops-clusters} below. A special point must be made in connection with the present work: there is sometimes a confusion related with the type of  object one may wish to consider as part of ``the'' Potts model CFT. By such a  CFT we shall mean here the field theory  describing long-distance properties of observables which are  built locally in terms of Potts spins for $Q$ integer, then continued to $Q$ real using the FK expansion. Examples include the spins themselves  but also the energy and, of course, many more observables as discussed, for instance, in \cite{VJS12,VJ13,CJV17}. Other objects have been defined and studied in the literature, in particular those describing the properties of domain walls, boundaries of domains where the Potts spins take identical values \cite{DJS0,DJS00}.  These are not local with respect to the Potts spin variables, and we will not consider them further in this work.%
\footnote{Whether there is a ``bigger" CFT containing {\sl all} these observables at once remains an open question---see \cite{VJ11} for an attempt in this direction.}

To have a better idea of the observables pertaining to the Potts model CFT for $Q$ generic, one can start  with the torus partition function, which was determined in the continuum limit in \cite{dFSZ} and \cite{RS01,RJ07}.  Parametrizing\footnote{The values $x \in (0,1)$ correspond to the so-called unphysical self-dual case discussed in \cite{JS2006}. Note the negative determination of the square root $\sqrt{Q}$ in this case. There is no change of analytic behavior of the results for generic values $x \in (0,\infty]$.} 
\begin{equation}
\sqrt{Q}=2\cos\left({\pi\over x+1}\right)\,, \quad \mbox{with } x \in (0,\infty] \,,
\end{equation}
the central charge is 
\begin{equation}
c=1-{6\over x(x+1)} \,, \label{centralc}
\end{equation}
while the Kac formula reads
\begin{equation}
h_{r,s}={[(x+1)r-xs]^2-1\over 4x(x+1)} \,.
\end{equation}
The continuum-limit partition function is then given by
\begin{equation}
 {\cal Z}_Q = F_{0,\q^{\pm 2}} + {Q-1\over 2} F_{0,-1} + \sum_{j>0} \hat{D}_{j,0}'F_{j,1}+\sum_{\substack{j>0,M>1 \\ M|j}}\sum_{\substack{0<p<M \\ p\wedge M=1}}
 \hat{D}_{j,\pi p/ M}' F_{j,\e^{2\pi \i p/M}} \,. \label{decompPottsZ}
\end{equation}
The coefficients $\hat{D}_{j,K}'$ can be thought of as ``multiplicities,'' although of course, for $Q$ generic, they are not integers. Their interpretation in terms of symmetries is beyond the scope of this paper \cite{GRSV}.
They are given by
\begin{equation}
\hat{D}_{j,K}'={1\over j}\sum_{r=0}^{j-1} \e^{2\i Kr}w(j,j\wedge r) \,,
\end{equation}
where $j\wedge r$ is the greatest common divisor of $j$ and $r$ (with $j\wedge 0=j$ by definition), and 
\begin{equation} \label{wjd_Potts}
w(j,d)=\q^{2d}+\q^{-2d}+{Q-1\over 2}( \i^{2d}+\i^{-2d})=\q^{2d}+\q^{-2d}+(Q-1)(-1)^d \,,
\end{equation}
where we introduced the quantum group parameter $\q$ defined via 
\begin{equation}
\sqrt{Q}=\q+\q^{-1} \,.
\end{equation}
The $F_{j,\e^{\i \phi}}$ are the following sums
\begin{equation}\label{F-func}
 \FN_{j,\e^{\i\phi}} = \frac{q^{-c/24}\bar{q}^{-c/24}}{P(q)P(\bar{q})} \sum_{e \in \mathbb{Z}} q^{h_{e-e_\phi,-j}}\bar{q}^{\,h_{e-e_\phi,j}}
\end{equation}
in which  
\begin{equation}
\displaystyle P(q) =  \prod_{n=1}^{\infty} (1 - q^n) = q^{-1/24} \eta (q) \,,
\end{equation}
where $\eta(q)$ is the Dedekind eta function, and $e_\phi = \phi/2\pi$. As usual, $q,\bar{q}$ are the modular parameters of the torus. 

\bigskip

Expressions (\ref{decompPottsZ}) and (\ref{F-func}) encode the operator content of the $Q$-state Potts model CFT as defined earlier. The conformal weights arising from the last term in (\ref{decompPottsZ}) are of the form
\begin{equation}
(h_{e-p/M,j},h_{e-p/M,-j}) \,, \quad \mbox{with } e\in \mathbb{Z} \,.
\end{equation}
The first two terms must be handled slightly differently. Using the identity
\begin{equation}
F_{0,\q^{\pm 2}}-F_{1,1}=\sum_{n=1}^\infty K_{n,1}\bar{K}_{n,1}\equiv \bar{F}_{0,\q^{\pm 2}}
\end{equation}
with the Kac character
\begin{equation}
K_{r,s}=q^{h_{r,s}-c/24} \: {1-q^{rs}\over P(q)} \,,
\end{equation}
we see that we get the set of diagonal fields
\begin{equation}
(h_{n,1},h_{n,1}) \,, \quad \mbox{with } n\in\mathbb{N}^* \,.
\end{equation}
The partition function can then be rewritten as
\begin{equation}
 {\cal Z}_Q = \bar{F}_{0,\q^{\pm 2}} + {Q-1\over 2} F_{0,-1} +F_{1,1}+ \sum_{j>0} \hat{D}_{j,0}'F_{j,1}+\sum_{\substack{j>0,M>1 \\ M|j}}\sum_{\substack{0<p<M \\ p\wedge M=1}}
 \hat{D}_{j,\pi p/ M}' F_{j,\e^{2\pi \i p/M}} \,.\label{decompPottsZ1}
\end{equation}
We notice now that $\hat{D}_{1,0}'=\q^2+\q^{-2}-(Q-1)=Q-2-(Q-1)=-1$. Hence $F_{1,1}$ disappears, in fact, from the partition function. Note that $F_{1,1}$ corresponds geometrically to the so-called hull operator \cite{DS}---related to the indicator function that a point is at the boundary of an FK cluster---with corresponding conformal weights $(h_{0,1},h_{0,1})$. It should probably not come as a surprise that this operator is absent from the partition function, since the definition of the hull is not local with respect to the Potts spins. We will, nevertheless, consider $\AStTL{1}{1}$ throughout this paper, since  this module {\sl does} appear in  related models, such as the ``ordinary'' loop model or the ``$U(\m)$'' model, to be discussed in section \ref{currents} below. We note meanwhile that the higher hull operators---related to the indicator function that $j>1$ distinct hulls come close together at the scale of the lattice spacing---with conformal weights $(h_{0,j},h_{0,j})$ in $F_{j,1}$ do appear in the partition function, also in the Potts case.

The decomposition (\ref{decompPottsZ}) of the Potts-model partition function for generic $Q$ is in fact in one-to-one correspondence with an algebraic decomposition of the Hilbert space ${\cal H}_Q$ in terms of modules of the affine Temperley--Lieb algebra which is exact in finite size \cite{RS07-1}. This decomposition formally reads
\begin{equation}
 {\cal H}_Q = \bAStTL{ 0}{\q^{\pm 2}}\oplus {Q-1\over 2} \AStTL{0}{-1} \oplus \AStTL{1}{1}\oplus\bigoplus_{j>0}  \hat{D}_{j,0}'\AStTL{j}{1}\oplus
 \bigoplus_{\substack{j>0,M>1 \\ M|j}}\bigoplus_{\substack{0<p<M \\ p\wedge M=1}}\hat{D}_{j,\pi p/ M}' 
 \AStTL{j,\e^{2\pi \i p/M}}\,. \label{decompPottsH}
\end{equation}
Equation  (\ref{decompPottsH})  is only formal in the sense that, for $Q$ generic, the multiplicities are not integers, and ${\cal H}_Q$ cannot be interpreted as a proper vector space. In contrast, the $\AStTL{j}{\e^{\i\phi}}$ are well-defined spaces with integer dimension {\sl independent of $Q$}, as discussed in the following section. Also, in (\ref{decompPottsH}) we have not taken into account the fact that, for a finite lattice system, the sums must be properly truncated. 

The torus partition function (\ref{decompPottsZ}) is obtained by the trace over ${\cal H}_Q$,
\begin{equation}
\text{Tr}_{{\cal H}_Q} \,\e^{-\beta_R \HN}\e^{-\i\beta_I \PN} \,,
%\rightarrow \text{Tr}\,q^{L_0-c/24}\bar{q}^{\bar{L}_0-c/24}
\end{equation}
where the real parameters $\beta_{R} > 0$ and $\beta_{I}$ determine the size of the torus, while $\HN$ and $\PN$ denote respectively the lattice Hamiltonian
and momentum operators. Introducing the (modular) parameters
\begin{subequations}
\begin{eqnarray}
 q &=& \exp\left[-{2\pi\over N}(\beta_R + \i\beta_I)\right] \,, \\
 \bar{q} &=& \exp\left[-{2\pi\over N}(\beta_R - \i\beta_I)\right]
\end{eqnarray}
\end{subequations} 
we have, in the limit where  the size  of the system $N\to\infty$, with $\beta_R,\beta_I\to\infty$ so that  $q$ and $\bar{q}$ remain finite,
 \begin{equation}
\text{Tr}_{\AStTL{j}{\e^{\i\phi}}}\,\e^{-\beta_R \HN}\e^{-\i\beta_I \PN}  \xmapsto{\, N\to\infty\,}\; \FN_{j,\e^{\i\phi}} \,.
 \end{equation}

In order to understand better how $\VirN$ acts in the $Q$-state Potts model CFT, we now focus on the 
action of discrete versions of the Virasoro generators on the spaces $\AStTL{j}{\e^{\i\phi}}$.

% \hubert{Finite size: $N$ or $L$?}

\section{The Temperley--Lieb algebra in the periodic case}\label{sec:TL-alg-def}

This whole section contains material already discussed in our earlier work on the subject \cite{GRS1,GRS2,GRS3,GRSV}, especially in the companion paper \cite{GSJS}. We  reproduce it here for clarity, completeness, and in order to establish notations.

\subsection{The algebra $\ATL{N}(\m)$}
\label{sec:algATL}

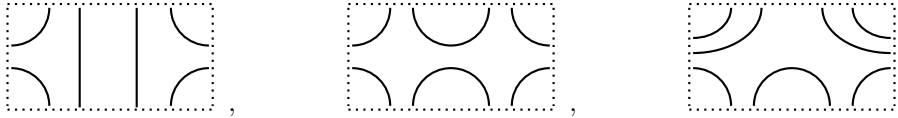
\begin{figure}
%	\usetikzlibrary{lindenmayersystems}
\begin{equation*}
 \begin{tikzpicture}
 %%%%%%%%% Frame %%%%%%%%
 	\draw[thick, dotted] (-0.05,0.5) arc (0:10:0 and -7.5);
 	\draw[thick, dotted] (-0.05,0.55) -- (2.65,0.55);
 	\draw[thick, dotted] (2.65,0.5) arc (0:10:0 and -7.5);
	\draw[thick, dotted] (-0.05,-0.85) -- (2.65,-0.85);
%%%%%%%%%%%%	
	\draw[thick] (0,0) arc (-90:0:0.5 and 0.5);
%	\draw[thick] (0.8,0.5) line (0:0.5:0.5 and 0);
	\draw[thick] (0.9,0.5) arc (0:10:0 and -7.6);
	\draw[thick] (1.65,0.5) arc (0:10:0 and -7.6);
	\draw[thick] (2.6,0) arc (-90:0:-0.5 and 0.5);

	\draw[thick] (0.5,-0.8) arc (0:90:0.5 and 0.5);
%	\draw[thick] (1.8,-0.8) arc (0:180:0.5 and 0.5);
	\draw[thick] (2.1,-0.8) arc (0:90:-0.5 and 0.5);
	\end{tikzpicture}\;\;,
	\qquad\qquad
 \begin{tikzpicture}
 %%%%%%%%% Frame %%%%%%%%
 	\draw[thick, dotted] (-0.05,0.5) arc (0:10:0 and -7.5);
 	\draw[thick, dotted] (-0.05,0.55) -- (2.65,0.55);
 	\draw[thick, dotted] (2.65,0.5) arc (0:10:0 and -7.5);
	\draw[thick, dotted] (-0.05,-0.85) -- (2.65,-0.85);
%%%%%%%%%%%%	
	\draw[thick] (0,0) arc (-90:0:0.5 and 0.5);
	\draw[thick] (0.8,0.5) arc (-180:0:0.5 and 0.5);
	\draw[thick] (2.6,0) arc (-90:0:-0.5 and 0.5);

	\draw[thick] (0.5,-0.8) arc (0:90:0.5 and 0.5);
	\draw[thick] (1.8,-0.8) arc (0:180:0.5 and 0.5);
	\draw[thick] (2.1,-0.8) arc (0:90:-0.5 and 0.5);

%	\draw[thick] (0.4,0) arc (-180:0:0.3 and 0.3);
%	\draw[thick] (0.6,0) arc (-180:0:0.1 and 0.18);
	\end{tikzpicture}\;\;,
	\qquad\qquad
 \begin{tikzpicture}
  %%%%%%%%% Frame %%%%%%%%
 	\draw[thick, dotted] (-0.05,0.5) arc (0:10:0 and -7.5);
 	\draw[thick, dotted] (-0.05,0.55) -- (2.65,0.55);
 	\draw[thick, dotted] (2.65,0.5) arc (0:10:0 and -7.5);
	\draw[thick, dotted] (-0.05,-0.85) -- (2.65,-0.85);
%%%%%%%%%%%%	
	\draw[thick] (0,0.1) arc (-90:0:0.5 and 0.4);
	\draw[thick] (0,-0.1) arc (-90:0:0.9 and 0.6);
	\draw[thick] (2.6,-0.1) arc (-90:0:-0.9 and 0.6);
	\draw[thick] (2.6,0.1) arc (-90:0:-0.5 and 0.4);
	
	\draw[thick] (0.5,-0.8) arc (0:90:0.5 and 0.5);
	\draw[thick] (1.8,-0.8) arc (0:180:0.5 and 0.5);
	\draw[thick] (2.1,-0.8) arc (0:90:-0.5 and 0.5);
	\end{tikzpicture}
\end{equation*}
%\begin{equation}
%\psset{xunit=2mm,yunit=2mm}
%\begin{pspicture}(0,0)(0.75,1)
%\psellipticarc[linecolor=black,linewidth=1.0pt]{-}(0,1.0)(0.5,1.42){180}{360}
%\end{pspicture}
%\end{equation}
\caption{Examples of affine diagrams for $N=4$, with the left and right sides of the framing rectangle identified. The first diagram represents the generator $e_4$, the second is $e_2 e_4$, and expressing the last one is left as an exercise for the reader. 
%it has rank $2$ as well as the second one. The third diagram has  rank $3$.
}
\label{fig:aff-diag}
\end{figure}

We are concerned here with the affine Temperley--Lieb algebra   $\ATL{N}$, which is spanned by particular diagrams
on an annulus.  A general basis element in the algebra of
diagrams corresponds to a diagram of $N$ sites on the inner boundary and $N$ on
the outer boundary of the annulus (we will always restrict in what follows to the case $N$ even, and we denote $N=2L$).
The sites are connected in pairs,
and only configurations that can be represented using simple curves inside the
annulus that do not cross are allowed. Such
diagrams  are commonly called \textit{affine diagrams}.  Examples of affine diagrams are shown in Fig.~\ref{fig:aff-diag}, where we draw them in a slightly different geometry: we cut the annulus and transform it to a rectangle, which we call \textit{framing}, with the sites labeled from left to right. The left and right sides of the framing rectangle are understood to be identified by the periodic boundary conditions.

An important parameter is the number of {\em through-lines}, which we denote by $2j$; each through-line is a simple curve connecting
a site on the inner and a site on the outer boundary of the
annulus. The $2j$ sites on the inner boundary attached to a through-line we call {\em free} or
{\em non-contractible}. The inner (resp.\ outer) boundary of the annulus corresponds to the bottom (resp.\ top) side of the framing rectangle.

Multiplication of two affine diagrams, $a$ and $b$, is defined in a natural
way, by joining the inner boundary of the annulus containing $a$ to the outer boundary of the annulus containing $b$, and
removing the interior sites. Accordingly, $ab$ is obtained by joining the bottom side of $a$'s framing rectangle to the top side of $b$'s framing
rectangle, and removing the corresponding joined sites. Whenever a closed contractible loop is
produced when diagrams are multiplied together, this loop must be
replaced by a numerical factor~$\m$.

In terms of generators and relations, 
the algebra $\ATL{N}$ is generated by the $e_j$'s together with the identity, subject to the usual Temperley--Lieb relations \cite{TL71}
\begin{subequations}
\label{TL}
\begin{eqnarray}
e_j^2&=&\m e_j \,, \\
e_je_{j\pm 1}e_j&=&e_j \,, \\
e_je_k&=&e_ke_j\qquad(\mbox{for } j\neq k,~k\pm 1) \,,
\end{eqnarray}
\end{subequations}
where $j=1,\ldots,N$ and the indices are now interpreted modulo $N$.
Moreover $\ATL{N}$ contains the elements
 $u$ and $u^{-1}$ which are generators of translations by one site to the right
and to the left, respectively.  The following additional defining
relations are then obeyed,
\begin{subequations}
\label{TL-u}
\begin{eqnarray}
ue_ju^{-1}&=&e_{j+1} \,, \label{TL-u1} \\
u^2e_{N-1}&=&e_1 \cdots e_{N-1} \,,
%u^{N}&=&1\label{PTL}
\end{eqnarray}
\end{subequations}
and $u^{\pm N}$ is a central element.
 The algebra generated by the $e_i$ and $u^{\pm1}$ together with these  relations is usually called the \textit{affine} Temperley--Lieb algebra $\ATL{N}$. 

\subsection{Loops and clusters}
\label{sec:loops-clusters}

The FK formulation of the $Q$-state Potts model leads to the following expansion of the  partition function %
\begin{equation}
 Z = \sum_{A \subseteq E} v^{|A|} Q^{k(A)} \,,
\end{equation}
where the underlying lattice (or graph) $G=(V,E)$ is defined by its vertex set $V$ and edge set $E$, and $k(A)$ denotes the number of connected
components (or clusters) in the subgraph $G_A = (V,A)$. For the purpose of defining a corresponding transfer matrix, it is most convenient to take $G$ to
be the square lattice wrapped on a cylinder with a circumference of $L$ lattice sites. In this construction, the transfer matrix then enjoys periodic transverse boundary conditions, while the conditions at the extremities of the cylinder can be left free or unspecified, and accordingly, $G$ can be supposed planar. Using the Euler
relation one then has equivalently
\begin{equation} \label{Zloop}
 Z = Q^{|V|/2} \sum_{A \subseteq E} \left( \frac{v}{\sqrt{Q}} \right)^{|A|} Q^{\ell(A)/2} \,,
\end{equation}
where the sum is now over loops on the medial lattice---another square lattice, rotated through 45 degrees, with vertices being the midpoints of the edges $E$.
These loops bounce off the edges in $A$ and cut
through those in the complement $E \setminus A$; see Figure~1 of \cite{HGJS} for an illustration. Configurations in these two formulations are completely equivalent: given a cluster configuration, the loops
surround each connected component as well as its inner cycles; and conversely each loop touches a cluster on its inside and a dual cluster on its outside, or vice versa.
For this reason, we henceforth refer to either of these formulations as \underline{the} {\em loop/cluster formulation}. The critical point on the square lattice is $v_{\rm c} = \sqrt{Q}$, implying a simplification in \eqref{Zloop}. 
Note that the equivalence between loop and cluster formulations must be handled with care on the torus: there are subtle differences between the two, which are manifest in the decompositions (\ref{decompPottsH}) and (\ref{decomploopH}) below. 

The loop/cluster formulation gives rise to a representation---in the technical sense of a representation of an associative algebra---of $\ATL{N}$, as we now explain. In practice, 
states in the transfer matrix must be defined so as to allow the book-keeping of the non-local quantities $k(A)$ or $\ell(A)$. In the cluster picture, a state is a set partition of the $L$ sites in a row, with two vertices belonging to the same block in the partition if and only if they are connected via the part of the FK clusters seen below that row. Equivalently, in the loop picture, a state is a pairwise matching of $N = 2L$ medial sites in a row, with each site seeing
either a vertex of $V$ on its left and a dual vertex on its right, or conversely. The above bijection between cluster and loop configurations provides as well a bijection between the corresponding cluster and loop states. The transfer matrix evolves the loop states by the relations \eqref{TL}--\eqref{TL-u} of the affine Temperley--Lieb algebra $\ATL{N}$, and to match the loop weights between \eqref{Zloop} and \eqref{TL-u1} we must identify
\begin{equation}
 \m = \sqrt{Q} \,.
\end{equation}
To account also for the computation of correlation functions, a few modifications must be made. The case of four-point functions has been expounded in \cite{JS}, but in the present paper it is enough to consider the simpler case of two-point functions. These can be computed in the cylinder geometry by placing one point at each extremity of the cylinder. The issue is then ensuring the propagation of $j$ distinct clusters between the two extremities in a setup compatible with the transfer matrix formalism. This can be done, on one hand, in the cluster picture by letting the states be $L$-site set partitions including $j$ marked blocks, and on the other hand, in the loop picture by letting the states be $N$-site pairwise matchings including $2j$ defect lines---which are precisely the through-lines already encountered in the discussion of $\ATL{N}(\m)$.
The sum over states must then be restricted so as to ensure that the marked clusters or defect loop-lines propagate all along the cylinder. Moreover, it turns out to be necessary to keep track of the windings of either type of marked object around the periodic direction of the cylinder. Fortunately, in the loop picture, these considerations lead directly to the definition of a type of representation---the affine Temperley--Lieb standard module---which is well-known in the algebra literature. We therefore proceed to define it precisely, keeping in mind that the $\ATL{N}$ diagrams are nothing but a graphical rendering of the loops resulting from \eqref{Zloop}.

\subsection{Standard modules} \label{sec:stdmod}

With the defining relations \eqref{TL}--\eqref{TL-u}
the algebra $\ATL{N}(\m)$ is infinite-dimensional. However, we will only be concerned in this work with lattice models involving a finite number of degrees of freedom per site and their description involves some finite-dimensional representations of $\ATL{N}$, the so-called {\em standard modules} $\AStTL{j}{\e^{\i\phi}}$, which depend on two parameters.  In terms of diagrams, the first defines the number of
through-lines $2j$, with $j=0,1,\ldots, N/2$. Using the natural action of the algebra---the stacking of diagrams discussed in section~\ref{sec:algATL}---we
% The action of the algebra $\ATL{N}(m)$ is defined
%in a natural way on these diagrams, by joining their outer boundary
%to an inner boundary of a diagram from $\ATL{N}(m)$, and removing the
%interior sites. As usual, a closed contractible loop is replaced by
%$m$. 
also stipulate that the result of  this action is zero in the standard modules whenever the affine diagrams  obtained have a number of
through-lines strictly less than $2j$, i.e., whenever the action contracts two or more free sites. Furthermore, for a given
nonzero value of $j$, it is possible, using the  action of the algebra,  to cyclically
permute the free sites: this gives rise to the introduction of a
{\em pseudomomentum}, which we parametrize by $\phi$. By definition,
whenever $2j$ through-lines wind counterclockwise around
the annulus $l$ times, we can unwind them at the price of a factor
$\e^{\i jl\phi}$; similarly, for clockwise winding, the phase is $\e^{-\i jl\phi}$ \cite{MartinSaleur,MartinSaleur1}. Stated more simply, there is a phase $\e^{\pm \i\phi/2}$ per winding through-line.
For technical reasons, we shall later ``smear out'' this phase, so that there is a phase $\e^{\pm \i\phi/2N}$ for each step a through-line moves left or right. This is equivalent, and is done in order to preserve invariance under the usual translation operator.

%\jesper{Moved up and expanded:}
A slightly more convenient formulation of this representation $\AStTL{j}{\e^{\i\phi}}$ can be obtained via the following consideration. Since the free sites are not allowed to be contracted, the pairwise connections between non-free sites on the inner boundary cannot be changed by the algebra action. This part of the diagrammatic information is thus nugatory and can be omitted. It is then enough to concentrate on the upper halves of the affine diagrams, obtained by cutting the affine diagrams across its $2j$ through-lines. Each upper half is then called a {\em link state}, and for simplicity the ``half'' through-lines attached to the free sites on the outer boundary (or top side of the framing rectangle) are still called through-lines. The phase $\e^{\i\phi/2}$ (resp.\ $\e^{-\i\phi/2}$) is now attributed each time one of these through-lines moves through the periodic boundary condition of the framing rectangle in the rightward (resp.\ leftward) direction. With these conventions, it is readily seen that the Temperley--Lieb algebra action obtained by stacking the affine diagrams on top of the link states gives rise to exactly the same representations $\AStTL{j}{\e^{\i\phi}}$ as defined above.
 
% in front of the (so-called standard) diagram with the same outer boundary (of $2L$ sites) and no through lines winding the annulus.
 The dimensions of these modules $\AStTL{j}{\e^{\i\phi}}$ over $\ATL{N}(\m)$ are then easily found by counting the link states. They are given by 
 \begin{equation}\label{eq:dj}
 \hat{d}_{j}=
 \binom{N}{\frac{N}{2}+j}%,\qquad \mbox{for } j>0 \,,
 \end{equation}
for the case $j>0$,
and we shall come back to the case $j=0$ below.
Note that these dimensions do not depend on $\phi$ (but representations with
different $\e^{\i\phi}$ are not isomorphic).
% We use in the following the notation 
%% $r_{j,\e^{2iK}}$
These standard modules $\AStTL{j}{\e^{\i\phi}}$ are known also as
{\em cell $\ATL{N}(\m)$-modules}~\cite{GL}.
%  with fixed number $2j$ of through lines and pseudomomentum $\e^{2iK}$. 
 
We now parametrize $\m=\q+\q^{-1}$. The standard modules $\AStTL{j}{\e^{\i\phi}}$ are irreducible for generic values of $\q$ and $\phi$.
However, degeneracies appear whenever the following {\em resonance criterion} is satisfied \cite{MartinSaleur1,GL}:\footnote{In \cite{GL} this criterion appears with some extra liberty in the form of certain $\pm$ signs, but we shall not need these signs here.}
% \linnea{from what I remember, this condition is a bit simplified? Jonathan had a ref with another version (Graham-Lehrer conditions)}
%
 \begin{eqnarray}\label{deg-st-mod}
 \e^{\i\phi}&=&\q^{2j+2k},\qquad
 \mbox{for } k > 0 \mbox{ integer} \,.
 \end{eqnarray}
 The representation $\AStTL{j}{\q^{2j+2k}}$ then becomes reducible, and contains a submodule isomorphic to 
 $\AStTL{j+k}{\q^{2j}}$. The quotient is generically irreducible, with
 dimension %\lawrence{$j,k>0$?}
 \begin{equation} \label{bard}
  \bar{d}_j \equiv \hat{d}_j-\hat{d}_{j+k} %\,, \qquad \mbox{for } j > 0  \,.
 \end{equation}
for the case $j>0$. When $\q$ is a root of unity, there are infinitely many solutions to \eqref{deg-st-mod}, leading to a complex pattern of degeneracies the discussion of which we postpone to another paper \cite{fullynongenericpaper}.

The case $j=0$ is particular. There is no pseudomomentum, but representations are still characterized by a parameter other than $j$, which now specifies the weight given to non-contractible loops. (Non-contractible loops are not possible for $j > 0$.) Parametrizing this weight as $z+z^{-1}$, the corresponding standard module of  $\ATL{N}(\m)$ is denoted  $\AStTL{0}{z^2}$. This module is isomorphic to $\AStTL{0}{z^{-2}}$. If we make the identification $z = \e^{\i\phi/2}$, the resonance criterion \eqref{deg-st-mod} still applies.

%For  the Jones--Temperley--Lieb algebra $\rJTL{N}(m)$, the rule that winding through-lines can simply be unwound means that the pseudomomentum must satisfy 
%$jK\equiv 0~\hbox{mod}~\pi$ \cite{Jones}.
% All possible values of the parameter $z^2=\e^{2iK}$ are thus $j$-th
% roots of unity ($z^{2j}=1$,~\cite{Green}).
It is natural to require that $z+z^{-1}=\m$, so that contractible and non-contractible loops get the same weight. Imposing this
leads to the module $\AStTL{0}{\q^2}$ which is reducible
even for generic $\q$. Indeed, \eqref{deg-st-mod} is satisfied with $j=0$, $k=1$, and hence $\AStTL{0}{\q^2}$ contains a submodule isomorphic to
$\AStTL{1}{1}$. Taking the quotient $\AStTL{0}{\q^2}/\AStTL{1}{1}$ leads to a simple module  for generic $\q$ which we denote 
by~$\bAStTL{\!\! 0}{\q^2}$. This module is isomorphic to $\bAStTL{\!\! 0}{\q^{-2}}$. It has dimension
\begin{equation}
 \bar{d}_0=\binom{N}{\frac{N}{2}}-\binom{N}{\frac{N}{2}+1} \,,
\end{equation}
in agreement with the general formula \eqref{bard} for $k=1$, using also \eqref{eq:dj} for $j=0$.

The difference between $\AStTL{0}{\q^2}$ and $\bAStTL{\!\! 0}{\q^2}$ has a simple geometrical meaning: in the second case,  one only keeps track of which sites are connected to which in the diagrams, while in the first case, one also keeps information of how the connectivities wind around the periodic direction of the annulus (the ambiguity does not arise when there are through-lines propagating). Formally, this corresponds to the existence of a surjection $\psi$ between different quotients of the $\ATL{N}$ algebra:
\begin{equation}\label{psi-ex}
  \xymatrix@C=8pt@R=1pt@M=-5pt@W=-2pt{
  &&	\mbox{}\quad\xrightarrow{{\mbox{}\quad\psi\quad} }\quad &\\
  & {
 \begin{tikzpicture}
  %%%%%%%%% Frame %%%%%%%%
 	\draw[thick, dotted] (-0.05,0.5) arc (0:10:0 and -7.5);
 	\draw[thick, dotted] (-0.05,0.55) -- (2.65,0.55);
 	\draw[thick, dotted] (2.65,0.5) arc (0:10:0 and -7.5);
	\draw[thick, dotted] (-0.05,-0.85) -- (2.65,-0.85);
%%%%%%%%%%%%	
	\draw[thick] (0,0.1) arc (-90:0:0.5 and 0.4);
	\draw[thick] (0,-0.1) arc (-90:0:0.9 and 0.6);
	\draw[thick] (2.6,-0.1) arc (-90:0:-0.9 and 0.6);
	\draw[thick] (2.6,0.1) arc (-90:0:-0.5 and 0.4);
%%%%%%%%%%%%%	
	\draw[thick] (0.5,-0.8) arc (0:90:0.5 and 0.5);
	\draw[thick] (1.8,-0.8) arc (0:180:0.5 and 0.5);
	\draw[thick] (2.1,-0.8) arc (0:90:-0.5 and 0.5);
	\end{tikzpicture}
\quad}&
	%\ar[r]
%	\xrightarrow{{\mbox{}\quad\psi\quad} }
%	\ar@{->}[r]^(0.45){\mbox{}\qquad\psi\qquad} 
	& {	\quad
  \begin{tikzpicture}
 %%%%%%%%% Frame %%%%%%%%
 	\draw[thick, dotted] (-0.05,0.5) arc (0:10:0 and -7.5);
 	\draw[thick, dotted] (-0.05,0.55) -- (2.65,0.55);
 	\draw[thick, dotted] (2.65,0.5) arc (0:10:0 and -7.5);
	\draw[thick, dotted] (-0.05,-0.85) -- (2.65,-0.85);
%%%%%%%%%%%%	
	\draw[thick] (0.5,0.5) arc (-180:0:0.8 and 0.56);
%	\draw[thick] (1.0,0.5) arc (0:10:-44.5 and -7.6);
%	\draw[thick] (1.65,0.5) arc (0:10:38.5 and -7.6);
	\draw[thick] (0.8,0.5) arc (-180:0:0.5 and 0.4);
%	\draw[thick] (2.6,0.2) arc (-90:0:-0.5 and 0.3);
%%%%%%%%
	\draw[thick] (2.1,-0.8) arc (0:180:0.8 and 0.56);
	\draw[thick] (1.8,-0.8) arc (0:180:0.5 and 0.4);
	\end{tikzpicture}}
%  &&&
  }
\end{equation}
The definition of link patterns as the upper halves of the affine diagrams also makes sense for $j=0$. The representation $\AStTL{0}{\q^2}$ requires keeping
track of whether each pairwise connection between the sites on the outer boundary (or top side of the framing rectangle) goes through the periodic boundary condition, whereas in the quotient module $\bAStTL{\!\! 0}{\q^2}$ this information is omitted. In both cases, it is easy to see that the number of link states coincides with the dimension $\hat{d}_0$ %\lawrence{$\hat d_0$? The notation $d_0$ is not introduced anywhere. Neither is $\hat d_0$, for that matter.} 
or $\bar{d}_0$, respectively.

\subsection{A note on indecomposability and $\bAStTL{\!\! 0}{\q^{\pm 2}}$}\label{indecomposability}

%Note that in this section we shall not smear out the twist.

%In the loop representation, or ``link state representation,'' the states correspond to the different possible upper halves of the affine diagrams defined above, and the Temperley--Lieb algebra action is given by stacking the affine diagrams on top of the states. 

Consider the standard module $\AStTL{0}{\q^{\pm 2}}$ for $N=2$, i.e., the loop model for a two-site system, in the sector with no through-lines and with non-contractible loops given the same weight $\m=\q+\q^{-1}$ as contractible ones. We emphasize that since $\q$ only enters in the combination $\q+\q^{-1}$, the sign of the exponent ($\q^2$ versus $\q^{-2}$) is immaterial, motivating the notation $\AStTL{0}{\q^{\pm 2}}$.

Let us first write the two elements of the Temperley--Lieb algebra in the basis of the two link states $\loopU$ and $\loopJL$:
\begin{equation}
 e_1=\m\begin{pmatrix}
1&1\\
0&0
\end{pmatrix},~e_2=\m\begin{pmatrix}
0&0\\
1&1
\end{pmatrix}.
\end{equation}
Clearly $e_1(\loopU-\loopJL)=e_2(\loopU-\loopJL)=0$. Meanwhile, at $N=2$ the action of $e_1 $ and $e_2$ on the single state $\loopII$ in $\AStTL{1}{1}$ is zero by definition, since the number of through-lines would decrease. By comparison we see that $\AStTL{0}{\q^{\pm 2}}$ admits a submodule, generated by $(\loopU-\loopJL)$, that is isomorphic to $\AStTL{1}{1}$. Pictorially (using what is technically called a Loewy diagram) we have 
\begin{equation}
\label{structstdmod}
\text{$\AStTL{0}{\q^{\pm 2}}$ :} \quad
\begin{tikzpicture}[auto, node distance=0.6cm, baseline=(current  bounding  box.center)]
  \node (node1)[align=center] {$\bAStTL{\!\! 0}{\q^{\pm 2}}$\\ } ;
  \node (node2) [below  = of node1,align=center]	{\\$\AStTL{1}{1}$};
    \draw[->] (node1) edge  (node2); 
\end{tikzpicture},
\end{equation}
where the bottom is a  submodule and the top a  quotient module. The arrow indicates that within the standard module $\AStTL{0}{\q^{\pm 2}}$ a state in $\AStTL{1}{1}$ can be reached from a state in $\bAStTL{\!\! 0}{\q^{\pm 2}}$ through the action of the Temperley--Lieb algebra, but the opposite is impossible.

\section{Discrete Virasoro algebra in the Potts model}\label{sec:disVir}

\subsection{Hamiltonian and representations}

While the Potts model is often defined as an isotropic lattice model on the square lattice (we have taken this point of view in section~\ref{sec:loops-clusters}), it is well known that the corresponding universality class extends to a critical manifold with properly related horizontal and vertical couplings. The case  of an infinitely large vertical coupling (we take the vertical direction as imaginary time) leads to the Hamiltonian limit where the model dynamics  is described by a Hamiltonian instead of a transfer matrix. This is the limit we shall restrict to in the following, in order to match as closely as possible  the lattice model to the formalism of radial quantization of the continuum CFT.  

The Hamiltonian describing the $Q$-state Potts model can be expressed  using Temperley--Lieb generators \cite{KooSaleur} 
\begin{equation}\label{H_phi}
\HN = -\frac{\gamma}{\pi \sin \gamma} \sum^{N}_{j=1} (e_j-e_\infty) 
\end{equation}
for $N$ even. Here, the prefactor is chosen to ensure relativistic invariance at low energy (see the next section), and we recall that $\gamma\in [0,\pi)$ %\lawrence{$\gamma\in(\pi/2,\pi]$ requires $x\in[0,1)$, which seems to be excluded by equation (4). See also equation (1), which does not have this restriction either.} 
is defined through $\q=\e^{\i\gamma}$, so $\m = \sqrt{Q} \in (-2,2]$.   $e_\infty$ is a constant energy density 
 added to cancel out extensive contributions to the ground state. Its value 
% (as discussed below) 
 is given by
\begin{equation}\label{e_inf}
e_\infty = \sin \gamma \: I_0,
\end{equation}
with $I_0$ being given by the integral 
\begin{equation}
I_0=\int^\infty_{-\infty} \frac{\sinh(\pi-\gamma)t }{\sinh(\pi t)\cosh(\gamma t)}\mathrm{d}t.
\end{equation}
In (\ref{H_phi}), the $e_j$ can be  taken to act in different representations of the $\ATL{N}(\m)$ algebra. The original representation, used for $Q$ integer, uses matrices $Q^L\times Q^L$, 
%\jesper{shouldn't that be $Q^L \times Q^L$?} 
corresponding to a chain of $L = N/2$ Potts spins. The Fortuin--Kasteleyn formulation of the Potts model for $Q$ real can be obtained by using instead the loop formulation discussed in the previous section. 

It is also known that the XXZ or vertex model representation of  $\ATL{N}$ could be used instead of the loop representation with ``very similar results.'' This point has to be considered with a lot of caution however: while the algebra is always the same $\ATL{N}$, the representations (i.e., using loops/clusters or spins/arrows in the transfer matrix)  are not necessarily isomorphic. The following subsection discusses this point in more detail.

Note that when taking one of the standard modules  $\AStTL{j}{\e^{\i\phi}}$ as the representation of choice, the value of the energy density $e_\infty$ is independent of $\phi$.

\subsection{A note on the XXZ representation}

In the 
XXZ representation, the $e_j$ act on $(\mathbb{C}^{2})^{\otimes N}$ with 
\begin{equation}
e_j = -\sigma_j^{-}\sigma_{j+1}^{+}-\sigma_j^{+}\sigma_{j+1}^{-}
-\frac{\cos\gamma}{2}\sigma_j^{z}\sigma_{j+1}^{z} -\frac{\i\sin\gamma}{2}(\sigma_j^{z}-\sigma_{j+1}^{z})+\frac{\cos\gamma}{2} \,,
\label{TLspin}
\end{equation}
where the $\sigma_j$ are the usual Pauli matrices, so the Hamiltonian is the familiar XXZ spin chain 
\begin{equation}\label{Pauliham}
\HN = \frac{\gamma}{2 \pi \sin \gamma}\sum^{N}_{j=1}   \left[ \sigma^x_j \sigma^x_{j+1} + \sigma^y_j \sigma^y_{j+1} + \cos\gamma \, (\sigma^z_j \sigma^z_{j+1} - 1 ) + 2 e_\infty \right]
\end{equation}
with anisotropy parameter
\begin{equation} \label{anisotropydelta}
 \Delta = \cos \gamma \,.
\end{equation}

In the usual basis where $\left[ \begin{smallmatrix} 1\\ 0 \end{smallmatrix} \right]$ corresponds to spin up in the $z$-direction at a given site, the Temperley--Lieb generator $e_j$ acts on spins $j, j+1$ (with periodic boundary conditions) as 
\begin{equation} e_j = \cdots\otimes\mathbf{1}\otimes
\begin{pmatrix}
0 & 0 & 0 & 0 \\
0 & \q^{-1} & -1 & 0\\
0& -1 & \q & 0\\
0 & 0 & 0 & 0 
\end{pmatrix} \otimes \mathbf{1}\otimes\cdots .
\end{equation}

It is also possible to introduce a twist in the spin chain without changing the expression (\ref{H_phi}), 
 by modifying the expression of the Temperley--Lieb generator acting between first and last spin with a twist parametrized by $\phi$.  
In terms of the Pauli matrices, this twist imposes the boundary conditions $\sigma^z_{N+1}=\sigma^z_1$ and $\sigma^{\pm}_{N+1}=\e^{\mp \i \phi} \sigma^{\pm}_1$.
%
%
%
%For technical reasons,  we will later on ``smear out'' the twist by taking $\phi/N$ for \emph{each} Temperley--Lieb generator:
%\begin{equation}\label{TLmatrix_spin} e_j = \cdots \otimes\mathbf{1}\otimes
%\begin{pmatrix}
%0 & 0 & 0 & 0 \\
%0 & \q^{-1} & -\e^{\i\phi/N} & 0\\
%0 & -\e^{-\i\phi/N} & \q & 0\\
%0 & 0 & 0 & 0 
%\end{pmatrix}\otimes \mathbf{1}\otimes \cdots \,.
%\end{equation}
%This is equivalent, and is done in order to preserve invariance under the usual translation operator, which will be useful in the sections below.  Note that the value of the energy density $e_\infty$ is independent of $\phi$ and remains given by (\ref{e_inf}). 
In the generic case, the XXZ model with magnetization $S_z=j$ and twist $\e^{\i\phi}$ provides a representation of the module $\AStTL{j}{\e^{\i\phi}}$. This is not true in the non-generic case---see below.

The  XXZ and the loop representations share many common features. Most importantly, the value of the ground-state energy is the same for both, and so is the value of the ``sound velocity'' determining the correct multiplicative normalization of the Hamiltonian in (\ref{H_phi}). This occurs because the ground state is found in the same module $\AStTL{j}{\e^{\i\phi}}$ for both models, or in closely related modules for which the extensive part of the ground state-energy  (and thus, the constant $e_\infty$) is the same. In general, of course, the XXZ and loop  representations involve mostly {\sl different modules}. For the XXZ chain, the modules appearing in the spin chain depend on the twist angle $\phi$. For the loop model, the modules depend on the rules one wishes to adopt to treat non-contractible loops, or lines winding around the system. If everything were always both generic and non-degenerate, a study of the physics in each irreducible module $\AStTL{j}{\e^{\i\phi}}$ would be enough to answer all questions about all $\ATL{N}(\m)$ models (as well as the corresponding Virasoro modules obtained in the scaling limit, see below). It turns out, however, that degenerate cases are always relevant to the physical problems at hand, and  the modules  can now ``break up'' or ``get glued'' differently. 

To illustrate the latter point, we  consider instead the XXZ representation with $S_z=0$ and twisted boundary conditions $\e^{\i\phi}=\q^{-2}$, here without ``smearing'' of the twist. 
We chose the basis of this sector as $u=|\! \uparrow\downarrow\rangle$ and $v=|\! \downarrow\uparrow\rangle$. We have then 
\begin{eqnarray}\label{stmod}
e_1=\begin{pmatrix}
\q^{-1}&-1\\
-1&\q
\end{pmatrix}\,, \quad
e_2=\begin{pmatrix}
\q &-\q^2\\
-\q^{-2}&\q^{-1}
\end{pmatrix} \,. \end{eqnarray}
We find that $e_1(u+\q^{-1}v)=e_2(u+\q^{-1}v)=0$, while $e_1(u-\q v)=(\q+\q^{-1})(u-\q v)$ and $e_2(u-\q v)=(\q+\q^{-1})(u-\q v)+(\q^3-\q^{-1})(u+\q^{-1}v)$. Now consider the module $\AStTL{1}{1}$, which is the spin $S_z=1$ sector with no twist, where $e_1=e_2=0$. By comparison, we see that $u+\q^{-1}v$ generates a module isomorphic to $\AStTL{1}{1}$. Meanwhile, $u-\q v$ does not generate a submodule, since $e_2$ acting on this vector yields a component along $u+\q^{-1}v$. However, if we quotient by $u+\q^{-1}v$, we get a one-dimensional module where $e_1$ and $e_2$ act as $\q+\q^{-1}$, which is precisely the module $\bAStTL{0}{\q^{\pm2}}$. We thus get the same result as for the loop model, i.e., the structure \eqref{structstdmod} of the standard module.

Considering instead $\e^{\i\phi}=\q^2$, we have  
\begin{eqnarray}
\nonumber\\
~e_1=\begin{pmatrix}
\q^{-1}&-1\\
-1&\q
\end{pmatrix}\,, \quad e_2=\begin{pmatrix}
\q &-\q^{-2}\\
-\q^{2}&\q^{-1}
\end{pmatrix} \,.
\end{eqnarray}
We see that  $e_1(u-\q v)=e_2(u-\q v)=(\q+\q^{-1})(u-\q v)$, while $e_1(u+\q^{-1}v)=0$ and $e_2(u+\q^{-1}v)=(\q-\q^{-3})(u-\q v)$. Hence this time we get a proper $\bAStTL{\!\! 0}{\q^{\pm 2}}$ module, while we only get $\AStTL{1}{1}$ as a quotient module. The corresponding structure can be represented as 
\begin{equation}\label{costmod}
\text{$\coAStTL{0}{\q^{\pm 2}}$ :} \quad
\begin{tikzpicture}[auto, node distance=0.6cm, baseline=(current  bounding  box.center)]
  \node (node1)[align=center] {$\bAStTL{0}{\q^{\pm 2}}$\\ } ;
  \node (node2) [below  = of node1,align=center]	{\\$ \AStTL{1}{1}$};
    \draw[<-] (node1) edge  (node2); 
\end{tikzpicture}.
\end{equation}
Observe that the shapes  in \eqref{structstdmod}
%\eqref{stmod}\lawrence{\eqref{structstdmod}?} 
and \eqref{costmod} are related by inverting the (unique in this case) arrows; the module in (\ref{costmod}) is referred to as  ``co-standard,'' and we indicate this dual nature by placing a tilde on top of the usual $\AStTL{0}{\q^{\pm 2}}$ notation for the standard module.

%To emphasize that in the XXZ chain the standard module $\AStTL{0}{\q^{\pm2}}$ corresponds to the twisted boundary condition $\e^{\i\phi}=\q^{-2}$, while the co-standard module $\coAStTL{0}{\q^{\pm2}}$ corresponds to the twisted boundary condition $\e^{\i\phi}=\q^{2}$, we introduce the notations $\AStTL{0}{\q^{-2}} \equiv \AStTL{0}{\q^{\pm2}}$ and $\AStTL{0}{\q^{2}} \equiv \coAStTL{0}{\q^{\pm2}}$. Later on, we shall write diagrams of the type above as
%\begin{equation}\label{XXZ_W0q2}
%\text{$\AStTL{0}{\q^{- 2}}$:}
%\begin{tikzpicture}[auto, node distance=0.5 and 0.3cm, baseline=(current  bounding  box.center)]
%  \node (node1) [align=center]{$[0,\q^{-2}]$};
%  \node (node2) [below = of node1, align=center]{$[1,1]$};
%  \draw[-latex]  (node1) edge  (node2);
%\end{tikzpicture},
%\hspace*{1cm}
%\text{$\AStTL{0}{\q^{ 2}}$:}
%\begin{tikzpicture}[auto, node distance=0.5 and 0.3cm, baseline=(current  bounding  box.center)]
%  \node (node1) [align=center]{$[0,\q^2]$};
%  \node (node2) [below = of node1, align=center]{$[1,1]$};
%  \draw[latex-]  (node1) edge  (node2);
%\end{tikzpicture},
%\end{equation}
%where it is implicit that any relevant quotients have been taken.

In summary, from this short exercise we see that while in the generic case the loop and spin representations are isomorphic, this equivalence  breaks down in the  non-generic case, where $\phi$ is such that the resonance criterion \eqref{deg-st-mod} is met. 
Only standard modules are encountered in the loop model while in the XXZ spin chain both standard and co-standard are encountered. This feature extends to larger $N$: see \cite{GSJS}. 
We note that in the case where $\q$ is also a root of unity, the distinction between the two representations becomes even more pronounced: in this case the modules in the XXZ chain are no longer isomorphic to standard \emph{or} co-standard modules. This will be further explored in a subsequent paper \cite{fullynongenericpaper}.

\subsection{The discrete Virasoro algebra}

Following \eqref{H_phi} we define the Hamiltonian density as %\jesper{Motivate by \eqref{H_phi}.} 
${\mathcal h}_j=-(\gamma/\pi\sin\gamma)e_j$. From the Hamiltonian density we then construct a lattice momentum density ${\mathcal p}_j = \i[{\mathcal h}_j,\mathcal{h}_{j-1}]=$ \phantom{ }%force a line break if needed
 $ -\i (\gamma/\pi\sin\gamma)^2 [e_{j-1},e_j]$ using energy conservation \cite{Vidal}. We can then introduce a momentum operator $\PN$ as 
\begin{equation}\label{P_phi}
\PN = -\i \left( \frac{\gamma}{ \pi \sin \gamma} \right)^2  \sum^{N}_{j=1} [e_j, e_{j+1} ] \,.
\end{equation}
%\linnea{Notational collision with $h_j$ in Appendix \ref{UnitaryXXZ} to be resolved}

From ${\mathcal h}_j$  and ${\mathcal p}_j$ we build components of a discretized stress tensor as
\begin{subequations}
\begin{eqnarray}
  {\mathcal T}_j=\frac{1}{2}(\mathcal{h}_j+\mathcal{p}_j) \,, \\
  \bar{\mathcal T}_j=\frac{1}{2}(\mathcal{h}_j-\mathcal{p}_j) \,,
 \end{eqnarray}
\end{subequations}
from which we wish to construct discretized versions of the Virasoro generators as the Fourier modes \cite{Vidal}. This construction gives rise to the Koo--Saleur generators%
\footnote{In this paper we consistently use calligraphic fonts for the lattice analogues of some key quantities: the Hamiltonian ${\mathcal H}$, the momentum ${\mathcal P}$---with their corresponding densities ${\mathcal h}_j$ and ${\mathcal p}_j$---, the Virasoro generators $\KSgen_n$, $\bar{\KSgen}_n$ and the stress-energy tensor ${\mathcal T}$, $\bar{\mathcal T}$. The corresponding continuum quantities are denoted by Roman fonts: $H$, $P$ and $L_n$, $\bar{L}_n$, as well as $T$, $\bar{T}$. The question of whether we have the convergence $\KSgen_n, \bar{\KSgen}_n \mapsto L_n, \bar{L}_n$ in the continuum limit $N \to \infty$---and if so, what is the precise nature of this convergence---will be dealt with in more detail in the companion paper \cite{GSJS}. }

\begin{subequations}
\label{generators}
\begin{eqnarray}
\KSgen_n[N] &=& \frac{N}{4\pi} \left[ -\frac{\gamma}{\pi \sin \gamma}  \sum^{N}_{j=1} \e^{2\pi \i nj/N} \left( e_j - e_\infty + \frac{\i\gamma}{\pi \sin \gamma} [e_j, e_{j+1} ] \right)\right] + \frac{c}{24} \delta_{n,0} \,, \\
\bar{\KSgen}_n[N] &=& \frac{N}{4\pi} \left[ -\frac{\gamma}{\pi \sin \gamma}  \sum^{N}_{j=1} \e^{-2\pi \i nj/N} \left( e_j - e_\infty - \frac{\i\gamma}{\pi \sin \gamma} [e_j, e_{j+1} ] \right)\right] + \frac{c}{24} \delta_{n,0} \,.
\end{eqnarray}
\end{subequations}
which were first derived via other means in \cite{KooSaleur}.
Here, the crucial additional ingredient is the central charge, which is given by \eqref{centralc}.
%\linnea{discussion edited here}
%
%
%\begin{equation}
%c=1-{6\over x(x+1)} \,, \label{c_value}
%\end{equation}
%
%where we recall the parametrization \eqref{gammaparam}.
%The choice (\ref{c_value}) is known to be relevant for models with Hamiltonian (\ref{H_phi}), such as the ferromagnetic $Q$-state Potts model with $Q=\m^2$. 
Note that the identification of the central charge is actually a subtle question, and may be affected by boundary conditions, as discussed further in \cite{GSJS}. %We discuss this in detail in the next section. \linnea{do we, though?}

\subsection{$\VirN$ modules in the Potts model CFT:  the non-degenerate case}

We recall once more that throughout this paper $\q$ is assumed to take generic values (not a root of unity).
Whenever $\phi$ is such that the resonance criterion \eqref{deg-st-mod} is not met we say that $\phi$ is {\em generic};
and when \eqref{deg-st-mod} is satisfied $\phi$ is referred to as {\em non-generic}.

Since $\q$ is generic throughout, both $c$ and its parametrization $x$ from \eqref{centralc} take generic, irrational values.
The conformal weights may be degenerate or not, depending on the lattice parameters. In the non-degenerate case, which corresponds to generic lattice parameters (the opposite does not always hold) it is natural to expect that the Temperley--Lieb module decomposes accordingly into a direct sum of Verma modules,
\begin{equation}\label{identcontl}
\AStTL{j}{\e^{\i\phi}}\mapsto \bigoplus_{e\in\mathbb{Z}} \Verma{e-e_\phi,-j}\otimes\Vermab{e-e_\phi,j} \,. 
\end{equation}
The symbol $\mapsto$ means that action of the lattice Virasoro generators {\sl restricted to scaling states} on $\AStTL{j}{\e^{\i\phi}}$ corresponds to the decomposition on the right-hand side when $N\to\infty$. This statement is discussed in considerable detail in our paper \cite{GSJS}. Throughout this paper we systematically place a bar above the right tensorand in expressions of the form ${\sf V} \otimes \bar{\sf V}$, as a reminder that this refers to the $\overline{\rm Vir}$ algebra.
%
%Note that to make notation lighter, we are not indicating explicitly that in $\Verma{}\otimes\Vermab{}$ the right tensorand is for the $\overline{\Vir}$ algebra: this should always be obvious from the context. 
%\jesper{Well, I thought that the new notation was actually meant to use the bars systematically!}

Recall that a Verma module is a highest-weight representation of the Virasoro algebra
\begin{equation}\label{Virasoro}
[L_m,L_n]=(m-n)L_{m+n}+\frac{c}{12}m(m^2-1)\delta_{m+n,0} \,,
\end{equation}
generated by a highest-weight 
vector $|h\rangle$ satisfying $L_n|h\rangle=0,n>0$,  and for which all the descendants
\begin{equation}
 L_{-n_1}\ldots L_{-n_k}|h\rangle \,, \mbox{ with } 0<n_1\leq n_2\leq \cdots \leq n_k \mbox{ and } k>0
\end{equation}
are considered as independent, subject only to the commutation relations \eqref{Virasoro}. %
In the non-degenerate case where the Verma module is irreducible, it is the only kind of module that can occur, motivating the identification in \eqref{identcontl}. We note that this identification is independent of whether we consider the loop model or the XXZ spin chain.

\subsection{The choices of metric. Duality}

It is observed in  \cite{GSJS} how the  XXZ chain can be considered in a precise way as a lattice analogue of the twisted free boson theory. It is well known in the latter case that two natural scalar products can be defined. The first one---which is positive definite---corresponds to the continuum limit of the ``native'' %\lawrence{``na{\"\inodot}ve''?} 
positive-definite scalar product for the spin chain, and, in terms of the free boson current modes, 
corresponds to choosing 
$a_n^\dagger = a_{-n}$. A crucial observation is that for this scalar product  $L^\dagger_n \neq L_{-n}$.  This means that norm squares of descendants cannot be obtained using  Virasoro algebra commutation relations. 

The second scalar product corresponds to the conjugation $\loopdagger$ with $L_n^\loopdagger$ simply given by $L^\loopdagger_n = L_{-n}$. This ``conformal scalar product'' is known to correspond \cite{DJS,VJS,VGJS}, on the lattice, to a modified scalar product in the XXZ spin chain where $\q$ is treated as a formal, self-conjugate parameter \cite{CJS}.

The loop model can be naturally equipped with two scalar products as well. Choosing basic loop states to be mutually orthogonal and of unit norm-square defines a ``native'' %\lawrence{``na{\"\inodot}ve''?} 
positive-definite scalar product for which the Temperley--Lieb generators, the transfer matrix and the Hamiltonian are not self-adjoint, while for  the lattice Virasoro generators $\KSgen_n^\dagger\neq \KSgen_{-n}$: we will denote this scalar product by $\spininner{-}{-}$ (whenever necessary, will use the same notation for lattice and continuum quantities).  

Meanwhile, we can also introduce the ``loop scalar product'' $\loopinner{-}{-}$, obtained by gluing the mirror image of one link state on top of the other and evaluating the result according to certain rules that we now describe. First, unless all through-lines connect through from bottom to top the result is zero. Considering a smeared-out phase we also take into account the weight of straightening the connected through-lines: a through-line that has moved to the right (left) is assigned the weight $\e^{\i\phi/2N}$ ($\e^{-\i\phi/2N}$) for each step. Each contractible loop carries the weight $\m=\q+\q^{-1}$, while each non-contractible loop carries the weight $\e^{\i\phi/2}+\e^{-\i\phi/2}$.   To illustrate this scalar product we take the following examples, where the solid lines around the rightmost diagrams signify that we assign them a value according to the aforementioned rules:
\begin{equation}
\begin{split}
\left(\vcenter{\hbox{\begin{tikzpicture}
 \newcommand{\dist}{0.2}
	\draw[thick] (0,0) arc (-180:0:\dist);
	\draw[thick] (4*\dist,0) -- (4*\dist,-2*\dist);
	\draw[thick] (6*\dist,0) -- (6*\dist,-2*\dist);
%%%%%%%%% Frame %%%%%%%%
\draw[thick, dotted] ($(current bounding box.north east) + (0.05+\dist,0.05)$) rectangle ($(current bounding box.south west)+ (-0.05-\dist,-0.05)$);
\end{tikzpicture}	} } 	
	,
\vcenter{\hbox{\begin{tikzpicture}
 \newcommand{\dist}{0.2}
	\draw[thick] (4*\dist,0) arc (-180:0:\dist);
	\draw[thick] (2*\dist,0) -- (2*\dist,-2*\dist);
	\draw[thick] (0,0) -- (0,-2*\dist);
%%%%%%%%% Frame %%%%%%%%
\draw[thick, dotted] ($(current bounding box.north east) + (0.05+\dist,0.05)$) rectangle ($(current bounding box.south west)+ (-0.05-\dist,-0.05)$);
\end{tikzpicture}	} } 	
	\right)
	&=    \;
	\vcenter{\hbox{\begin{tikzpicture}
 \newcommand{\dist}{0.2}
 \begin{scope}[yscale=-1,xscale=1]
	\draw[thick] (0,0) arc (-180:0:\dist);
	\draw[thick] (4*\dist,0) -- (4*\dist,-2*\dist);
	\draw[thick] (6*\dist,0) -- (6*\dist,-2*\dist);
 \end{scope}
	\draw[thick] (4*\dist,0) arc (-180:0:\dist);
	\draw[thick] (2*\dist,0) -- (2*\dist,-2*\dist);
	\draw[thick] (0,0) -- (0,-2*\dist);
%%%%%%%%% Frame %%%%%%%%
\draw ($(current bounding box.north east) + (0.05+\dist,0.05)$) rectangle ($(current bounding box.south west)+ (-0.05-\dist,-0.05)$);
\end{tikzpicture}	} } 
= 0 .\\
%%%%%%%%%%%%%%%
\left(\vcenter{\hbox{\begin{tikzpicture}
 \newcommand{\dist}{0.2}
	\draw[thick] (2*\dist,0) arc (-180:0:\dist);
	\draw[thick] (0,0) arc (-180:0:3*\dist);
	\draw[thick] (8*\dist,0) -- (8*\dist,-3*\dist);
	\draw[thick] (10*\dist,0) -- (10*\dist,-3*\dist);
%%%%%%%%% Frame %%%%%%%%
\draw[thick, dotted] ($(current bounding box.north east) + (0.05+\dist,0.05)$) rectangle ($(current bounding box.south west)+ (-0.05-\dist,-0.05)$);
\end{tikzpicture}	} } 	
	,
\vcenter{\hbox{\begin{tikzpicture}
 \newcommand{\dist}{0.2}
	\draw[thick] (2*\dist,0) arc (-180:0:\dist);
	\draw[thick] (6*\dist,0) arc (-180:0:\dist);
	\draw[thick] (0,0) -- (0,-2*\dist);
	\draw[thick] (10*\dist,0) -- (10*\dist,-2*\dist);
%%%%%%%%% Frame %%%%%%%%
\draw[thick, dotted] ($(current bounding box.north east) + (0.05+\dist,0.05)$) rectangle ($(current bounding box.south west)+ (-0.05-\dist,-0.05)$);
\end{tikzpicture}	} } 	
	\right)
	&=    \;
	\vcenter{\hbox{\begin{tikzpicture}
 \newcommand{\dist}{0.2}
 \begin{scope}[yscale=-1,xscale=1]
 	\draw[thick] (2*\dist,0) arc (-180:0:\dist);
	\draw[thick] (0,0) arc (-180:0:3*\dist);
	\draw[thick] (8*\dist,0) -- (8*\dist,-3*\dist);
	\draw[thick] (10*\dist,0) -- (10*\dist,-3*\dist);
 \end{scope}
	\draw[thick] (2*\dist,0) arc (-180:0:\dist);
	\draw[thick] (6*\dist,0) arc (-180:0:\dist);
	\draw[thick] (0,0) -- (0,-2*\dist);
	\draw[thick] (10*\dist,0) -- (10*\dist,-2*\dist);
%%%%%%%%% Frame %%%%%%%%
\draw ($(current bounding box.north east) + (0.05+\dist,0.05)$) rectangle ($(current bounding box.south west)+ (-0.05-\dist,-0.05)$);
\end{tikzpicture}	} } 
= \e^{4\i\phi/2N} \m .\\
%
%%%%%%%%%%%%%%%%%
%
\left(
\vcenter{\hbox{\begin{tikzpicture}
 \newcommand{\dist}{0.2}
	\draw[thick] (\dist,-\dist) arc (-90:0:\dist);
	\draw[thick] (4*\dist,0) arc (-180:0:\dist);
	\draw[thick] (9*\dist,-\dist) arc (-90:-180:\dist);
%%%%%%%%% Frame %%%%%%%%
\draw[thick, dotted] ($(current bounding box.north east) + (0.05,0.05)$) rectangle ($(current bounding box.south west)+ (-0.05,-0.05-1*\dist)$);
\end{tikzpicture}	} } 	
,
\vcenter{\hbox{\begin{tikzpicture}
 \newcommand{\dist}{0.2}
	\draw[thick] (2*\dist,0) arc (-180:0:\dist);
	\draw[thick] (0,0) arc (-180:0:3*\dist);
%%%%%%%%% Frame %%%%%%%%
\draw[thick, dotted] ($(current bounding box.north east) + (0.05+\dist,0.05)$) rectangle ($(current bounding box.south west)+ (-0.05-\dist,-0.05)$);
\end{tikzpicture}	} } 	
	\right)
	&=    \;
	\vcenter{\hbox{\begin{tikzpicture}
 \newcommand{\dist}{0.2}
 \begin{scope}[xshift=2*\dist cm]
	\draw[thick] (2*\dist,0) arc (-180:0:\dist);
	\draw[thick] (0,0) arc (-180:0:3*\dist);
 \end{scope}
 \begin{scope}[yscale=-1,xscale=1]
	\draw[thick] (\dist,-\dist) arc (-90:0:\dist);
	\draw[thick] (4*\dist,0) arc (-180:0:\dist);
	\draw[thick] (9*\dist,-\dist) arc (-90:-180:\dist);
 \end{scope}
%%%%%%%%% Frame %%%%%%%%
\draw ($(current bounding box.north east) + (0.05,0.05)$) rectangle ($(current bounding box.south west)+ (-0.05,-0.05)$);
\end{tikzpicture}	} } 
= (\e^{\i\phi/2}+\e^{-\i\phi/2}) \m .\\
\end{split}
\end{equation}
This ``loop scalar product'' is then extended by sesquilinearity to the whole space of loop states. The adjoint $U^\loopdagger$ of a word $U$ in the Temperley--Lieb algebra can be defined similarly by flipping the diagram representing it about a horizontal line, as in the following example:
\begin{equation}
\vcenter{\hbox{\begin{tikzpicture}
 \newcommand{\dist}{0.2}
 	\draw[thick] (2*\dist,0) arc (-180:0:\dist);
	\draw[thick] (0,0) arc (-180:0:3*\dist);
	\draw[thick] (8*\dist,0) -- (8*\dist,-3*\dist);
	\draw[thick] (10*\dist,0) -- (10*\dist,-3*\dist);
 \begin{scope}[xshift=0cm,yshift=-5*\dist cm]
 \begin{scope}[yscale=-1,xscale=1]
 	\draw[thick] (4*\dist,0) arc (-180:0:\dist);
	\draw[thick] (0,0) arc (-180:0:\dist);
	\draw[thick] (8*\dist,0) -- (8*\dist,-3*\dist);
	\draw[thick] (10*\dist,0) -- (10*\dist,-3*\dist);
 \end{scope}
 \end{scope}
%%%%%%%%% Frame %%%%%%%%
\draw[thick,dotted] ($(current bounding box.north east) + (0.05+\dist,0.05)$) rectangle ($(current bounding box.south west)+ (-0.05-\dist,-0.05)$);
\end{tikzpicture}	} } 	
	^\ddag
	=  \;\;
	\vcenter{\hbox{\begin{tikzpicture}
 \newcommand{\dist}{0.2}
 \begin{scope}[yscale=-1,xscale=1]
 	\draw[thick] (2*\dist,0) arc (-180:0:\dist);
	\draw[thick] (0,0) arc (-180:0:3*\dist);
	\draw[thick] (8*\dist,0) -- (8*\dist,-3*\dist);
	\draw[thick] (10*\dist,0) -- (10*\dist,-3*\dist);
 \begin{scope}[xshift=0cm,yshift=-5*\dist cm]
 \begin{scope}[yscale=-1,xscale=1]
 	\draw[thick] (4*\dist,0) arc (-180:0:\dist);
	\draw[thick] (0,0) arc (-180:0:\dist);
	\draw[thick] (8*\dist,0) -- (8*\dist,-3*\dist);
	\draw[thick] (10*\dist,0) -- (10*\dist,-3*\dist);
 \end{scope}
 \end{scope}
 \end{scope}
 %%%%%%%%% Frame %%%%%%%%
\draw[thick,dotted] ($(current bounding box.north east) + (0.05+\dist,0.05)$) rectangle ($(current bounding box.south west)+ (-0.05-\dist,-0.05)$);
 \end{tikzpicture}	} } 	.
\end{equation}
From this definition it is clear that the generators $e_i$ themselves are self-adjoint, and consequently $L^\loopdagger_n = L_{-n}$. It is well known that the loop scalar product is invariant with respect to the Temperley--Lieb action: 
$\loopinner{x}{Uy}=\loopinner{U^\loopdagger x}{y}$.
% $\loopinner{x}{Uy}=\loopinner{Ux}{y}$. \jesper{From what we just defined, shouldn't it be $\loopinner{x}{Uy}=\loopinner{U^\loopdagger x}{y}$?}
The loop scalar product is of course not positive definite. It is however not degenerate (provided $\m\neq 0$). Moreover, it  is known to go over to the conformal scalar product in the continuum limit \cite{DJS}. 

For a given module $\AStTL{}{}$ we can define the dual (conjugate)  module $\tAStTL{}{}$ by the map $u\to \loopinner{u}{-}$, i.e., by taking mirror images. %\jesper{This notation is not very clear, nor is it helpful for the reader to see something being defined ``as usual''. Maybe we could just say ``by taking mirror images''.} 
In general, %\jesper{add: ``we have the\lawrence{``an''?} isomorphism''?}
we have an isomorphism $\tAStTL{j}{\e^{\i\phi}}\approx \AStTL{j}{\e^{-\i\phi}}$. When $\AStTL{j}{\e^{\i\phi}}$ is reducible but indecomposable, the corresponding Loewy diagram has its arrows reversed, as illustrated in subsection (\ref{indecomposability}). The modules $\AStTL{j}{1}$ are self-dual. 

An important point is that, if a Temperley--Lieb module is self-dual, then since the Hamiltonian itself is, as well as the definition of scaling states, the action of the continuum limit of the Koo--Saleur generators should define an action on the scaling limit of the module that is also invariant under duality in the CFT. If both the Temperley--Lieb module and the $\VirN$ module are irreducible, this has no useful consequences. We shall soon see however that the $\AStTL{j}{1}$ modules, while irreducible, have a continuum limit which is not so. Self-duality of the $\AStTL{j}{1}$ implies invariance of the Loewy diagrams for the continuum limit with respect to reversal of the $\VirN$ arrows, with very interesting consequences.

\section{$\VirN$ modules in the Potts model CFT: the  degenerate case. Evidence from the lattice}\label{sec:degenerate}

In the degenerate cases
the conformal weights may take degenerate values $h=h_{r,s}$ with $r,s\in\mathbb{N}^*$, in which case a singular vector appears in the Verma module. By definition, a singular vector is a vector that is both a descendant and a highest-weight state. For instance, starting with $|h_{1,1}=0\rangle$ we see, by using the commutation relations \eqref{Virasoro}, that
\begin{equation}
L_1(L_{-1}|h_{1,1}\rangle)=2L_0|h_{1,1}\rangle=0 \,,
\end{equation}
while of course $L_nL_{-1}|h_{1,1}\rangle=0$
%$L_n|h_{1,1}\rangle=0$ \lawrence{$L_nL_{-1}|h=h_{1,1}\rangle=0$?} 
for $n>1$. Hence $L_{-1}|h_{1,1}\rangle$ is a singular vector. Under the action of the Virasoro algebra (recall that we are interested here in the full  $\VirN$ action, not just a single Virasoro action)  this vector generates a submodule. For $\q$ generic, this submodule is irreducible, and thus we have the decomposition
\begin{equation}\label{Vermadecomp}
\text{$\Verma{1,1}^{\rm (d)}$ :} \quad
\begin{tikzpicture}[auto, node distance=0.6cm, baseline=(current  bounding  box.center)]
  \node (node1)[align=center] {$\IrrV{1,1}$\\ } ;
  \node (node2) [below  = of node1,align=center]	{\\$\Verma{1,-1}$};
  
  \draw[->] (node1) edge  (node2); 
\end{tikzpicture},
\hspace*{1.5cm}
\end{equation}
%
%\hubert{I think a figure for the dual module would also be useful to help the reader}
where we have introduced the notation $\Verma{}^{\rm (d)}$ to denote the degenerate Verma module, and we also denote by $\IrrV{r,s}$ the irreducible Virasoro module (in this case, technically a ``Kac module''), with generating function of levels 
\begin{equation}
\label{Kac-module}
K_{r,s}=q^{h_{r,s}-c/24} \: {1-q^{rs}\over P(q)} \,.
\end{equation}
The subtraction of the singular vector at level $r s$ gives rise to a quotient module.%, and corresponds to the use of an open circle in the diagram \eqref{Vermadecomp}.

In cases of degenerate conformal weights, there is more than one possible module that could appear, and the identification in \eqref{identcontl} may no longer hold. Furthermore the identification now depends on the  representation of  $\ATL{N}(\m)$ one considers. We restrict here to the loop/cluster representation, while corresponding results about the XXZ representation can be found in  \cite{GSJS}.

\subsection{The loop-model case: without through-lines}

For the modules $\AStTL{0}{\q^{\pm 2}}$, this Verma  structure is seen even at finite size---see equation \eqref{structstdmod}.\footnote{Recall that in the loop representation, the loop weight is $\mathfrak{m} = \q + \q^{-1}$, with $\q^2 = \e^{\i \phi}$ adjusting the weight of non-contractible loops, so the sign of the twist $\phi$ is immaterial. In contrast,   in the XXZ case, one finds \cite{GSJS} that only one sign $\q^{-2}$ corresponds to a standard module, while the other $\q^{2}$ is a co-standard module.}
%Some numerical results for $\AStTL{0}{\q^{\pm 2}}$ in the loop-model case are shown in Appendix \ref{loop_twisted}: 
Using the numerical methods described in Appendix \ref{KS_Appendix}
we find that the corresponding loop states are never annihilated by the $A_{n,1}$ or $\bar{A}_{n,1}$ combinations of Virasoro generators. 

We recall now from Section~\ref{sec:stdmod} that the module $\AStTL{0}{\q^{\pm2}}$ appears in the loop model by keeping track of how points are connected across the periodic boundary condition. However, the Potts model where non-contractible loops have the same weight $\m$ as contractible ones naturally involves the quotient $\bAStTL{\!\! 0}{\q^{\pm 2}}$ for which there are no degenerate states on the lattice. The spectrum generating function for this module in the continuum limit is then
\begin{equation}
\bar{\FN}_{0,\q^{\pm 2}}=\FN_{0,\q^{\pm 2}}-\FN_{1,1}=\sum_{n=1}^\infty K_{n,1}\bar{K}_{n,1} \,, \label{MainRes4}
\end{equation}
where $K_{r,s}$ was defined in \eqref{Kac-module}.
It involves only Kac modules, so we have:

\vspace*{5pt}
\noindent\fbox{
\hspace*{5pt}\begin{minipage}{\linewidth-20pt}\em
\vspace*{5pt}
{\bf Quotient loop-model module without through-lines:} We have the scaling limit
\begin{equation}\label{result1}
\bAStTL{\! 0}{\q^{\pm 2}}\mapsto\bigoplus_{n=1}^\infty \IrrV{n,1}\otimes\IrrVb{n,1} \,.
\end{equation}
\vspace*{5pt}
\end{minipage}\hspace*{5pt}}
\vspace*{5pt}

\noindent
Note that this implies that the corresponding highest-weight states $|h,\bar{h}\rangle$ are now annihilated:
\begin{equation}
A_{n,1}|h_{n,1},h_{n,1}\rangle=\bar{A}_{n,1}|h_{n,1},h_{n,1}\rangle=0 \,.
\end{equation}
In particular, the ground state at central charge \eqref{centralc} is indeed annihilated by $L_{-1}$ and $\bar{L}_{-1}$, a satisfactory situation physically. Results for $\bAStTL{\!\! 0}{\q^{\pm 2}}$ are shown in Appendix \ref{loop_twisted}.

\subsection{The loop model case: $j>0$}
\label{loop_Wj}

For the modules $\AStTL{j}{1}$ with $j>0$, the numerical results in Appendix \ref{loop_lines} indicate that the  highest-weight states with conformal weight $h_{e,j}$ and $e>0$ are {\sl never} annihilated by the corresponding $A_{e,j}$ operators, whether in the chiral or antichiral sector. It would be tempting to conclude that the modules are now systematically of Verma type, but this is not possible. Indeed, recall that for $\q$ generic, the ATL (affine Temperley--Lieb) modules $\AStTL{j}{1}$ are irreducible and thus self-dual. The Virasoro generators being obtained as continuum limits of ATL generators should also obey this self-duality (see the discussion in section 4.3 of \cite{GRSV}).%
\footnote{This is a point well known in axiomatic CFT as well. Quoting \cite{GaberdielRunkel}: ``It is also worth mentioning that a non-degenerate bulk two-point function requires that ${\cal H}_{\rm bulk}$ is isomorphic to its conjugate representation ${\cal H}_{\rm bulk}^*$.  A necessary condition for this is that the composition series does not change when reversing all arrows [\ldots].''} Verma modules clearly do not, as their structure is not  invariant under reversal of the $\VirN$ action. To understand what might happen, let us discuss in more detail, as an example,  the case $j=2$. The generating function of levels shows a pair of primary fields 
\begin{subequations}
\begin{eqnarray}
\Phi_{1,2}&\equiv& \phi_{1,2}\otimes \bar{\phi}_{1,-2} \,, \\
\bar{\Phi}_{1,2}&\equiv&   \phi_{1,-2}\otimes \bar{\phi}_{1,2}
\end{eqnarray}
\end{subequations}
with conformal weights  $(h_{1,2},h_{1,-2})$ and $(h_{1,-2},h_{1,2})$. Note that here by $\phi_{r,s}$ we simply mean a chiral primary field with conformal weight $h_{r,s}$: the structure of the associated Virasoro module  will be discussed below. This means in particular that $\phi_{r,s}=\phi_{-r,-s}$. 

By expanding the factor $1/ P(q)P(\bar{q})$ in the spectrum generating functions, we see that model also has four descendants at level two, that is with  conformal weights $(h_{1,-2},h_{1,-2})$, where we have used that $h_{1,-2}=h_{1,2}+2$. 
Now, if the modules generated by $\Phi_{1,2}$ and $\bar{\Phi}_{1,2}$  in the continuum limit were a product of two Verma modules, these four descendants would be  the two independent fields, $L_{-2} \Phi_{1,2}$ and $L_{-1}^2\Phi_{1,2}$, as well as the two fields obtained by swapping chiral and antichiral components, $\bar{L}_{-2} \bar{\Phi}_{1,2}$ and $\bar{L}_{-1}^2\bar{\Phi}_{1,2}$. The chiral/antichiral symmetry corresponds to exchanging right and left (i.e., exchanging momentum $p$ for momentum $-p$) and is present on the lattice as well, by reflecting the site index $i \to N+1-i$ \cite{GSJS}. This means one would expect to observe, in the finite-size transfer matrix, two eigenvalues, both converging (once properly scaled)  to $h_{1,-2}=h_{1,2}+2$, and corresponding  to two linear combinations  \cite{GSJS} of $L_{-2} \Phi_{1,2}$ and  $L_{-1}^2\Phi_{1,2}$ and their conjugates---hence both appearing in the form of doublets. This is however {\sl not} what is observed numerically (see Appendix~\ref{app:singlets}). Instead, we see one doublet and two singlets, which means that the module in the continuum limit and at level two does not have, as a basis, a pair of independent states and their chiral/antichiral conjugates. 

Introducing 
\begin{subequations}
\begin{eqnarray}
A_{1,2}&=&L_{-2}-{3\over 2+4h_{1,2}}L_{-1}^2 \,, \\
 \bar{A}_{1,2}&=&\bar{L}_{-2}-{3\over 2+4h_{1,2}}\bar{L}_{-1}^2
\end{eqnarray}
\end{subequations}
we now claim that, in the continuum limit,  the identity
\begin{equation}
A_{1,2}\Phi_{1,2}=\bar{A}_{1,2}\bar{\Phi}_{1,2}
\end{equation}
 is satisfied. Note that both sides of the equation are primary fields---i.e., they are annihilated by $\VirN$ generators $L_n,\bar{L}_n$ with $n>0$. They are also of vanishing norm $\loopinner{-}{-}$. 
 Corresponding numerical results are given in Appendix \ref{loop_lines}.
 
 We have therefore identified part of the module as a quotient of $ (\Verma{1,2}^{\rm (d)} \otimes \Vermab{1,-2})\oplus (\Verma{1,-2}\otimes \Vermab{1,2}^{\rm (d)})$, corresponding to the following diagram
 for the degenerate fields:
%%
%\begin{equation}
% \begin{array}{ccccc}
%      \Phi_{12}&&&\hskip-1cm \bar{\Phi}_{12} \\
%      &\hskip-.2cm\searrow&\swarrow&\\
%      &&\hskip-1.3cm \bar{A}_{12}\Phi_{12}=A_{12}\bar{\Phi}_{12}&&
%\end{array}
%\label{eq_general_proj}
%\end{equation}
%%

\begin{equation}
\begin{tikzpicture}[>=stealth]
%\coordinate (PSI) at (0, 1.5) {};
\coordinate (XY) at (-2, 0) {};
\coordinate (YX) at (2, 0) {};
\coordinate (AXY) at (0, -1.5) {};
\node(axy) at (AXY) {$A_{1,2}\Phi_{1,2}=\bar{A}_{1,2}\bar{\Phi}_{1,2}$};
\node(yx) at (YX) {$\bar{\Phi}_{1,2}=\phi_{1,-2}\otimes \bar{\phi}_{1,2}$};
\node(xy) at (XY) {$\Phi_{1,2}=\phi_{1,2}\otimes \bar{\phi}_{1,-2}$};
%\node(psi) at (PSI) {$\Psi$};
%\draw[->] (psi) -- (yx);
%\draw[->] (psi) -- (xy);
\draw[->] (xy) -- (axy);
\draw[->] (yx) -- (axy);
%\draw[->] (psi) -- (axy);
%\node(L0) at ([xshift=6ex]$(psi)!0.5!(axy)$) {\small$L_0-h_{-1,2}$};
%\node(A+) at ([yshift=3ex]$(psi)!0.5!(xy)$) {\small$\frac{A^{\dagger}}{\kappa^{-1}\nu}$};
%\node(A+bar) at ([yshift=3ex]$(psi)!0.5!(yx)$) {\small$\frac{\bar{A}^{\dagger}}{\kappa^{-1}\nu}$};
\node(A) at ([yshift=-2ex]$(xy)!0.4!(axy)$) {\small$A$};
\node(Abar) at ([yshift=-2ex]$(yx)!0.4!(axy)$) {\small$\bar{A}$};
\end{tikzpicture}
\end{equation}

\medskip

%\lawrence{In this paragraph and the next: bars over the right Virasoro tensorand?}
Note we have the quotient modules (obtained by quotienting by the  submodule generated by the bottom field), $\IrrV{1,2}\otimes \Vermab{1,-2}$ and $\Verma{1,-2}\otimes \IrrVb{1,2}$ and  with generating functions $(q^{h_{1,-2}-c/24}/ P(q))\times \bar{K}_{1,2}$ and $K_{1,2}\times 
(\bar{q}^{h_{1,-2}-c/24}/ P(\bar{q}))$. The bottom field generates a product of  Verma modules $\Verma{1,-2}\otimes \Vermab{1,-2}$  with generating function $(q^{h_{1,-2}-c/24}/ P(q))\times (\bar{q}^{h_{1,-2}-c/24}/ P(\bar{q}))$. 

This cannot, however, be the end of the story, since the quotient identified so far is not self-dual---nor does it account for the proper multiplicity of fields. Invariance of the diagram under reversal of the arrow demands that there exists a field ``on top,'' with  a quotient which is also a product of  Verma modules  $\Verma{1,-2}\otimes \Vermab{1,-2}$. This should give rise, in terms of fields, to the diagram 
\begin{equation} \label{diamondPhi12}
\begin{tikzpicture}[>=stealth]
\coordinate (PSI) at (0, 1.5) {};
\coordinate (XY) at (-3.5, 0) {};
\coordinate (YX) at (3.5, 0) {};
\coordinate (AXY) at (0, -1.5) {};
\node(axy) at (AXY) {$A_{1,2}\Phi_{1,2}=\bar{A}_{1,2}\bar{\Phi}_{1,2}$};
\node(yx) at (YX) {$\bar{\Phi}_{1,2}=\phi_{1,-2}\otimes \bar{\phi}_{1,2}$};
\node(xy) at (XY) {$\Phi_{1,2}=\phi_{1,2}\otimes \bar{\phi}_{1,-2}$};
\node(psi) at (PSI) {$\Psi_{1,2}$};
\draw[->] (psi) -- (yx);
\draw[->] (psi) -- (xy);
\draw[->] (xy) -- (axy);
\draw[->] (yx) -- (axy);
\draw[->] (psi) -- (axy);
\node(L0) at ([xshift=6ex,yshift=-2ex]$(psi)!0.5!(axy)$) {\small$(L_0-h_{-1,2})$};
\node(A+) at ([yshift=3ex]$(psi)!0.5!(xy)$) {\small$A^{\dagger}$};
\node(A+bar) at ([yshift=3ex]$(psi)!0.5!(yx)$) {\small$\bar{A}^{\dagger}$};
\node(A) at ([yshift=-2ex]$(xy)!0.4!(axy)$) {\small$A$};
\node(Abar) at ([yshift=-2ex]$(yx)!0.4!(axy)$) {\small$\bar{A}$};
\end{tikzpicture}
\end{equation}
with $\Psi_{1,2}$ a field to be determined---see below. 

The same construction seems to apply to all cases in the $F_{j,1}$ characters. 
The simplest example occurs, in fact, in $\AStTL{1}{1}$---even though this module does not appear in the Potts model, as discussed around \eqref{decompPottsZ1}---with 
$
\Phi_{1,1}\equiv \phi_{1,1}\otimes \bar{\phi}_{1,-1}$ and 
$\bar{\Phi}_{1,1}\equiv   \phi_{1,-1}\otimes \bar{\phi}_{1,1}$.  In this case, the quotient is simply given by $L_{-1} \Phi_{1,1}=\bar{L}_{-1}\bar{\Phi}_{1,1}$. 

%\jesper{Added:}
The indecomposable structure for arbitrary positive integer values of $e,j$ can then be conjectured to be
\begin{equation} \label{diamondPhiej}
\begin{tikzpicture}[>=stealth]
\coordinate (PSI) at (0, 1.5) {};
\coordinate (XY) at (-3.5, 0) {};
\coordinate (YX) at (3.5, 0) {};
\coordinate (AXY) at (0, -1.5) {};
\node(axy) at (AXY) {$A_{e,j}\Phi_{e,j}=\bar{A}_{e,j}\bar{\Phi}_{e,j}$};
\node(yx) at (YX) {$\bar{\Phi}_{e,j}=\phi_{e,-j}\otimes \bar{\phi}_{e,j}$};
\node(xy) at (XY) {$\Phi_{e,j}=\phi_{e,j}\otimes \bar{\phi}_{e,-j}$};
\node(psi) at (PSI) {$\Psi_{e,j}$};
\draw[->] (psi) -- (yx);
\draw[->] (psi) -- (xy);
\draw[->] (xy) -- (axy);
\draw[->] (yx) -- (axy);
\draw[->] (psi) -- (axy);
\node(L0) at ([xshift=6ex,yshift=-2ex]$(psi)!0.5!(axy)$) {\small$(L_0-h_{e,-j})$};
\node(A+) at ([yshift=3ex]$(psi)!0.5!(xy)$) {\small$A^{\dagger}$};
\node(A+bar) at ([yshift=3ex]$(psi)!0.5!(yx)$) {\small$\bar{A}^{\dagger}$};
\node(A) at ([yshift=-2ex]$(xy)!0.4!(axy)$) {\small$A$};
\node(Abar) at ([yshift=-2ex]$(yx)!0.4!(axy)$) {\small$\bar{A}$};
\end{tikzpicture}
\end{equation}
The validity of (\ref{diamondPhiej}) in general  comes from strong numerical evidence for small values of 
$e,j$. It is also the simplest structure we can imagine solving the problems of poles in the OPEs, based on our independent knowledge of the spectrum of the theory. More complete evidence should come from the construction of four-point functions using the  corresponding regularized conformal blocks \cite{SylvainNextPaper}.

It is  interesting to draw the corresponding structure of Virasoro modules defining the quotient modules ${\cal L }_{e,j}$:
\begin{equation}\label{moduleL}
\begin{tikzpicture}[>=stealth]
\coordinate (PSI) at (0, 1.5) {};
\coordinate (XY) at (-2, 0) {};
\coordinate (YX) at (2, 0) {};
\coordinate (AXY) at (0, -1.5) {};
\coordinate (SPE) at (-6.5,0) {};
\node(axy) at (AXY) {$\Verma{e,-j}\otimes \Vermab{e,-j}$};
\node(yx) at (YX) {$\Verma{e,-j}\otimes \IrrVb{e,j}$};
\node(xy) at (XY) {$\IrrV{e,j}\otimes \Vermab{e,-j}$};
\node(psi) at (PSI) {$\Verma{e,-j}\otimes \Vermab{e,-j}$};
\node(spe) at (SPE) {${\cal L}_{e,j}={\cal Q}[(\Verma{e,j}^{\rm (d)} \otimes \Vermab{e,-j})\oplus (\Verma{e,-j}\otimes \Vermab{e,j}^{\rm (d)})]\equiv$};
\draw[->] (psi) -- (yx);
\draw[->] (psi) -- (xy);
\draw[->] (xy) -- (axy);
\draw[->] (yx) -- (axy);
%\draw[->] (psi) -- (axy);
%\node(L0) at ([xshift=6ex]$(psi)!0.5!(axy)$) {\small$(L_0-h_{-1,2})$};
%\node(A+) at ([yshift=3ex]$(psi)!0.5!(xy)$) {\small$A^{\dagger}$};
%\node(A+bar) at ([yshift=3ex]$(psi)!0.5!(yx)$) {\small$\bar{A}^{\dagger}$};
%\node(A) at ([yshift=-2ex]$(xy)!0.4!(axy)$) {\small$A$};
%\node(Abar) at ([yshift=-2ex]$(yx)!0.4!(axy)$) {\small$\bar{A}$};
\end{tikzpicture}
\end{equation}

\noindent
Accordingly we have the result:

%\vspace*{5pt}
%\noindent\fbox{
%\hspace*{5pt}\begin{minipage}{\linewidth-20pt}\em
%\vspace*{5pt}
%{\bf Loop-model modules with through-lines:} For $j>0$ and $2j$ through-lines we have the scaling limit
%%
%\begin{equation}
%\AStTL{j}{1}\mapsto {\color{red} Quotient of} \left( \bigoplus_{e>0} \Verma{e,-j}\otimes \Verma{e,j}^{\rm (d)} \right) \oplus \big(  \Verma{0,-j}\otimes\Verma{0,j}  \big)   \oplus \left( \bigoplus_{e<0} \Verma{e,-j}^{\rm (d)} \otimes \Verma{e,j} \right)  \,.
%\label{MainRes3}
%\end{equation}
%%
%\vspace*{5pt}
%\end{minipage}\hspace*{5pt}}
%\vspace*{5pt}
%
\vspace*{5pt}
\noindent\fbox{
\hspace*{5pt}\begin{minipage}{\linewidth-20pt}\em
\vspace*{5pt}
{\bf Loop-model modules with through-lines:} For $j>0$ and $2j$ through-lines we have the scaling limit
\begin{equation}
\AStTL{j}{1}\mapsto  \big(  \Verma{0,-j}\otimes\Vermab{0,j}  \big) \oplus  \bigoplus_{e>0} {\cal L}_{e,j}   \,.
\label{MainRes3}
\end{equation}
\vspace*{5pt}
\end{minipage}\hspace*{5pt}}
\vspace*{5pt}

As already mentioned, an important piece of evidence for the correctness of the structure \eqref{diamondPhiej} is based on the numerical observation of a pair
of singlet states in the transfer matrix spectrum. In Appendix~\ref{app:singlets} we identify this pair of singlets precisely in the cases
$(e,j) = (1,1)$, $(2,1)$, $(1,2)$ and $(1,3)$. These observations in turn lend credence to the general result \eqref{MainRes3}.

\section{Modules for the loop model in the degenerate case: the OPE point of view}\label{OPEsection}

%\hubert{Edited by Yifei and H, please do not modify}

\newcommand{\yifei}[1]{\textbf{\color{red}\footnotesize[Y: #1]}}
\newcommand{\yifeinote}[1]{\textbf{\color{blue}\footnotesize[Y: #1]}}

As in the early works on logarithmic CFTs \cite{GurarieLudwig,VJS}, it is possible to understand the appearance of indecomposable modules in the continuum limit of $\AStTL{j}{1}$ by carefully examining the OPEs and their potential divergences when one of the fields in the $s$-channel has a degenerate conformal weight.

\smallskip

To start, imagine that we have some OPE of a field of dimension $\Delta$ with itself where a field with conformal weights $(h_{1,2},h_{1,2})$ appears. In ordinary CFT, the descendants  of this field at level two in the chiral and in the antichiral sector would not be independent: this fact is crucial  to cancel the divergence arising in the OPE coefficients from  to the fact that $h_{1,2}$ is in the Kac table, resulting in a finite OPE such as the ones arising in the minimal-model CFTs \cite{BPZ}. Let us now see what happens if the null descendants are not zero, and the divergences potentially remain.  To proceed, we  factor out the 
 $(z\bar{z})^{-2\Delta}$, with $\Delta=\bar{\Delta}$ denoting the conformal weight of the fields being fused, and analyze the potential divergences by slightly shifting the conformal weights of the field on the right-hand side of the OPE:
\begin{eqnarray}
C(\epsilon)\left\{(z\bar{z})^{h_{1+\epsilon,2}}\left[\left(X_\epsilon+{z\over 2}\partial X_\epsilon+\alpha^{(-2)}(\Delta,h_{1+\epsilon,2})z^2L_{-2}X_\epsilon+
\alpha^{(-1,-1)}(\Delta,h_{1+\epsilon,2})z^2L_{-1}^2X_\epsilon\right)\times \hbox{h.c.}\right]+\ldots\right\}
\label{OPE0}
\end{eqnarray}
where $C(\epsilon)$ is a number to be determined, the dots stand for other fields, and we have used the short-hand notations
\begin{subequations}
\begin{eqnarray}
X_\epsilon&=&\phi_{1+\epsilon,2}(z)\;,\\
\bar{X}_\epsilon&=&\bar{\phi}_{1+\epsilon,2}(\bar{z})\;.
\end{eqnarray}
\end{subequations}
The coefficients $\alpha$ in \eqref{OPE0} are fully determined by conformal invariance 
\begin{subequations}
\begin{eqnarray}
\alpha^{(-2)}(\Delta,h)&=&{(h-1)h+2\Delta(1+2h)\over 16 (h-h_{1,2})(h-h_{2,1})}\;,\label{kac}\\
\alpha^{(-1,-1)}(\Delta,h)&=&{(1+h)(c+8h)-12(\Delta+h)\over 64 (h-h_{1,2})(h-h_{2,1})}\;,
\end{eqnarray}
\end{subequations}
and note that we have
\begin{equation}\label{separate}
\alpha^{(-1,-1)}(\Delta,h)L_{-1}^2+\alpha^{(-2)}(\Delta,h)L_{-2}=\alpha^{(-2)}(\Delta,h)A(h)+\alpha_0^{(-1,-1)}(h)L_{-1}^2,
\end{equation}
%\begin{equation}
%\alpha^{(-1,-1)}L_{-1}^2+\alpha^{(-2)}L_{-2}=\alpha^{(-2)}\left(L_{-2}-{3\over 2+4h}L_{-1}^2\right)+{1+h\over 4(1+2h)}L_{-1}^2,
%\end{equation}
% 
where
\begin{equation}
A(h)\equiv L_{-2}-{3\over 2+4h}L_{-1}^2,\;\;
\alpha_0^{(-1,-1)}(h)\equiv {1+h\over 4(1+2h)}.
\end{equation}
It is important to notice that in writing \eqref{separate}, the dependence on the external field $\Delta$ only appears in the coefficient $\alpha^{(-2)}$, i.e., the operator $A$ which will turn out to give rise to the Jordan cell structure is independent of the external field. This point will become more clear below.

\smallskip

Going back to  $h=h_{1+\epsilon,2}$ with $\epsilon\to 0$, and writing $A_{\epsilon}\equiv A(h_{1+\epsilon,2})$, it is convenient to define
\begin{equation}\label{gammadef}
\gamma(\epsilon)\equiv\langle X_\epsilon| A_\epsilon^\dagger A_\epsilon|X_\epsilon\rangle
={8(h-h_{1,2})(h-h_{2,1})\over (1+2h)}=\nu\epsilon \,,
\end{equation}
with
\begin{equation}\label{nu}
\nu=-\frac{2(1-2\beta^2-\beta^4+2\beta^6)}{\beta^6} \,,
%4{\left(\beta^2-{1\over 2}\right)(1-\beta^4)\over \beta^6}
\end{equation}
where we have used the parametrization $\beta^2=x/(x+1)$. On the other hand, notice that as $\epsilon\to 0$, the coefficient $\alpha^{(-2)}(\Delta,h_{1+\epsilon,2})$ has a simple pole, since the denominator is proportional to the Kac determinant, as is obvious from equation~\eqref{kac}.
This means that the OPE potentially presents  singularities, which must be properly canceled by the contribution  of other fields with the proper dimensions---a point well understood since the works \cite{GurarieLudwig,DJS,VJS,VGJS}. Obviously, the leading singularity in the OPE is a second-order pole coming from the descendants at level two of $X_\epsilon\bar{X}_\epsilon$. 
Keeping in mind that $h_{1,2}+2=h_{-1,2}$, and of course $h_{r,s}=h_{-r,-s}$, we therefore introduce the other fields 
\begin{subequations}
\begin{eqnarray}
Y_\epsilon&=&\phi_{-1+\epsilon,2}(z) \,, \\
\bar{Y}_\epsilon&=&\bar{\phi}_{-1+\epsilon,2}(\bar{z})
\end{eqnarray}
\end{subequations}
in order to cancel such singularities, and we complete the OPE as follows:
\begin{equation}
\begin{aligned}
C(\epsilon)&\bigg\{(z\bar{z})^{h_{1+\epsilon,2}}\left[\left(X_\epsilon+{z\over 2}\partial X_\epsilon+
\alpha_0^{(-1,-1)}(\epsilon)z^2L_{-1}^2X_\epsilon+\alpha^{(-2)}(\epsilon)z^2A_\epsilon X_\epsilon\right)\otimes \hbox{h.c.}\right]\\
&\qquad {}+(z\bar{z})^{h_{-1+\epsilon,2}}
a(\epsilon)Y_\epsilon\otimes\bar{Y}_\epsilon\bigg\} \,, \label{OPE1}
\end{aligned}
\end{equation}
where we have adopted the short-hand notations $\alpha^{(-1,-1)}_0(\epsilon)$, $\alpha^{(-2)}(\epsilon)$, and the new coefficient $a(\epsilon)$ is yet to be determined. 

\medskip

To study the necessary cancellation of singularities, we focus on the most divergent term at level 2:
\begin{equation}\label{mostdiv}
\begin{aligned}
&C(\epsilon)\Big\{(z\bar{z})^{h_{1+\epsilon,2}+2}[\alpha^{(-2)}(\epsilon)]^2A_\epsilon X_\epsilon\otimes\bar{A}_\epsilon\bar{X}_\epsilon+a(\epsilon)(z\bar{z})^{h_{-1+\epsilon,2}}Y_\epsilon\otimes\bar{Y}_\epsilon\Big\}\\
&\qquad{}=C(\epsilon)\Big\{\epsilon\kappa\ln(z\bar{z})(z\bar{z})^{h_{-1,2}}[\alpha^{(-2)}(\epsilon)]^2A_\epsilon X_\epsilon\otimes\bar{A}_\epsilon\bar{X}_\epsilon+{1\over\sqrt{\epsilon}}(z\bar{z})^{h_{-1+\epsilon,2}}\Phi_\epsilon\Big\} \,,
\end{aligned}
\end{equation}
where we have defined
\begin{equation}\label{kappa}
\kappa\equiv \frac{h_{1+\epsilon,2}+2-h_{-1+\epsilon,2}}{\epsilon}={1\over \beta^2}
\end{equation}
and introduced the new field
\begin{equation}
\Phi_\epsilon\equiv \sqrt{\epsilon}\left([\alpha^{(-2)}(\epsilon)]^2A_\epsilon X_\epsilon\otimes\bar{A}_\epsilon\bar{X}_\epsilon+a(\epsilon)Y_\epsilon\otimes\overline{Y}_\epsilon\right).
\end{equation}
The two-point function of this field is given by
\begin{equation}\label{Phi2pt}
\langle \Phi(w,\bar{w})\Phi(0,0)\rangle=\epsilon \left\{[\alpha^{(-2)}(\epsilon)]^4\gamma(\epsilon)^2 (w\bar{w})^{-2h_{1+\epsilon,2}-4}+a^2 (\epsilon) 
(w\bar{w})^{-2h_{-1+\epsilon,2}}\right\}.
\end{equation}
Recall equation~\eqref{gammadef} and that $\alpha^{(-2)}(\epsilon)$ has a simple pole in $\epsilon$. One can write
\begin{equation}\label{rs}
[\alpha^{(-2)}(\Delta,h_{1+\epsilon,2})]^2\gamma\equiv {r\over \epsilon}+s+O(\epsilon) \,.
\end{equation}
It is then clear that the coefficient of the first term in \eqref{Phi2pt} has a double pole which must be canceled by the divergence from the second term. This requires $a^2(\epsilon)$ to be of the form
%For instance, if $\Delta=h_{1/2,0}$ is the dimension of the order operator, we find 
%%
%\begin{equation}
%[\alpha^{(-2)}(\Delta,h_{1+\epsilon,2})]^2\gamma=
% {(1-\beta^4)\over 1024 }{\beta^2\over (\beta^2-{1\over 2})\epsilon}+
%{28-112\beta^2+91\beta^4+96\beta^6-99\beta^8\over 3072 \beta^2(-1+2\beta^2)^2}
%\end{equation}
%
\begin{equation}\label{adef}
a^2(\epsilon)= {\lambda\over \epsilon^2}+{\mu\over\epsilon}+O(1) \,.
\end{equation}
Such behavior can in fact be established using that $\phi_{2,1}$ is degenerate in the theory, as we will see in more detail in section \ref{orderpara} below. 
The singularity cancellation condition then reads
\begin{equation}\label{rlambda}
\lambda=-r^2 \,,
\end{equation}
and the two point function \eqref{Phi2pt} becomes
\begin{equation}
\langle \Phi(w,\bar{w})\Phi(0,0)\rangle={-2\kappa r^2\ln(w\bar{w})+2rs+\mu\over (w\bar{w})^{2h_{-1,2}}} \,.
\end{equation}
Taking into account the $1/\sqrt{\epsilon}$ factor in \eqref{mostdiv}, we must therefore take $C(\epsilon)=\sqrt{\epsilon}$, such that the contribution of $\Phi_{\epsilon}$ in the OPE is of $O(1)$.

%while the first term reads 
%%
%\begin{equation}
%\kappa\epsilon^{3/2} \ln(z\bar{z})(z\bar{z})^{h_{-1,2}}[\alpha^{(-2)}]^2A_\epsilon X_\epsilon\otimes\bar{A}_\epsilon\bar{X}_\epsilon\label{rewrite1}
%\end{equation}
%

At this point, it is natural to introduce the normalized field
\begin{equation}\label{Xhat}
\hat{X}_{\epsilon}\equiv\frac{1}{\sqrt{\gamma}}A_{\epsilon}X_{\epsilon} \,,
%\;\;\bar{\hat{X}}_{\epsilon}\equiv\frac{1}{\sqrt{\gamma}}\bar{A}_{\epsilon}\bar{X}_{\epsilon},
\end{equation}
%
%\begin{equation}
%|\hat{X}\rangle\equiv\sqrt{{1+2h\over 8(h-h_{12})(h-h_{21})}}\left(L_{-2}-{3\over 2+4h}L_{-1}^2\right)|X\rangle\end{equation}
%%
%so 
%%
%\begin{equation}
%A_\epsilon X_\epsilon=\sqrt{\gamma}\hat{X}_\epsilon
%\end{equation}
%
and identify it as another copy of $Y_{\epsilon}$ in the limit $\epsilon\to 0$, since both have dimension $h_{-1,2}$ and are annihilated by $L_1$ and $L_2$.
%As $\epsilon\to 0$ we have that $\hat{X}$ has dimension $h_{-1,2}$, and is annihilated by both $L_1$ and $L_2$ (the divergence of the factor is $O(1/\sqrt{\epsilon})$ while the action of the generators gives a factor $O(\epsilon)$). It seems reasonable to  identify it with another copy of $Y$ in this limit ($Y$ itself disappears anyhow).
The first term in second line of \eqref{mostdiv} is then given by:
\begin{eqnarray}
{\kappa r\over2\sqrt{\nu}}(z\bar{z})^{h_{-1,2}}\ln(z\bar{z})(AX\otimes\bar{Y}+Y\otimes \bar{A}\bar{X}).
\end{eqnarray}
%
%\begin{eqnarray}
%\kappa\epsilon^{3/2}(z\bar{z})^{h_{-1,2}}\ln(z\bar{z})[\alpha^{(-2)}]^2A_\epsilon X_\epsilon\otimes\bar{A}_\epsilon\bar{X}_\epsilon\nonumber\\
%={\kappa r\over2\sqrt{\nu}}(z\bar{z})^{h_{-1,2}}\ln(z\bar{z})\left(\hat{X}\otimes \bar{A}\bar{X}+AX\otimes\bar{\hat{X}}\right)
%\end{eqnarray}
%
Combining with the remaining terms in the OPE \eqref{OPE1}, i.e.,
\begin{equation}
\sqrt{\epsilon}(z\bar{z})^{h_{1+\epsilon,2}}\left[\left(X_\epsilon+{z\over 2}\partial X_\epsilon+
\alpha_0^{(-1,-1)}(\epsilon)z^2L_{-1}^2X_\epsilon\right)\otimes \alpha^{(-2)}(\epsilon)\sqrt{\gamma}\bar{z}^2\bar{\hat{X}}_{\epsilon}+ \hbox{h.c.}\right],
\end{equation}
%
%The only term arising from this will come from 
%%
%\begin{equation}
%\sqrt{\epsilon} ...\otimes{ \sqrt{r}\over\sqrt{\epsilon}}\bar{z}^2 \bar{\hat{X}}
%\end{equation}
%
and recalling \eqref{rs}, we have then the full OPE as $\epsilon\to 0$:\footnote{After factoring out a global factor of $\sqrt{r}$.}
\begin{equation}\label{fullOPE}
\begin{aligned}
&z^{h_{1,2}}\bar{z}^{h_{-1,2}}
\left[\left(X+{z\over 2}\partial X+
\alpha_0^{(-1,-1)}(\epsilon)z^2L_{-1}^2X\right)\otimes\bar{Y}\right]+\hbox{h.c.}\\
&\qquad{}+(z\bar{z})^{h_{-1,2}}{\kappa \sqrt{r}\over\sqrt{\nu}}\left(\frac{1}{2}\ln(z\bar{z})(AX\otimes\bar{Y}+\hbox{h.c.})+\frac{\sqrt{\nu}}{\kappa r}\Phi\right)\\
&\quad{}=z^{h_{1,2}}\bar{z}^{h_{-1,2}}
\left[\left(X+{z\over 2}\partial X+
\alpha_0^{(-1,-1)}(\epsilon)z^2L_{-1}^2X\right)\otimes\bar{Y}\right]+\hbox{h.c.}\\
&\quad\qquad{}+(z\bar{z})^{h_{-1,2}}{\kappa \sqrt{r}\over\sqrt{\nu}}\left(\ln(z\bar{z})(AX\otimes\bar{Y})+\frac{\sqrt{\nu}}{\kappa r}\Phi\right).
\end{aligned}
\end{equation}
In the last line of \eqref{fullOPE} we have set
\begin{equation}
AX\otimes \bar{Y}=\sqrt{\gamma} \hat{X}\otimes \bar{Y}=\sqrt{\gamma} Y\otimes\bar{\hat{X}}=Y\otimes \bar{A}\bar{X} \,,
\end{equation}
using the identification of $\hat{X},\bar{\hat{X}}$ with $Y,\bar{Y}$ in the $\epsilon\to 0$ limit.
As will become obvious below, this has the interpretation that $L_0-\bar{L}_0$ is diagonalizable.

\medskip

We are interested in the logarithmic mixing at level 2, i.e., the last line of \eqref{fullOPE}. Inspecting the terms, it is natural to redefine the field
\begin{equation}\label{Psidef}
\Psi\equiv {\sqrt{\nu}\over \kappa r}\Phi
\end{equation}
which, as we shall see, becomes the logarithmic partner of $AX\otimes \bar{Y}=Y\otimes \bar{A}\bar{X}$. It is a simple exercise to calculate their two-point functions\footnote{In computing the two-point functions, one must keep in mind the distinction between $\hat{X}$ and $Y$ when $\epsilon\neq 0$, and take the definition \eqref{Psidef} at $\epsilon\neq 0$, i.e., $\Psi_{\epsilon}\equiv (\sqrt{\nu}/ \kappa r)\Phi_{\epsilon}$.} and one arrives at
\begin{subequations}\label{2ptfunctions}
\begin{eqnarray}
\langle (AX\otimes \bar{Y})(w,\bar{w})(AX\otimes \bar{Y})(0,0)\rangle&=&0,\\
\langle \Psi(w,\bar{w})(AX\otimes \bar{Y})(0,0)\rangle&=&{\kappa^{-1}\nu\over (w\bar{w})^{2h_{-1,2}}},\\
\langle \Psi(w,\bar{w})\Psi(0,0)\rangle&=&{-2{\kappa^{-1}\nu}\ln(w\bar{w})+{\nu\over\kappa^2r^2}(2rs+\mu)\over (w\bar{w})^{2h_{-1,2}}}.\label{twopt2}
\end{eqnarray}
\end{subequations}
We recognize the usual logarithmic structure of a rank-2 Jordan cell \cite{Gurarie}.

\medskip

As a final step, we compute the action of Virasoro algebra on the pair $(AX\otimes \bar{Y},\Psi)$:
\begin{subequations}
\begin{eqnarray}
L_0(AX\otimes\bar{Y})&=&h_{-1,2}(AX\otimes\bar{Y}),\\
L_0\Psi&=&h_{-1,2}\Psi+\frac{\sqrt{\nu\epsilon}}{\kappa r}[\alpha^{(-2)}(\epsilon)]^2(h_{1+\epsilon,2}+2-h_{-1+\epsilon,2})A_{\epsilon}X_{\epsilon}\otimes \bar{A}_{\epsilon}\bar{X}_{\epsilon}\nonumber\\
&=&h_{-1,2}\Psi+AX\otimes\bar{Y},
\end{eqnarray}
\end{subequations}
and similarly for $\bar{L}_0$.
%We also give the two-point function:
%%
%\begin{equation}
%\langle \Phi(w,\bar{w})(AX\otimes \bar{Y})(0,0)\rangle={r\sqrt{\nu}\over (w\bar{w})^{2h_{-1,2}}}
%\end{equation}
%%
%Finally, we recall of course that 
%%
%\begin{equation}
%\langle (AX\otimes \bar{Y})(w,\bar{w})(AX\otimes \bar{Y})(0,0)\rangle=0
%\end{equation}
%%
%since it is proportional to $\gamma$, the Kac determinant. 
%
%Let us now calculate 
%%
%\begin{equation}
%L_0\Phi=h_{-1,2}\Phi+\sqrt{\epsilon} [\alpha^{(-2)}]^2\left(h_{1+\epsilon,2}+2-h_{-1+\epsilon,2}\right) A_\epsilon X_\epsilon\otimes \bar{A}_\epsilon\bar{X}_\epsilon
%\end{equation}
%%
%The second term is 
%%
%\begin{equation}
%\sqrt{\epsilon} [\alpha^{(-2)}]^2 \kappa\epsilon\sqrt{\gamma} AX\otimes \hat{X}={\kappa r\over \sqrt{\nu}} AX\otimes \bar{Y}
%\end{equation}
%%
%Hence
%%
%\begin{eqnarray}
%L_0\Phi=h_{-1,2}\Phi+{\kappa r\over\sqrt{\nu}} AX\otimes \bar{Y}\nonumber\\
%\bar{L}_0\Phi=h_{-1,2}\Phi+{\kappa r\over \sqrt{\nu}} Y\otimes \bar{A}\bar{X}
%\end{eqnarray}
%
%\begin{eqnarray}
%L_0\Psi=h_{-1,2}\Psi+ AX\otimes \bar{Y}\nonumber\\
%\bar{L}_0\Psi=h_{-1,2}\Psi+Y\otimes \bar{A}\bar{X}\label{JordanCell}
%\end{eqnarray}
%
Therefore we see that in the basis $(AX\otimes \bar{Y},\Psi)=(Y\otimes \bar{A}\bar{X},\Psi)$ we have 
\begin{equation}
L_0=\begin{pmatrix}
h_{-1,2}&1\\
0&h_{-1,2}\end{pmatrix}=\bar{L}_0\label{Jordan}
\end{equation}
forming a rank-2 Jordan cell.
%and the OPE (we factored out a global $\sqrt{r}$ term)
%%
%\begin{eqnarray}
%z^{h_{1,2}}\bar{z}^{h_{-1,2}}
%\left[\left(X+{1\over 2}z\partial X+
%\alpha_0^{(-1,-1)}z^2L_{-1}^2X\right)\otimes\bar{Y}\right]+\hbox{h.c.}\nonumber\\
%+{\kappa \sqrt{r}\over2\sqrt{\nu}}(z\bar{z})^{h_{-1,2}}\left[\ln(z\bar{z})\left(AX\otimes\bar{Y}+\hbox{h.c.}\right)+2\Psi\right]\label{finalOPE}
%\end{eqnarray}
%%
%In the following we will set
%%
%\begin{equation}
%g\equiv \kappa\sqrt{r\over\nu}
%\end{equation}
%
In addition we find
%\begin{equation}
%A^\dagger \Phi=\sqrt{\epsilon} [\alpha^{(-2)}]^2 \gamma X\otimes \bar{A}\bar{X}={r\over\sqrt{\epsilon}}\sqrt{\gamma} X\otimes \bar{Y}=r\sqrt{\nu} X\otimes \bar{Y}
%\end{equation}
\begin{equation}\label{Adagger}
A^\dagger \Psi=\frac{\sqrt{\nu\epsilon}}{\kappa r}[\alpha^{(-2)}]^2\gamma X\otimes \bar{A}\bar{X}=\kappa^{-1}\nu X\otimes\bar{Y} \,,
\end{equation}
where we have used \eqref{rs}, \eqref{gammadef} and \eqref{Xhat}. Note also that $L_1\Psi=0$. Hence, the module is depicted as
%\begin{equation}
% \begin{array}{ccccc}
%      &&\hskip-1.4cm \Psi &&\\
%      &\hskip-.2cm\swarrow&\searrow&\\
%      X\otimes \bar{Y} &&&\hskip-1cm Y\otimes \bar{X} \\
%      &\hskip-.2cm\searrow&\swarrow&\\
%      &&\hskip-1.3cm AX\otimes \bar{Y}=Y\otimes\bar{A}\bar{X}&&
%\end{array}
%\label{eq_general_proj}
%\end{equation}
%
\begin{equation}\label{logmodule}
\begin{tikzpicture}[>=stealth]
\coordinate (PSI) at (0, 1.5) {};
\coordinate (XY) at (-3, 0) {};
\coordinate (YX) at (3, 0) {};
\coordinate (AXY) at (0, -1.5) {};
\node(axy) at (AXY) {$AX\otimes\bar{Y}=Y\otimes\bar{A}\bar{X}$};
\node(yx) at (YX) {$Y\otimes\bar{X}$};
\node(xy) at (XY) {$X\otimes\bar{Y}$};
\node(psi) at (PSI) {$\Psi$};
\draw[->] (psi) -- (yx);
\draw[->] (psi) -- (xy);
\draw[->] (xy) -- (axy);
\draw[->] (yx) -- (axy);
\draw[->] (psi) -- (axy);
\node(L0) at ([xshift=6ex,yshift=-2ex]$(psi)!0.5!(axy)$) {\small$L_0-h_{-1,2}$};
\node(A+) at ([yshift=3ex]$(psi)!0.5!(xy)$) {\small$\frac{A^{\dagger}}{\kappa^{-1}\nu}$};
\node(A+bar) at ([yshift=3ex]$(psi)!0.5!(yx)$) {\small$\frac{\bar{A}^{\dagger}}{\kappa^{-1}\nu}$};
\node(A) at ([yshift=-2ex]$(xy)!0.4!(axy)$) {\small$A$};
\node(Abar) at ([yshift=-2ex]$(yx)!0.4!(axy)$) {\small$\bar{A}$};
\end{tikzpicture}
\end{equation}
a structure that coincides with \eqref{diamondPhi12}.

\medskip

As we have briefly commented before, the logarithmic coupling $\kappa^{-1}\nu$ in (\ref{twopt2}) which characterizes the Jordan-cell structure does not depend on the dimension $\Delta$ of the external fields. More explicitly, from \eqref{nu} and \eqref{kappa}, we have
\begin{equation}
\kappa^{-1}\nu=-\frac{2(1-2\beta^2-\beta^4+2\beta^6)}{\beta^4} \,,
\end{equation}
which is entirely determined by the Kac formula and the Kac determinant. In contrast, the coefficient $\kappa \sqrt{r}/\sqrt{\nu}$ in the OPE \eqref{fullOPE} does depend on $\Delta$ through $r$, due to \eqref{rs}. Similarly, the constant in the two-point function \eqref{twopt2} also depends on $\Delta$. This is however compatible with the Jordan cell structure, since the field $\Psi$ always admits a shift by a multiple of the null field \cite{Gurarie},
\begin{equation}
\Psi\to\Psi+\hbox{const.} \times AX\otimes \bar{Y}=\Psi+\hbox{const.} \times Y\otimes \bar{A}\bar{X} \,,
\end{equation}
which does not change (\ref{Jordan}). 

\bigskip

The construction also generalizes to the case of operators $\phi_{r,s}$ and $\phi_{r,-s}$. In general, the module has the structure in \eqref{gencell} with $X=\phi_{r,s}$, $Y=\phi_{r,-s}$, and $A$ replaced by the proper combination of Virasoro generators. Setting 
\begin{equation}
\langle \phi_{r+\epsilon,s}|A_{r,s}^\dagger A_{r,s}|\phi_{r+\epsilon,s}\rangle=\nu_{r,s}\epsilon
\end{equation}
and observing that
\begin{equation}
h_{r+\epsilon,s}+rs-h_{-r+\epsilon,s}=\kappa_{r,s}\epsilon \,, \quad \mbox{with } \kappa_{r,s}={r\over\beta^2} \,,
\end{equation}
we find that the free parameter of the module (the so-called logarithmic coupling, or indecomposability parameter) is 
\begin{equation}
b_{r,s}=\kappa_{r,s}^{-1}\nu_{r,s} \,,
\end{equation}
so that
\begin{subequations}
\begin{eqnarray}
(L_0-h_{-r,s})\Psi_{r,s}=(\bar{L}_0-h_{r,-s})\Psi_{r,s}&=&A_{r,s}\phi_{r,s}\otimes \bar{\phi}_{r,-s}=\phi_{r,-s}\otimes \bar{A}_{r,s}\bar{\phi}_{r,s} \,, \\
A_{r,s}^\dagger \Psi_{r,s}&=&b_{r,s} \phi_{r,s}\otimes \bar{\phi}_{r,-s} \,, \\
\bar{A}_{r,s}^\dagger \Psi_{r,s}&=&b_{r,s} \phi_{r,-s}\otimes \bar{\phi}_{r,s}
\end{eqnarray}
\end{subequations}
with the structure:
\begin{equation}
\begin{tikzpicture}
\coordinate (PSI) at (0, 1.5) {};
\coordinate (XY) at (-3, 0) {};
\coordinate (YX) at (3, 0) {};
\coordinate (AXY) at (0, -1.5) {};
\node(axy) at (AXY) {$A_{r,s}\phi_{r,s}\otimes\bar{\phi}_{r,-s}=\phi_{r,-s}\otimes\bar{A}_{r,s}\bar{\phi}_{r,s}$};
\node(yx) at (YX) {$\phi_{r,-s}\otimes\bar{\phi}_{r,s}$};
\node(xy) at (XY) {$\phi_{r,s}\otimes\bar{\phi}_{r,-s}$};
\node(psi) at (PSI) {$\Psi_{r,s}$};
\draw[->] (psi) -- (yx);
\draw[->] (psi) -- (xy);
\draw[->] (xy) -- (axy);
\draw[->] (yx) -- (axy);
\draw[->] (psi) -- (axy);
\node(L0) at ([xshift=6ex]$(psi)!0.5!(axy)$) {\small$L_0-h_{-r,s}$};
\node(A+) at ([yshift=3ex]$(psi)!0.5!(xy)$) {\small$\frac{A_{r,s}^{\dagger}}{b_{r,s}}$};
\node(A+bar) at ([yshift=3ex]$(psi)!0.5!(yx)$) {\small$\frac{\bar{A}_{r,s}^{\dagger}}{b_{r,s}}$};
\node(A) at ([yshift=-2ex]$(xy)!0.4!(axy)$) {\small$A_{r,s}$};
\node(Abar) at ([yshift=-2ex]$(yx)!0.4!(axy)$) {\small$\bar{A}_{r,s}$};
\end{tikzpicture}\label{gencell}
\end{equation}
%
%\begin{equation}
% \begin{array}{ccccc}
%      &&\hskip-2.cm \Psi_{rs} &&\\
%      &\hskip-.4cm\swarrow&\searrow&\\
%      X_{rs}\otimes \bar{Y}_{rs} &&&\hskip-2cm Y_{rs}\otimes \bar{X}_{rs} \\
%      &\hskip-.2cm\searrow&\swarrow&\\
%      &&\hskip-1.3cm A_{rs}X_{rs}\otimes \bar{Y}_{rs}=Y_{rs}\otimes\bar{A}_{rs}\bar{X}_{rs}&&
%\end{array}
%\label{eq_general_proj}
%\end{equation}
%
in agreement with \eqref{diamondPhiej}.

For the special case $r=s=1$, for instance, we find that  $\nu_{1,1}=-1+1/ \beta^2
$
%
%\begin{equation}
%\gamma_{1,1}=2h_{1+\epsilon,1}=\left(-1+{1\over \beta^2}\right)\epsilon
%\end{equation}
and therefore
\begin{equation}
b_{1,1}=1-\beta^2 \,.
\end{equation}

\section{The particular case of the order operator and conformal blocks}\label{orderpara}

%\hubert{Edited by Yifei and H. please do not modify}

In the case where the external field is given by the order operator $\Delta=h_{1/2,0}$, we can construct the $s$-channel expansion of conformal blocks by combining the OPEs of two pairs of external fields, and compare with the results obtained in \cite{HJS}.

\subsection{Constructing logarithmic conformal blocks from OPEs}
Our basic ingredients are the OPE \eqref{fullOPE} and the two-point functions \eqref{2ptfunctions}. Take the OPE of two order operators $\Delta=h_{1/2,0}$ and focus on the contributions involving the module \eqref{logmodule}:
\begin{equation}\label{OPEforblock}
\begin{aligned}
\Phi_{\Delta}(w,\bar{w})\Phi_{\Delta}(0,0)=(w\bar{w})^{-2\Delta}&\bigg\{w^{h_{1,2}}\bar{w}^{h_{-1,2}}
\left[\left(X+{w\over 2}\partial X+
\alpha_0^{(-1,-1)}(\epsilon)w^2L_{-1}^2X\right)\otimes\bar{Y}\right]+\hbox{h.c.}\\
&\qquad{}+(w\bar{w})^{h_{-1,2}}{\kappa \sqrt{r}\over\sqrt{\nu}}[\ln(w\bar{w})(AX\otimes\bar{Y})+\Psi]+\ldots\bigg\} \,,
\end{aligned}
\end{equation}
where $\ldots$ stands for other fields appearing in the OPE. The corresponding logarithmic conformal block can be constructed by combining two pair of fields $\Phi_{\Delta}(z_1,\bar{z}_1)\Phi_{\Delta}(z_2,\bar{z}_2)$ and $\Phi_{\Delta}(z_3,\bar{z}_3)\Phi_{\Delta}(z_4,\bar{z}_4)$ with cross-ratio $z=z_{12}z_{34}/z_{13}z_{24}$, and similarly for $\bar{z}$.

First, the usual calculations give the first few terms of the blocks
\begin{equation}\label{block1}
(z\bar{z})^{-2\Delta}\left[z^{h_{1,2}}\bar{z}^{h_{1,-2}}+z^{h_{1,-2}}\bar{z}^{h_{1,2}}+{h_{1,2}\over 2}
\left(z^{h_{1,2}+1}\bar{z}^{h_{1,-2}}+z^{h_{1,-2}}\bar{z}^{h_{1,2}+1}\right)\right].
\end{equation}
Now, focus on the terms at level 2. The two-point function of $\alpha_0^{(-1,-1)}L_{-1}^2X\otimes\bar{Y}+\text{h.c.}$ in the first line of \eqref{OPEforblock} gives contribution to the blocks with
\begin{equation}\label{block2}
4h(1+2h)[\alpha^{(-1,-1)}_0]^2(z\bar{z})^{-2\Delta}\left(z^{h_{1,2}+2}\bar{z}^{h_{1,-2}}+z^{h_{1,-2}}\bar{z}^{h_{1,2}+2}\right) \,,
\end{equation}
where we have used
\begin{equation}
L_{1}^2L_{-1}^2 |h\rangle=4h(1+2h)|h\rangle \,.
\end{equation}
The last line of \eqref{OPEforblock} then contributes to the conformal block as (factoring out $(z\bar{z})^{-2\Delta}$)
\begin{equation}
\begin{aligned}
&(z_{12}\bar{z}_{12})^{h_{-1,2}}(z_{34}\bar{z}_{34})^{h_{-1,2}}\frac{\kappa^2 r}{\nu}\Big\{\big(\ln(z_{12}\bar{z}_{12})+\ln(z_{34}\bar{z}_{34})\big)\langle(AX\otimes\bar{Y})\Psi\rangle+\langle\Psi\Psi\rangle\Big\}\\
&\qquad{}=\Big[\Big(\frac{z_{12}z_{34}}{z_{13}z_{24}}\Big)^{h_{-1,2}}\times\text{h.c.}\Big]\frac{\kappa^2 r}{\nu}\Big\{\kappa^{-1}\nu\big[\ln(z_{12}\bar{z}_{12})+\ln(z_{34}\bar{z}_{34})\big]-2\kappa^{-1}\nu\ln(z_{13}\bar{z}_{13})+\frac{\nu}{\kappa^2 r^2}(2rs+\mu)\Big\} \,,
\end{aligned}
\end{equation}
where we have used the two-point functions \eqref{2ptfunctions}. Simplifying expressions, we have the following term in the conformal block:
\begin{equation}\label{block3}
(z\bar{z})^{-2\Delta}(z\bar{z})^{h_{-1,2}}\Big(\kappa r\ln(z\bar{z})+2s+\frac{\mu}{r}\Big).
\end{equation}

\medskip

To summarize, \eqref{block1}, \eqref{block2} and \eqref{block3} assemble to the following logarithmic conformal block:
\begin{equation}\label{blocksfromOPE}
\begin{aligned}
{\cal F}^{\text{log}}(z,\bar{z})=(z\bar{z})^{-2\Delta}&\bigg\{z^{h_{1,2}}\bar{z}^{h_{1,-2}}+z^{h_{1,-2}}\bar{z}^{h_{1,2}}+{h_{1,2}\over 2}
\left(z^{h_{1,2}+1}\bar{z}^{h_{1,-2}}+z^{h_{1,-2}}\bar{z}^{h_{1,2}+1}\right)\\
&\qquad{}+\Big[\frac{\kappa r}{2}\ln(z\bar{z})+s+\frac{\mu}{2r}+\frac{h_{1,2}(1+h_{1,2})^2}{4(1+2h_{1,2})}\Big]\left(z^{h_{1,2}+2}\bar{z}^{h_{1,-2}}+z^{h_{1,-2}}\bar{z}^{h_{1,2}+2}\right)+\ldots
\bigg\} \,.
\end{aligned}
\end{equation}
Note that by construction $[L_{-1},A]=0$, so the correlations between $\Psi$ and the terms in the first part of the OPE vanish.
Note also that by construction we have $L_1AX=0$, and since $A^\dagger AX=0$, $L_2AX=0$ as well, so $X\otimes \bar{Y}$ is a primary field.

\subsection{Input from ordinary conformal blocks}

In this section, we obtain the logarithmic block \eqref{blocksfromOPE} using input from the ordinary conformal blocks as a consistency check. 

Recall the ordinary $s$-channel expansion of the ordinary conformal blocks
\begin{equation}\label{ordblock}
{\cal F}_h=z^{h-2\Delta}\left[1+{h\over 2} z+{z^2\over 16 (h-h_{1,2})(h-h_{2,1})} \begin{pmatrix}
h(h+1)&h+\Delta\end{pmatrix}\begin{pmatrix}
2+{c\over 4h}&-3\\
-3&4h+2\end{pmatrix}\begin{pmatrix}
h(h+1)\\h+\Delta\end{pmatrix}+\ldots\right]
\end{equation}
and similarly for $\bar{z}$. We focus on the four-point function of the fields with conformal weight $\Delta=h_{1/2,0}$ (the Potts-model order operator) and consider the conformal block in the case of $\Phi_{1,2}$. As discussed in depth in \cite{HJS}, the amplitudes associated with the fields with weight $h_{1+\epsilon,2}$ and $h_{-1+\epsilon,2}$ are related by recursions resulting from the degeneracy of $\phi_{2,1}$. We then consider the combinations (first mentioned in \cite{SantaViti})
\begin{equation}\label{comb}
\tilde{C}(\epsilon)\bigg\{{\cal F}_{h_{1+\epsilon,2}}(z){\cal F}_{h_{1+\epsilon,2}}(\bar{z})+{A_{-1+\epsilon,2}\over A_{1+\epsilon,2}}{\cal F}_{h_{-1+\epsilon,2}}(z){\cal F}_{h_{-1+\epsilon,2}}(\bar{z})\bigg\} \,,
\end{equation}
where $A_{-1+\epsilon,2}/ A_{1+\epsilon,2}$ is a known function; see \cite{HJS} for more details. To make connections with the OPE discussed in section \ref{OPEsection}, we recognize that this ratio should be identified with $a^2(\epsilon)$ in \eqref{adef} and thus has the expansion 
\begin{equation}
{A_{-1+\epsilon,2}\over A_{1+\epsilon,2}}={\lambda\over\epsilon^2}+{\mu\over\epsilon}+O(1) \,.
\end{equation}
More explicitly, taking $\Delta=h_{1/2,0}$, one finds
\begin{equation}\label{lambda}
\lambda=- \left({1-\beta^4\over 512}{\beta^2\over 2\beta^2-1}\right)^2.
\end{equation}
This results in the following contribution from the second term of \eqref{comb}:
\begin{equation}\label{b1}
\begin{aligned}
&\tilde{C}(\epsilon)(z\bar{z})^{-2\Delta}(z\bar{z})^{h_{-1+\epsilon,2}}\Big[\Big(\frac{\lambda}{\epsilon^2}+\frac{\mu}{\epsilon}\Big)+\ldots\Big],\\
&\qquad{}=\tilde{C}(\epsilon)(z\bar{z})^{-2\Delta}(z\bar{z})^{h_{-1,2}}\Big[\Big(\frac{\lambda}{\epsilon^2}+\frac{\mu}{\epsilon}-\frac{\lambda}{\epsilon}\frac{(2\beta^2+1)}{2\beta^2}\ln(z\bar{z})\Big)+\ldots\Big]\,,
\end{aligned}
\end{equation}
where $\ldots$ stands for higher powers in $z,\bar{z}$ and $O(1)$ terms.

Now focus on the first term in \eqref{comb}. As $\epsilon\to 0$, \eqref{ordblock} has a simple pole for $h=h_{1,2}$. Explicit calculations then give
\begin{equation}
{1\over 16 (h-h_{1,2})(h-h_{2,1})} \begin{pmatrix}
h(h+1)&h+\Delta\end{pmatrix}\begin{pmatrix}
2+{c\over 4h}&-3\\
-3&4h+2\end{pmatrix}\begin{pmatrix}
h(h+1)\\h+\Delta\end{pmatrix}={\rho\over\epsilon}+\sigma+O(\epsilon)\label{singular}
\end{equation}
with 
\begin{subequations}
\begin{eqnarray}
\rho&=&\frac{\beta^2(1-\beta^4)}{512(2\beta^2-1)} \,,\label{rho}\\
\sigma&=&{-12+16\beta^2+121\beta^4-216 \beta^6-129\beta^8+288\beta^{10}\over 1024 \beta^2(-1+2\beta^2)^2}\,.\label{sigma}
\end{eqnarray}
\end{subequations}
The first term in \eqref{comb} then gives the contribution
\begin{equation}\label{b2}
\begin{aligned}
&\tilde{C}(\epsilon)(z\bar{z})^{-2\Delta}(z\bar{z})^{h_{1+\epsilon,2}}\Big|\Big(1+\frac{h_{1+\epsilon,2}}{2}z+z^2\big(\frac{\rho}{\epsilon}+\sigma\big)+\ldots\Big)\Big|^2\\
&{}=\tilde{C}(\epsilon)(z\bar{z})^{-2\Delta}(z\bar{z})^{h_{1,2}}\bigg\{\frac{\rho^2}{\epsilon^2}(z\bar{z})^2+\frac{\rho}{\epsilon}(z^2+\bar{z}^2)+\frac{\rho}{\epsilon}\frac{h_{1,2}}{2}(z\bar{z}^2+z^2\bar{z})+\frac{\rho}{\epsilon}\Big(2\sigma+\frac{\rho(1-2\beta^2)}{2\beta^2}\ln(z\bar{z})\Big)(z\bar{z})^2+\ldots\bigg\} \,,
\end{aligned}
\end{equation}
where again $\ldots$ stands for higher powers in $z,\bar{z}$ and $O(1)$ terms.

Combining \eqref{b1} and \eqref{b2}, we see first that the double poles cancel due to
\begin{equation}\label{rholambda}
\lambda+\rho^2=0 \,,
\end{equation}
as is evident from \eqref{lambda} and \eqref{rho}. On the other hand, it is natural to take $\tilde{C}(\epsilon)=\frac{\epsilon}{\rho}$. Therefore the combination \eqref{comb} reduces to
\begin{equation}\label{blocklog}
\begin{aligned}
(z\bar{z})^{-2\Delta}&\bigg\{z^{h_{1,2}}\bar{z}^{h_{1,-2}}+z^{h_{1,-2}}\bar{z}^{h_{1,2}}+{h_{1,2}\over 2}
\left(z^{h_{1,2}+1}\bar{z}^{h_{1,-2}}+z^{h_{1,-2}}\bar{z}^{h_{1,2}+1}\right)\\
&\qquad{}+\Big(\sigma+\frac{\mu}{2\rho}+\frac{\rho}{2\beta^2}\ln(z\bar{z})\Big)\left(z^{h_{1,2}+2}\bar{z}^{h_{1,-2}}+z^{h_{1,-2}}\bar{z}^{h_{1,2}+2}\right)+\ldots\bigg\} \,,
\end{aligned}
\end{equation}
where we have used $h_{1,2}+2=h_{-1,2}$ and \eqref{rholambda}.

\medskip

We now compare \eqref{blocklog} with the logarithmic block \eqref{blocksfromOPE} that we obtained previously. 

First, it is obvious that the first lines of \eqref{blocklog} and \eqref{blocksfromOPE} agree. To compare the level-2 coefficients, we need $r,s$ as defined in \eqref{rs}. As discussed above, these quantities depend on the external fields and in this case we take $\Delta=h_{1/2,0}$ in \eqref{rs}. First we find
\begin{equation}\label{rrho}
r=\frac{\beta^2(1-\beta^4)}{512(2\beta^2-1)}=\rho \,.
\end{equation}
Recall that in the OPE study in section \ref{OPEsection}, we have obtained the singularity cancellation condition \eqref{rlambda}. Now we see that for the four-point function of the order operator we focus on here, this is the same as \eqref{rholambda}. On the other hand, it is a simple exercise to check that the following identity holds:
\begin{equation}
\sigma=s+\frac{h_{1,2}(1+h_{1,2})^2}{4(1+2h_{1,2})} \,,
\end{equation}
using \eqref{rs} and \eqref{sigma}. Therefore we have seen that the constant terms in the second lines of \eqref{blocklog} and \eqref{blocksfromOPE} agree. Finally, the coefficients for the $\ln(z\bar{z})$ terms are easily matched using \eqref{kappa} and \eqref{rrho}.

\subsection{Numerical amplitudes and Jordan cells}

%We can finally find evidence for the existence of rank-two Jordan cells at generic $Q$ by studying the size-dependency of the amplitudes directly measured using the technology developed in our earlier papers.
%The easiest amplitude to measure is related with the four-point correlation function of the order operator in the Potts model.

In Appendix~\ref{app:singlets} we have identified some singlet levels in the transfer matrix of the loop model that confirm the existence of the indecomposable structure \eqref{diamondPhiej}. To go further and find numerical evidence for the existence of the expected Jordan cell for $L_0,\bar{L}_0$ (or the conjectured values of the logarithmic couplings) is more difficult, since it turns out that the Hamiltonian and transfer matrices of the Potts model for $Q$ generic remain, for the levels we are interested in,  completely diagonalizable in finite size. In other words, {\sl the $(L_0,\bar{L}_0)$ Jordan cells appear only in the continuum limit}. While this possibility was foreseen in \cite{GRSV}, it makes the problem quite different from the one studied in  \cite{DJS,VJS}, where Jordan cells were present for finite systems as a result of Temperley--Lieb representation theory, with the indecomposable structures in the continuum limit being identical to those observed in the lattice model. Luckily, we shall see that it is nonetheless possible for the case at hand to observe the ``build-up'' of Jordan cells in the lattice model.

To that end, we now go back to the four-point functions of the order operator in the Potts model. In lattice terms,  they are of the form $P_{a_1a_2a_3a_4}$, where a label $a_i$ is associated with each of the four insertion points $z_i$ (with $i=1,2,3,4$), the convention being that points are required to belong to the same FK cluster if and only if their corresponding labels are identical. For instance, $P_{abab}$ denotes the four-point function in which $z_1$ and $z_3$ belong to the same cluster, while $z_2$ and $z_4$ belong to a different cluster (see Figure~2 of \cite{JS}). To study such correlation functions on the lattice by the transfer matrix technique, it is convenient to place points $z_1, z_2$ on the same time slice (i.e., lattice row) and points $z_3, z_4$ on a different, distant slice (see Figure~1 of \cite{JS}). This geometric arrangement amounts to performing the $s$-channel expansion of the correlation function\cite{JS,HGJS,HJS}. The simplest example of the structure \eqref{diamondPhiej} involves the fields $(\Phi_{e,j},\bar{\Phi}_{e,j})$ from the standard module $\AStTL{j}{z^2}$ with $j=1$, but we have seen in \eqref{decompPottsZ} that these fields decouple from the Potts-model partition function, and the results of \cite{JS} show that they also decouple from the correlation functions of the order parameter.

It is therefore natural to turn to the next available case, $j=2$, and thus the representation $\AStTL{2}{z^2}$. The results of \cite{JS} show that $P_{abab}$ and $P_{abba}$ both have the property of coupling to $\AStTL{2}{1}$ and $\AStTL{2}{-1}$ in their $s$-channel expansion, and they are the only four-point functions that contain these two representations as their {\em leading} contributions (other correlation functions couple to $\bAStTL{\!\! 0}{\q^{\pm 2}}$ and/or $\AStTL{0}{-1}$ as well). Moreover, the symmetric combination
\begin{equation}
 P_{\rm S} = P_{abab} + P_{abba}
\end{equation}
decouples from $\AStTL{2}{-1}$ for symmetry reasons, and since $\AStTL{2}{1}$ contains the fields $(\Phi_{e,2},\bar{\Phi}_{e,2})$ with integer $e \ge 0$, it transpires that $P_{\rm S}$ is the most convenient correlation function to investigate in the present context. Finally, the lowest-lying levels that can give rise to the structure \eqref{diamondPhiej} correspond to the case $e=1$. For all these reasons we henceforth focus on the case $(e,j) = (1,2)$.

Denoting the separation between the two groups of points $z_1,z_2$ and $z_3,z_4$ along the imaginary time direction%
\footnote{A shift between the two groups of points along the space-like direction was shown in \cite{JS} to be irrelevant. In the notations of Figure~1 in \cite{JS} one can therefore consider the two groups to be aligned, i.e., with a shift $x=0$.}
by $\ell$, the correlation function in
the cylinder geometry generically takes the form
\begin{equation}
 P_{\rm S} = \sum_i A_i \left( \frac{\Lambda_i}{\Lambda_0} \right)^\ell \,,
\end{equation}
where the sum is over the contributing eigenvalues $\Lambda_i$ (with $\Lambda_0$ referring to the ground state), and $A_i$ are the corresponding amplitudes. A  rank-2 Jordan cell for the transfer matrix on the lattice manifests itself  by a ``generalized amplitude,'' with $A_i$ of the form $a_i + \ell b_i$. This structure can be observed in many cases when $\q$ is a root of unity \cite{fullynongenericpaper}. In our problem, however, the Jordan cells {\em are not present} for $L$ finite, and only expected to appear in the limit $L\to\infty$. A natural scenario for how this might happen is as follows: we should have  two eigenvalues which become close as $L\to\infty$, with divergent and opposite amplitudes. Assuming that $\Lambda_1=\Lambda(1+a\epsilon)$ and $\Lambda_2=\Lambda(1-a\epsilon)$ appear with respective amplitudes $A_1=A+b/ \epsilon$ and $A_2=A-b/ \epsilon$, where the small parameter $\epsilon\to 0$ when $L\to\infty$, we have then
 \begin{eqnarray}
 A_1\left({\Lambda_1\over \Lambda_0}\right)^\ell+A_2\left({\Lambda_2\over\Lambda_0}\right)^\ell&\approx& \left(A+{b\over\epsilon}\right)\left({\Lambda\over\Lambda_0}\right)^\ell (1+a \ell \epsilon)+\left(A-{b\over\epsilon}\right)\left(
 {\Lambda\over\Lambda_0}\right)^\ell (1-a \ell \epsilon)\nonumber\\
 &=&2A\left({\Lambda\over\Lambda_0}\right)^\ell +2ab \ell \left({\Lambda\over\Lambda_0}\right)^\ell \label{scenario}
 \end{eqnarray}
reproducing as $L\to\infty$  the behavior expected from  the presence of a  Jordan cell for the continuum-limit Hamiltonian.
 
% While the generalized amplitudes $a_i$ and $b_i$ can indeed be determined in this form by
%the first numerical method described in section 4.3.1 of \cite{JS}, the available sizes $L$ are quite limited for computational reasons (in practice $L \le 7$).
The  method best adapted to identifying the scenario in (\ref{scenario})  is based on scalar products,
as discussed in section~4.3.2 of \cite{JS}. Notice that although this method measures the amplitudes $A_i$ directly in the $\ell \to \infty$ limit, the hypotheses leading to the scaling form can still be tested, and in particular the scaling of the amplitudes under the approach to the thermodynamic limit $L \to \infty$.
%takes the limit $\ell \to \infty$ before measuring $A_i$, and allows one to access
%larger sizes (in practice $L \le 11$). In the case of a generalized amplitude, this second method is however not capable of measuring $a_i$ and $b_i$
%independently. While one would formally expect $A_i = a_i + \ell b_i \to \infty$ in this case, it appears reasonable to expect that one would rather
%observe that the measured $A_i$ diverges when $L \to \infty$ in this case, instead of tending to a constant. 

We now investigate this issue in the context of the $(\Phi_{1,2},\bar{\Phi}_{1,2})$ structure, which is numerically the most accessible case for the reasons given above.

\begin{sidewaystable}
\begin{center}
\begin{tabular}{|c|lllllll|}
\cline{1-8}
  Line
    & \multicolumn{7}{|c|}{$L$} \\  \cline{2-8}
   $i_{13}$ & 5 & 6 & 7 & 8 & 9 & 10 & 11 \\
  \hline
 3 & \phantom{$-$}0.51584673 & \phantom{$-$}0.53739515 & \phantom{$-$}0.51435306 & \phantom{$-$}0.53469774 & \phantom{$-$}0.52426708 & \phantom{$-$}0.53949703 & \phantom{$-$}0.53338217 \\
 \textcolor{red}{24} & $-$0.0041836648 & $-$0.012473807 & $-$0.018607995 & $-$0.032601923 & $-$0.041773974 & $-$0.059633592 & \\
 25 & $-$0.011807194 &  $-$0.025268048 & $-$0.024113896 & $-$0.034228263 & $-$0.033153298 & $-$0.040478536 & \\
 \textcolor{red}{35} & \phantom{$-$}0.023005683 & \phantom{$-$}0.053207857 & \phantom{$-$}0.061027619 & \phantom{$-$}0.093065936 & \phantom{$-$}0.10297778 & \phantom{$-$}0.13439104 & \\
\hline
\end{tabular}
\end{center}
\caption{Amplitudes $A_i$ of the correlation function $P_{\rm S}$ corresponding to selected fields within $\AStTL{2}{1}$, in finite size $L$. The distance between the two points within each group is taken as $d=\lfloor L/2 \rfloor$. The lines of the table are labeled, as in Table~\ref{Tab_exp_V2}, by the index $i_{13}$.}
  \label{tab:amp}
\begin{center}
\begin{tabular}{|c|llllll|}
\cline{1-7}
  Line
    & \multicolumn{6}{|c|}{$L$} \\  \cline{2-7}
   $i_{13}$ & 5 & 6 & 7 & 8 & 9 & 10 \\
  \hline
 3 & \phantom{$-$}0.29631794 & \phantom{$-$}0.23327610 & \phantom{$-$}0.19042437 & \phantom{$-$}0.15916824 & \phantom{$-$}0.13539463 & \phantom{$-$}0.11679002 \\
 \textcolor{red}{24} & $-$0.000002260380 & $-$0.000005746649 & $-$0.000008815612 & $-$0.000010766671 & $-$0.000011635877 & $-$0.000011709765 \\
 25 & $-$0.0019402603 & $-$0.0026491875 & $-$0.0029769838 & $-$0.0030550303 & $-$0.0029897585 & $-$0.0028501860 \\
 \textcolor{red}{35} & \phantom{$-$}0.000038085542 & \phantom{$-$}0.000050876586 & \phantom{$-$}0.000053221816 & \phantom{$-$}0.000049738327 & \phantom{$-$}0.000043951008 & \phantom{$-$}0.000037766542 \\
\hline
\end{tabular}
\end{center}
\caption{Amplitudes $A_i$ of the correlation function $P_{\rm S}$ corresponding to selected fields within $\AStTL{2}{1}$, in finite size $L$. The distance between the two points within each group is now chosen the smallest possible, $d=1$.}
  \label{tab:amp1}
\end{sidewaystable}

The finite-size level corresponding to the pair of fields $(\Phi_{1,2},\bar{\Phi}_{1,2})$ has been identified in Appendix~\ref{app:singlets} as the line with $i_{13}=3$ in Table~\ref{Tab_exp_V2}. Note that this is a twice degenerate level (doublet) in the transfer matrix spectrum, because the fields $\Phi_{1,2}$ and $\bar{\Phi}_{1,2}$ are related by the exchange of chiral and antichiral components. The corresponding combined amplitude (i.e., summed over the doublet) for the contribution of this level to $P_{\rm S}$ is shown in the first line of Table~\ref{tab:amp}. The amplitudes are normalized by that of the leading contribution to $P_{\rm S}$, namely the amplitude of the line with $i_{13}=1$ in Table~\ref{Tab_exp_V2}.
To be precise, the table shows the amplitudes for cylinders of circumference $L=5,6,\ldots,11$, and in all cases the distance $d$ between the two points in each group ($z_1,z_2$ and $z_3,z_4$) is taken the largest possible: $d=L/2$ for $L$ even, and $d=(L-1)/2$ for $L$ odd. This choice (which was also used in the numerical work in \cite{JS,HJS}) corresponds to a fixed, finite distance between the two points in the continuum limit. Unfortunately, it also leads to parity effects in $L$, which are clearly visible from Table~\ref{tab:amp}. It is nevertheless clear that the amplitude of the line with $i_{13}=3$ converges to a finite constant, as expected for this non-logarithmic pair of fields, and this can be confirmed by independent fits of even and odd sizes.
Regrettably, the situation for the remaining lines of Table~\ref{tab:amp} is less clear. Naively the amplitude for each one of the last three lines appears to grow
with $L$, but our attempts to quantify this have not been very compelling, due to fact that we only have three sizes of each parity at our disposal.

We therefore turn to another strategy, in which the same amplitudes are measured with the smallest possible distance $d=1$ between the two points in each group.
This will eliminate the parity effects, so that more reliable fits can be studied.
Note that the choice $d=1$ corresponds to a vanishing distance in the continuum limit, so one might expect the finite-size amplitudes to pick up an extra factor of $1/L$. In particular, the amplitude of a generic, non-logarithmic field contributing to $P_{\rm S}$ is then expected to vanish as $L^{-1}$ in the $L \to \infty$ limit. Indeed, the amplitude of the line with $i_{13}=3$ in Table~\ref{tab:amp1} fits very nicely to $c_0 + c_1 L^{-1} + c_2 L^{-2} + \cdots$, and the absolute value of the constant term $c_0$ can be determined to be at least 80 times smaller than the data point with $L=10$. We therefore conjecture that, in this case, $c_0 = 0$ indeed.

For the line with $i_{13} = 24$ (a singlet level) we attempt a fit of the form $c_0 + c_1 L^{-\delta} + c_2 L^{-2\delta} + c_3 L^{-3\delta}$. This matches the data nicely with $\delta \simeq 1.005$, indicating that $\delta=1$ might be the exact value of the exponent. But we find now that the absolute value of the constant term $c_0$ is about 3 times {\em larger} than the data point with $L=10$, which is strongly indicative of $c_0$ being nonzero in this case. We therefore conjecture that this line should be identified with one of the two fields in the Jordan cell \eqref{2ptfunctions}.

\begin{figure}
\begin{center}
\includegraphics[width=8cm]{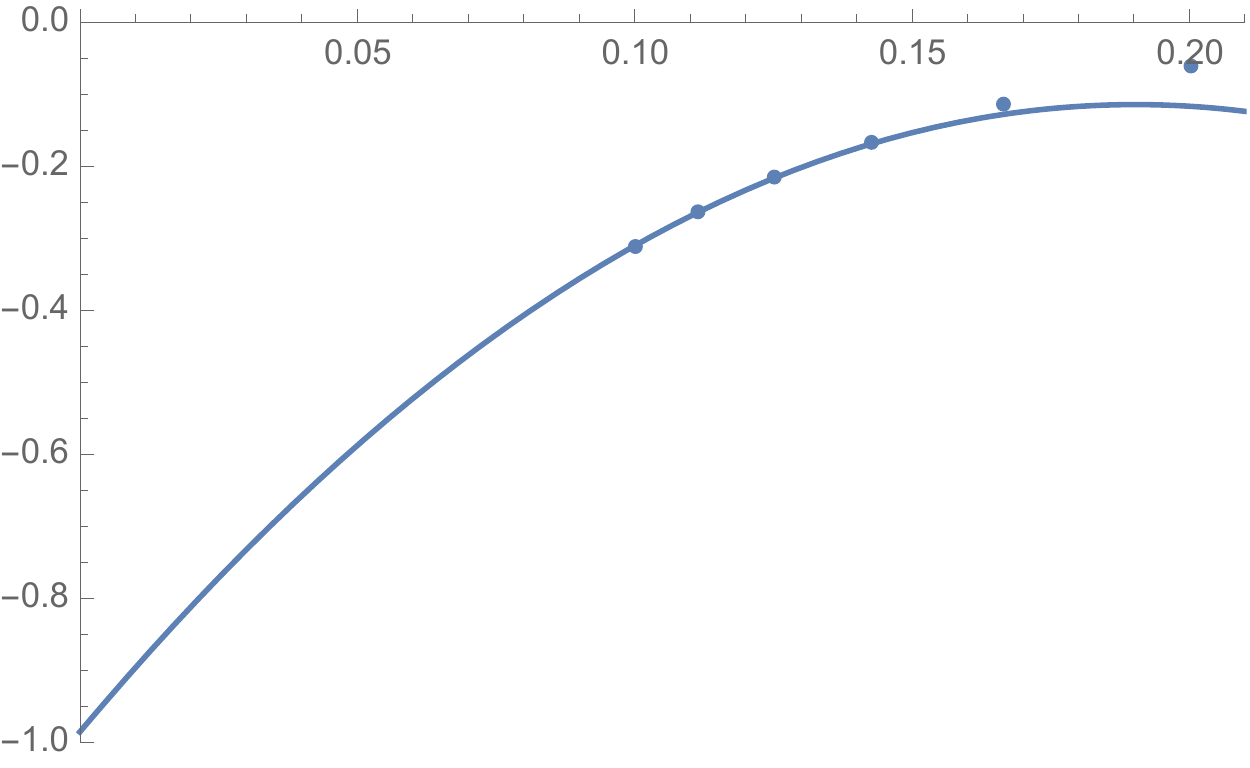}
\end{center}
\caption{Ratio $A_1/A_2$ between the amplitudes of the two singlet fields (see Table~\ref{tab:amp1}), corresponding to the lines with $i_{13} = 24$ and $i_{13}=35$ (see Table~\ref{Tab_exp_V2}), plotted against $1/L$. The curve is a second-order polynomial fit to the last three data points.}
\label{fig:logfits}
\end{figure}

The same type of fit for the line with $i_{13}=35$ (the other singlet level) yields $\delta \simeq 2.05$ and a constant term $c_0$ which is about 4 times {\em smaller} than the $L=10$ data point. Finally, the line with $i_{13} = 25$ (a doublet) matches the fit with $\delta \simeq 1.3$ and $c_0$ about 3 times smaller than the data point with $L=10$. Seen in isolation, these fits do not permit us to convincingly conclude whether the value of $c_0$ is finite or zero for those two lines. However, structural considerations provide more compelling evidence. According to the argument given in \eqref{scenario}, the logarithmic singlet with $i_{13}=24$ needs to be accompanied by another singlet field with an opposite and diverging (for finite conformal distance) amplitude. Being a singlet, the line with $i_{13}=35$ is the only possible candidate for such a logarithmic partner.

As a decisive test, we therefore plot in Figure~\ref{fig:logfits} the ratio between the amplitudes of the two singlets. A second-order polynomial in $1/L$ fits the data nicely and gives an extrapolated value of the ratio of $-0.985$, very close to the exact ratio of $-1$ expected from \eqref{scenario}. We believe that this settles the issue, showing that the two singlets correspond to the conformal fields $A_{1,2} \Phi_{1,2} + \bar{A}_{1,2} \bar{\Phi}_{1,2}$ and $\Psi$, and that the indecomposable structure \eqref{diamondPhi12} builds up only in the $L\to\infty$ limit. On the other hand, Figure~\ref{fig:logfits} vividly illustrates that a maximum size of $L=10$ is still quite far from the thermodynamic limit, and with hindsight it is therefore hardly surprising that only a combination of arguments can reveal the true nature (logarithmic or non-logarithmic) of the four fields from Table~\ref{Tab_exp_V2} having conformal weights $(h_{1,2}+2,h_{1,2}+2)$.

\section{Currents and the ``ordinary'' loop model}\label{currents}

The ``ordinary'' dense loop model is defined simply as a model of dense loops with fugacity $\m$ for {\sl all} loops. It can be considered  as a continuation to all values of $\m$ of  a $U(\m)$ model defined initially 
for $\m$ integer by introducing alternating fundamental and conjugate fundamental representations of $U(\m)$ on the edges of a square lattice, with a simple nearest neighbor spin-spin coupling \cite{RS01}. The continuum limit partition function is similar to the one of the Potts model, with subtle differences:
\begin{equation}
 {\cal Z}_\m = F_{0,\q^{\pm 2}} + \sum_{j>0} \hat{D}_{j,0}F_{j,1}+\sum_{\substack{j>0,M>1 \\ M|j}}\sum_{\substack{0<p<M \\ p\wedge M=1}}
 \hat{D}_{j,\pi p/ M} F_{j,\e^{2\pi \i p/M}} \,,\label{decomploopZ}
\end{equation}
where again  
\begin{equation}
\hat{D}_{j,K}={1\over j}\sum_{r=0}^{j-1} \e^{2\i Kr}w(j,j\wedge r) \,,
\end{equation}
but $w$ takes the form
\begin{equation}
w(j,d)=\q^{2d}+\q^{-2d} \,,
\end{equation}
to be compared with \eqref{wjd_Potts}.
This decomposition of the torus partition function corresponds to the exact decomposition of the Hilbert space over modules of $\ATL{}{}$ in finite size:
\begin{equation}
 {\cal H}_\m = \bAStTL{ 0}{\q^{\pm 2}}\oplus \AStTL{1}{1} \oplus\bigoplus_{j>0}  \hat{D}_{j,0}\AStTL{j}{1}
 \oplus\bigoplus_{\substack{j>0,M>1 \\ M|j}} \bigoplus_{\substack{0<p<M \\ p\wedge M=1}}\hat{D}_{j,\pi p/ M}
 \AStTL{j,\e^{2\pi \i p/M}}\,\label{decomploopH}
\end{equation}
to be compared with \eqref{decompPottsH}.
An interesting difference with the Potts model  is the module $\AStTL{1}{1}$ which now occurs with multiplicity $\hat{D}_{1,1}'+1=\m^2-1$. A remarkable thing about this module is that it contains fields with conformal weight $(h_{1,-1},h_{1,1})$ and $(h_{1,1},h_{1,-1})$ with $\m$-independent values $(1,0)$ and $(0,1)$, like for chiral  {\sl currents}. Of course, we do not expect to have currents in the Potts model, since the symmetry of the latter is only discrete: this is compatible with the fact that  $\AStTL{1}{1}$  disappears in this case, as observed earlier. In contrast, for the $U(\m)$ model, we find a multiplicity $D_{\rm adj}=\m^2-1$ which is precisely the dimension of the adjoint representation, as expected for models with continuous symmetries. 
 As discussed in \cite{RS01},  $D_{\rm adj}$ is half the  multiplicity of the fields with weight $(1,1)$: the number $2D_{\rm adj}$ simply counts the two fields with weights $(h_{1,-1},h_{1,-1})$ in the $L_0$ or $\bar{L}_0$ Jordan cell, and there are $D_{\rm adj}$ such cells. 
 
It is then interesting to compare our results with those obtained by Gorbenko and Zan \cite{GorbenkoZan} in their study of the related $O(n)$ model. Their model describes ``dilute loops'' instead of the ``dense loops'' described by the $U(\m)$ model discussed here.%
\footnote{These ``dense loops'' are sometimes referred to more correctly as ``completely packed loops,'' because the cover all the edges of the medial lattice (see Section~\ref{sec:loops-clusters}).}
On top of this, it also differs from the $U(\m)$ model in that the number of non-contractible loops can be odd or even, while for $U(\m)$ it is necessarily even. It is  nonetheless instructive to compare the Jordan-cell structure for the currents with the one obtained in \cite{GorbenkoZan}. To match their normalizations, we set 
\begin{equation}
A\equiv {\Psi_{1,1}\over \sqrt{-2\kappa_{1,1}^{-1}\nu_{1,1}}}={\Psi_{1,1} \over \sqrt{-2b_{1,1}}}
\end{equation}
(this $A$ from \cite{GorbenkoZan} should not be confused with the combination of Virasoro generators $A_{1,1}$ used earlier), so that
\begin{equation}
\langle A(w,\bar{w})A(0)\rangle={\ln(w\bar{w})\over (w\bar{w})^2} \,.
\end{equation}
To match their current two-point function, %\jesper{added:}
 reading in the notations of \cite{GorbenkoZan}
\begin{equation}
\langle J(w,\bar{w})J(0)\rangle=-{1\over w^2} \,,
\end{equation}
we set $X\otimes \bar{Y}=\i J$. We have then %\jesper{$\kappa_{1,1}$ and $\nu_{1,1}$ throughout?}
\begin{equation}
L_1A={L_1\Psi_{1,1} \over \sqrt{-2 b_{1,1}}}=\sqrt{- b_{1,1} \over 2}X\otimes \bar{Y}=
\sqrt{ b_{1,1} \over 2}J=\sqrt{1-\beta^2\over 2}J \,.
\end{equation}
Since $\beta^2=x/(x+1)$, we find finally
\begin{equation}
L_1A=\sqrt{1\over 2(x+1)}J \,,
\end{equation}
where we recall that $\m=2\cos(\pi/(x+1))$. This must be  compared with equations (5.24) and (5.31) from \cite{GorbenkoZan}, where a similar but different result $L_1A=J/\sqrt{2x}$ is found, with $n=2\cos ( \pi/ x )$ and the usual central charge (\ref{centralc}). The shift $x\to x+1$ is familiar in the context of the dilute/dense phases relationship. We believe that a lattice analysis similar to the one we have presented here---but carried out instead for the dilute critical loop model and the dilute Temperley--Lieb algebra---would fully reproduce the results in \cite{GorbenkoZan}. Conversely, their analysis could be extended to reproduce our result for the currents in the $U(\m)$ model.

\section{Conclusion}
\label{sec:conclusion}

One of the lessons of this paper is that Jordan cells for $L_0$ or $\bar{L}_0$ are expected to appear in the continuum limit of the $Q$-state Potts  model and the loop models (dense or dilute), even though there are no such Jordan cells in the finite-size lattice model. This possibility was already mentioned in \cite{GRSV} in the particular case $c=0$, but occurs quite generically, whenever fields with degenerate conformal weights $h_{r,s}$, with $r,s\in\mathbb{N}^*$, appear in the spectrum. It  is in fact a logical consequence of the self-duality of the modules $\AStTL{j}{1}$, and thus can be argued on very general grounds.\footnote{The absence of Jordan cells on the lattice makes measuring  the logarithmic couplings $b_{r,s}$ appearing in the indecomposable modules (\ref{gencell}) quite difficult, as there seems to be no simple way of normalizing the lattice version of the field $\Psi_{r,s}$.}

The  CFT for the XXZ spin chain seems well described by the somewhat mundane Dotsenko--Fateev twisted boson theory \cite{GSJS}. In contrast, the $Q$-state Potts model or loop model CFTs appear  to be  new objects, related to but not identical with the  $c<1$ Liouville theory \cite{DelfinoViti,PiccoViti,IJS}, and slowly getting under control thanks to this and other recent work. A possible direction for future progress in understanding these CFTs  better would be to revisit the bootstrap approach of 
\cite{HJS} by taking into account properly regularized conformal blocks \cite{SylvainNextPaper}. More pressing qualitative questions, perhaps, include a better understanding of the OPEs: in particular, the OPEs for  the hull operators, which should have some interesting geometrical \cite{DL} and algebraic \cite{GJS18} meanings, or the OPE of the currents, where logarithmic features should explain why  there are much fewer than $D_{\rm adj}^2$ fields with weights $(1,1)$---or the behavior when $\q$ approaches a root of unity, and more Jordan cells appear, probably of rank higher than two. We hope to get back to these questions soon. 

\subsection*{Acknowledgments}

We thank J.\ Cardy, A.\ Gainutdinov,  V.\ Gorbenko, S.\ Ribault, B.\ Zan, and A.\ Zamolodchikov  for discussions.
We are grateful to S.\ Ribault for commenting on the manuscript.
This work was supported in part by the ERC advanced grant NuQFT. Moreover, L. Liu benefited from a Chateaubriand Fellowship of the Office for Science and
Technology of the Embassy of France in the United States. Part of the computations described in this paper were supported by the University of Southern California Center for Advanced Research Computing ({\tt carc.usc.edu}).

\appendix
\section{Numerics for the Koo--Saleur generators}\label{KS_Appendix}

%\linnea{All general discussion moved here. The figures can be updated whenever new versions are made. k is already used for something else in appendix B, and I'm not a fan of m for momentum since it's used for the loop weight, so I'm going with $\Pscaled$ as in the Bethe draft. Will edit notation in the end to match main text.}

%\lawrence{I might suggest a lightened notation for $\KSgen_{-1}[N]\Phi_{1,1}[N]$, $\bar{\KSgen}_{-1}[N]\bar{\Phi}_{1,1}[N]$, $\mathcal{A}_{1,2}[N] \Phi_{1,2}[N]$, and $\bar{\mathcal{A}}_{1,2}[N]\bar{\Phi}_{1,2}[N]$ since we use them so frequently and they make for a fairly heavy reading. Maybe $\lambda$, $\bar\lambda$, $\alpha$, $\bar\alpha$?}

Within this Appendix we provide partial evidence for the main results  given in equations \eqref{result1}, \eqref{MainRes3} and \eqref{moduleL}, by acting directly with the Koo--Saleur generators \eqref{generators} on eigenstates of the lattice Hamiltonian \eqref{H_phi}. In these numerical studies we shall split our state space at each system size $N$ into eigenspaces of the translation operator, with eigenvalues $\{\e^{2\pi \i \Pscaled/N} | 0\le \Pscaled \le N - 1\}$.
As the Hamiltonian is manifestly invariant under translation we may diagonalize it independently within each such sector. 
The Koo--Saleur generators exactly reproduce the fact that the action of $\KSgen_n[N]$ (resp. $\bar{\KSgen}_n[N]$) on a state of momentum $\Pscaled$ produces a state of momentum $\Pscaled - n$ (resp. $\Pscaled + n$), at finite size.  
%
%In finite size we block-diagonalize the Hamiltonian \eqref{H_phi} by changing to a translationally invariant basis, composed of states whose eigenvalue under translation is one of $\{\e^{2\pi \i k/N}| {-N/2}+1\le k \le N/2\}$. 
%
%
For a state of eigenvalue $\epsilon$ of the Hamiltonian at a given system size $N$, we consider lattice precursors to its conformal weights,\footnote{Sometimes called ``effective conformal weights.'' We will omit the qualifiers and simply refer to ``conformal weights'' when the context makes it clear that the term is being applied to lattice quantities. Similarly, we will frequently assign conformal weights $h_{r,s}$ given by the Kac formula to finite-size states---by this we mean that following a state for increasing $N$ leads to an extrapolation $h = h_{r,s}$.} which we also denote $(h,\bar{h})$, defined as the solutions to
\begin{equation}
\begin{split}
\epsilon &= \frac{2\pi}{N}\left(h + \bar{h} - \frac{c}{12}\right), \\
\Pscaled &= h - \bar{h}.
\end{split}
\end{equation}
By ``following'' a state (say, the lowest-energy state within a given sector of lattice momentum) as $N$ increases, and extrapolating the values of $h,\bar{h}$, we  can identify the conformal weights in the continuum limit.\footnote{We refer to Appendix \ref{app:singlets} for more advanced ``state following.'' %\linnea{Jesper, is this roughly what you had in mind for a footnote?}
} 
To make the notation lighter, we shall in this Appendix exclude the explicit dependence on system size, and write $\KSgen_n,\mathcal{A}_{r,s}$ rather than $\KSgen_n[N], \mathcal{A}_{r,s}[N]$ for Koo--Saleur generators and the combinations thereof. For the fields $\Phi_{r,s}$ the context will indicate whether we are discussing the field in the continuum limit or the corresponding link state at finite size, since at finite size (resp. in the continuum limit) $\Phi_{r,s}$ is acted upon by calligraphic operators $\KSgen_n$ and $\mathcal{A}_{r,s}$ (resp. Roman operators $L_n$ and $A_{r,s}$). We will in practice only be able to access low values of $r,s$ on the lattice, since larger system sizes are needed to accommodate a larger lattice momentum (which governs $r$) and a larger number of through-lines (which governs $s$).

%To distinguish between states in the limit and states at finite size, we shall in the latter case write the dependence on system size explicitly. Thus, as an example, $\Phi_{1,1} $ is the lattice state corresponding to the field $\Phi_{1,1}$ in the continuum. The Koo--Saleur generators, and null vector operators built from these, are given in calligraphic font, also with explicit dependence on $N$, so that $\mathcal{A}_{1,1} =\KSgen_{-1} $ is the lattice operator corresponding to $A_{1,1}=L_{-1}$ in the continuum.

%When we assign conformal weights to eigenstates of the Hamiltonian at finite size, this procedure is followed. 

Before discussing details of the numerics we must eliminate an ambiguity that may arise in the results due to phase degrees of freedom. In the following sections we will discuss quantities of the form $\|Z - Z'\|_2$,\footnote{The subscript 2 refers to the Euclidean norm or 2-norm.} where $Z$ and $Z'$ are (descendants of) eigenstates of the Hamiltonian (e.g. $Z = \KSgen_{-1} \Phi_{1,1} $ and $Z' = \bar{\KSgen}_{-1} \bar{\Phi}_{1,1} $). In quantum mechanics the overall phase of a vector or wave function has no observable consequences and $\e^{\i\alpha}Z$ for any real $\alpha$ would serve just as well in computations of observables. Typically one chooses the phase of a state such that its components in some basis are entirely real, where possible. In the situation at hand, the eigenvectors of the Hamiltonian are generically complex\footnote{By this, we mean that no choice of phase can make all of the components real.} and there is no canonical way to fix the relative phase between eigenvectors. The measurement of $\|Z - \e^{\i\alpha}Z'\|_2$ thus takes on a continuum of values. Where this ambiguity occurs, we fix the relative phase by choosing the value of $\alpha$ that minimizes this quantity:
\begin{equation}
\|Z - Z'\|_{\underline 2} \equiv \inf_\alpha \|Z - \e^{\i\alpha}Z'\|_2.
\end{equation}
This optimization is succinctly denoted by the underlined 2 in the notation $\|Z - Z'\|_{\underline 2}$.

Our main goal shall be to establish certain identities by observing whether deviations from these identities at finite size decay to zero. Let us give two examples. In order to provide evidence for \eqref{result1} in the sector of $j=0$ we wish to see if $\KSgen_{-1}  \mathbf{1}  \to 0$ as $N\to\infty$, with $\mathbf{1}=|h_{1,1},h_{1,1} \rangle$ being the identity state. Meanwhile, to provide evidence for \eqref{MainRes3} we would like to establish that  $\KSgen_{-1} \Phi_{1,1}  \to \bar{\KSgen}_{-1}  \bar{\Phi}_{1,1}  $, or equivalently that $ \KSgen_{-1} \Phi_{1,1}  - \bar{\KSgen}_{-1}  \bar{\Phi}_{1,1}  \to 0 $ as $N \to \infty$. Using the positive-definite scalar product to define a norm $\|V\|_2^2 = \braket{V}{V}$ we equivalently examine whether 
$\|  \KSgen_{-1}  \mathbf{1}  \|_2 \to 0$ and
$\|\KSgen_{-1} \Phi_{1,1}  - \bar{\KSgen}_{-1}  \bar{\Phi}_{1,1}  \|_{\underline 2} \to 0$ as $N\to \infty$.
%, where $\|\cdot\|_2$ is the norm induced by the positive-definite scalar product, $\|V\|_2^2 = \braket{V}{V}$. 

As shall be seen in the tables below, this simple measurement is insufficient for our purposes. Indeed, as $N$ increases the values observed actually
%An inspection of Table \ref{level1} indicates that this does not happen, and this measurement is 
grow in magnitude in most cases.
An interpretation of this observation is the fact that, since the finite-size Koo--Saleur generators do not yet furnish a representation of the Virasoro algebra, the action of $\bar{\KSgen}_{-1} $ on $\Phi_{1,1} $, for instance, produces a state with nonzero components even in highly excited eigenstates of the Hamiltonian. While each such component would tend to zero on its own, the number of these so-called ``parasitic couplings'' grows rapidly, yielding a nonzero contribution in total. 

To avoid the issue of this rapid growth, we choose to project on the $d$ lowest-energy states within the relevant sector of lattice momentum, keeping $d$ fixed as $N\to\infty$. This will be the subject of the following section.

\subsection{Projectors $\Pi^{(d)}$ and scaling-weak convergence}

For the following discussion we shall consider a concrete example, namely the fields $L_{-1}\Phi_{1,1}$ and $\bar{L}_{-1}\bar{\Phi}_{1,1}$ in the loop model. In the continuum limit, these fields have conformal weights $(1,1)$. Their lattice analogues $\KSgen_{-1}\Phi_{1,1} $ and $\bar{\KSgen}_{-1}\bar{\Phi}_{1,1} $ both belong to the sector of lattice momentum $\Pscaled = N/2$. By following the energies $\epsilon$ of states within this sector for increasing lattice sizes $N$, we find that the two lowest-energy states will correspond to these conformal weights.

%In the continuum limit, the fields $\bar{L}_{-1}X$ and $L_{-1}\bar{X}$ should be found with conformal weights $(1,1)$. Thus we should first investigate this relation in the subspace spanned by states corresponding to those in the continuum limit. 

%Thus $\bar{\KSgen}_{-1} X$ is entirely contained in the $\Pscaled = 0$ sector. By following states for increasing lattice sizes $N$, we identify the lowest four states in the $k = 0$ sector as containing all fields through $(1,1)$. 

Let us write schematically
\begin{equation}
{\KSgen}_{-1}  {\Phi}_{1,1}  = u + v,
\end{equation}
where $u$ is a linear combination of these two lowest states and $v$ represents all other states in the sector of $\Pscaled=N/2$. In order to %restrict our study to those aspects which must be present in the continuum theory and
 exclude the consequences of the ``parasitic couplings'' described above, we wish to build a projection operator $\Pi$ such that $\Pi  {\KSgen}_{-1}  {\Phi}_{1,1}   = u$.

In the basis of link states, and with respect to the scalar product $\braket{\cdot}{\cdot}$ where distinct link states are declared to be orthogonal, the Hamiltonian is not Hermitian. We must therefore distinguish between left and right eigenstates: the usual right eigenstates $\ket{i,R}$ are determined by the familiar $\HN \ket{i,R} = \epsilon_i \ket{i,R}$ and the left eigenstates $\bra{i,L}$ are determined via $\HN^\dagger \ket{i,L} = \epsilon_i \ket{i,L} \iff \bra{i,L}\HN = \epsilon_i\bra{i,L}$. The projectors $\dyad{i}$ of Hermitian quantum mechanics are replaced by
\begin{equation}
\Pi_i = \frac{\dyad{i,R}{i,L}}{\braket{i,L}{i,R}}
\end{equation}
which satisfy the expected properties of projectors, orthogonality and idempotency: $\Pi_i \Pi_j = \delta_{ij}\Pi_i$. For our purposes, $\Pi_i$ picks out the $i$th component of a vector expressed in the basis of right eigenvectors:
\begin{equation}
\Pi_i\sum_j c_j\ket{j,R} = c_i\ket{i,R}.
\end{equation}
Thus, letting $\Pi_{1,2}$ denote the projectors to the two lowest states of the $\Pscaled=N/2$ sector, the projector $\Pi^{(2)} = \Pi_1 + \Pi_2$ accomplishes the desired goal of $\Pi^{(2)} {\KSgen}_{-1}  {\Phi}_{1,1}   = u$. Since $\bar{L}_{-1}\bar{\Phi}_{1,1}$ has conformal weights $(1,1)$ as well, the projector $\Pi^{(2)}$ also truncates the lattice quantity $\bar{\KSgen}_{-1}  \bar{\Phi}_{1,1}  $ to the same two states.

%%%%%%%%%%%%%%%%%%%%%%

As discussed above it is not necessary to restrict to only the components in $u$ (given by the projection to the lowest two states in the example at hand)---one could also include higher energy states. As long as the rank of the projection operator is kept fixed, we expect the influence of such parasitic couplings to vanish as $N\to\infty$. We call convergence of values in the context of this procedure ``scaling-weak convergence.''
To illustrate this type of convergence, we will apply projectors of different rank $d$ to $\KSgen_{-1} \Phi_{1,1}  - \bar{\KSgen}_{-1} \bar{\Phi}_{1,1} $.
%
%
%
%
%In Table \ref{level1proj} we show the results of this truncation procedure and report the values of $\|\Pi^4\bar{\KSgen}_{-1} X - \KSgen_{-1} \bar{X})\|_2/\|\Pi^4\bar{\KSgen}_{-1} X\|_2$. The convergence is much improved over Table \ref{level1}.
%
%As $N$ increases, more and more of the lowest states of $\HN$ approach their continuum values and the multiplicities of conformal weights stabilize and match those predicted by algebraic analyses. We should thus expect that, 
%
We expect that for any fixed projector rank $d$ independent of $N$, so long as $\Pi^{(d)}$ is composed of the lowest $d$ states,\footnote{In fact, it is not strictly necessary to take the lowest $d$ states, but it suffices to take $d$ states with fixed conformal weights, so long as all of the lower states are eventually included as $d \to \infty$. In practice, however, convergence happens the most quickly at the lowest states.}
\begin{equation}\label{scaling_weak}
\lim_{N\to\infty} \|\Pi^{(d)}(\KSgen_{-1} \Phi_{1,1}  - \bar{\KSgen}_{-1} \bar{\Phi}_{1,1} )\|_{\underline 2} = 0\,,\quad \forall d \in \mathbb N\,;
\end{equation}
i.e., scaling-weak convergence of the lattice values towards the identity $L_{-1}\Phi_{1,1} = \bar L_{-1}\bar{\Phi}_{1,1}$.

The notion of scaling-weak convergence is defined and discussed in greater detail in \cite{GSJS}, where it is shown that a crucial difference compared to weak convergence is that limits of products of Koo--Saleur generators are in certain cases different than products of limits of Koo--Saleur generators, necessitating the insertion of projectors. This difference is found to affect the products with dual operators that are induced by the positive-definite inner product, as in $\|\KSgen_{-1}\Phi_{1,1}\|_2 =  \langle \Phi_{1,1} |\KSgen_{-1}^\dag\KSgen_{-1}\Phi_{1,1}\rangle$, but not the product $\KSgen_{-1}^2$ inside the operator $\mathcal{A}_{1,2}$ used below.

%\thiswillfailtocompile

In general, for any of the fields $Z$ relevant below, we say that its lattice analogue $\mathcal Z$ scaling-weakly converges %\lawrence{In saying this I attempted to follow the laws of grammar. However, common practice does not always do so (e.g. ``well ordered''\smiley but ``well ordering''\lightning) and we might elect to use a less awkward option} 
to zero if
\begin{equation}
\lim_{N\to\infty}\|\Pi^{(d)}\mathcal Z[N] \| = 0\,,\quad \forall d \in \mathbb N\,,
\label{generic_swc}
\end{equation}
with $\|\cdot\|$ some positive-definite norm. The meaning of $\Pi^{(d)}$ is context-dependent, but should be built in such a way that $\lim_{d\to\infty}\Pi^{(d)}$ effectively functions as the identity operator:\footnote{``Effectively,'' since $\lim_{d\to\infty}\Pi^{(d)}$ does not necessarily have to equal the identity operator. For instance, in the discussion of scaling-weak convergence of $\KSgen_{-1} \Phi_{1,1}  - \bar{\KSgen}_{-1} \bar{\Phi}_{1,1} $ to zero, $\Pi^{(d)}$ is built from the lowest $d$ states of lattice momentum $\Pscaled = N/2$. Thus $\lim_{d\to\infty}\Pi^{(d)}$ is the identity operator in the subspace of momentum $N/2$ and zero elsewhere. However, $\KSgen_{-1} \Phi_{1,1}  - \bar{\KSgen}_{-1} \bar{\Phi}_{1,1} $ is zero in all momentum sectors save for $\Pscaled = N/2$. Thus $\lim_{d\to\infty}\Pi^{(d)}$ effectively functions as the identity in this measurement. It is also possible to construct $\Pi^{(d)}$ using the $d$ lowest states of the entire Hamiltonian, regardless of momentum. This does not affect the limit \eqref{projToIdentity}, but merely the rate of convergence. In this case $\lim_{d\to\infty}\Pi^{(d)}$ becomes the identity operator.} 
\begin{equation}
\lim_{d\to\infty} \Pi^{(d)}\mathcal Z[N] = \mathcal Z[N]\,.\label{projToIdentity}
\end{equation}

An analogous discussion applies to the demonstration of the identity $A_{1,2}{\Phi}_{1,2}=\bar{A}_{1,2}\bar{\Phi}_{1,2}$, mutatis mutandis. We present numerical evidence that $\mathcal{A}_{1,2}  \Phi_{1,2}  - \bar{\mathcal{A}}_{1,2} \bar{\Phi}_{1,2} $ scaling-weakly converges to zero.

\begin{figure}
\centering
%\begin{minipage}{0.7\linewidth}
\includegraphics[width=0.9\textwidth]{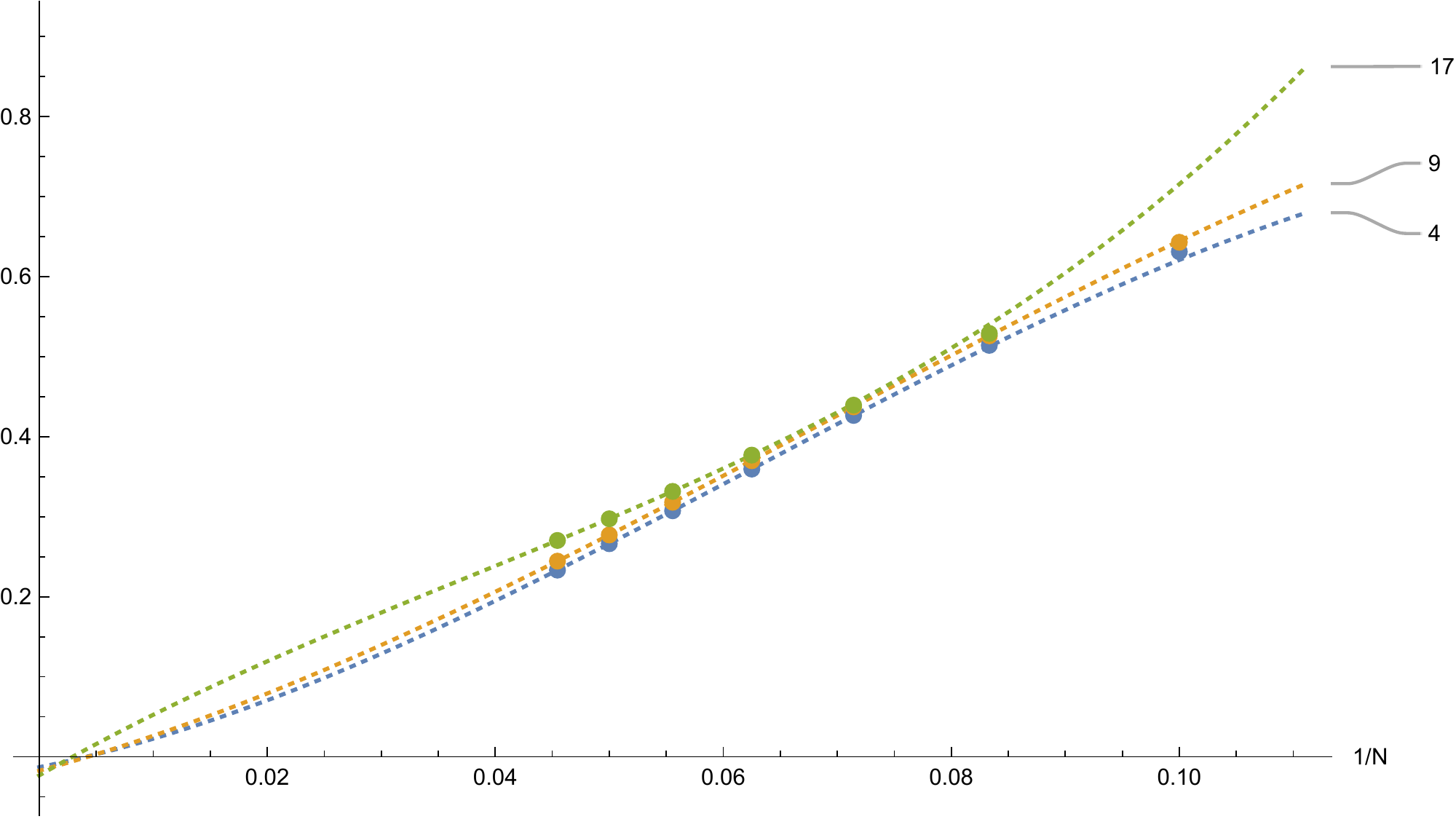}
%\end{minipage}
\caption{%$\bar A_{1,2} \Phi_{1,2} = \bar{A}_{1,2} \bar{\Phi}_{1,2}$ in the sense of scaling-weak convergence.
Comparison of lattice results using projectors of different rank, illustrating the concept of scaling-weak convergence \eqref{scaling_weak} at $j=2$.
The horizontal axis is $1/N$. The vertical axis is $\|\Pi^{(d)}( \mathcal{A}_{1,2} {\Phi}_{1,2}  -  \bar{\mathcal{A}}_{1,2} \bar{\Phi}_{1,2} ) \|_{\underline 2} / \| \Pi^{(d)} \mathcal{A}_{1,2} {\Phi}_{1,2}  \|_2$. The tags on the graphs indicate the rank $d$ of the projector $\Pi^{(d)}$. The dotted lines are third-order polynomial fits (in $1/N$) to the last four data points.}\label{x2vProj}
\end{figure}

\begin{figure}
\centering
%\begin{minipage}{0.7\linewidth}
\includegraphics[width=0.9\textwidth]{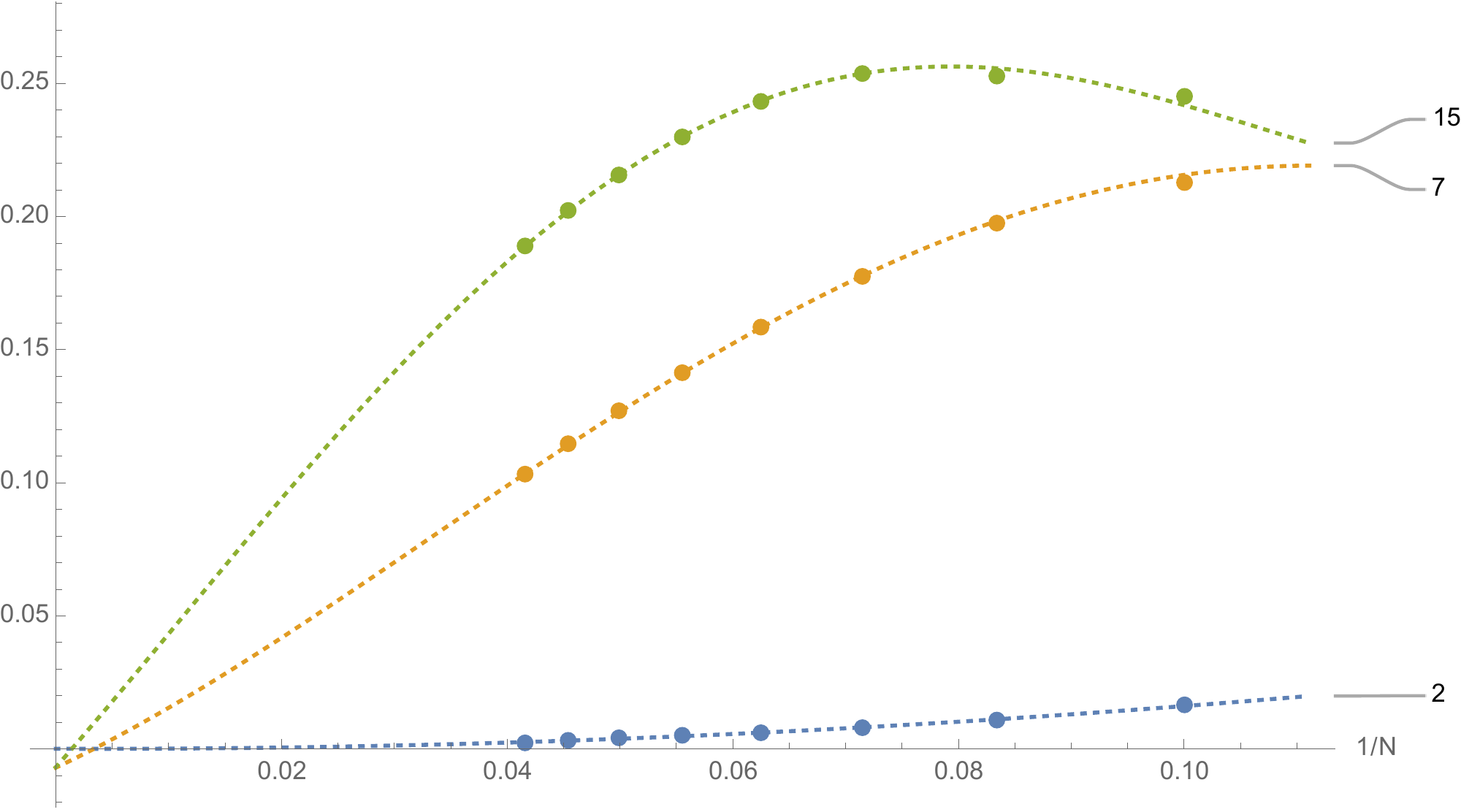}
%\end{minipage}
\caption{%$\bar L_{-1} \Phi_{1,1} = L_{-1} \bar{\Phi}_{1,1}$ in the sense of scaling-weak convergence. 
Comparison of lattice results using projectors of different rank, illustrating the concept of scaling-weak convergence \eqref{scaling_weak} at $j=1$.
The horizontal axis is $1/N$. The vertical axis is $\|\Pi^{(d)}(\KSgen_{-1} \Phi_{1,1}  - \bar{\KSgen}_{-1} \bar{\Phi}_{1,1} )\|_{\underline 2}/\|\Pi^{(d)}\KSgen_{-1} \Phi_{1,1}  \|_2$. The tags on the graphs indicate the rank $d$ of the projector $\Pi^{(d)}$. The dotted lines are fourth-order polynomial fits (in $1/N$) to the last five data points.}\label{x1vProj}
\end{figure}

 %\lawrence{Linnea, did you change this from $L_{-1}\Phi = \bar{L}_{-1}\bar{\Phi}$? I think it should be the unscripted one here since that is the identity we are trying to show with scaling-weak convergence and the finite size ones are not true}\linnea{good point, I've changed it back} 
We show in Figures \ref{x2vProj},\ref{x1vProj} that when applying projectors of different rank $d$, the numerical results extrapolate to almost the same value. We expect that the difference in the extrapolated values can be made arbitrarily small by including data points for large enough system sizes. 
% \linnea{text to be filled in once we have some extrapolation in the figures and know what to say. If the first figure has too much finite size effects for extrapolation, perhaps we can stick to the second one for j=1? If we keep it, could we move the tags to make it easier to see which curve is what?}


%\hubert{To discuss: the left and right eigenstates etc.}

\subsection{Numerical results for $\bAStTL{\!\!0}{\q^{\pm 2}}$} \label{loop_twisted}

%We show this in the case of the identity field $|0\rangle$ with weights $(h,\bar{h})=(0,0)$, which has a null state at level 1. Due to symmetry it is enough to consider the chiral action $L_{-1}|0\rangle$. 
%
%
%We first show that $\|L_{-1} |0\rangle \|_2$ is nonzero in Table \ref{Wbar_norm}. As mentioned above, however, there will be parasitic coupling influencing this value. We find that there is one state of the appropriate $(h,\bar{h})=(1,0)$ to be identified with the level 1 descendant. Projecting on this state%, which we call $|1,R\rangle$ as it is the lowest energy state within the sector of $\Pscaled=1$, we obtain Table \ref{W0q2_table_proj}. Within this table the values do not tend to zero either. 

Within the module $\AStTL{0}{\q^{\pm 2}}$, the link states corresponding to primary fields with degenerate conformal weights are never annihilated by the $A_{n,1}$ or $\bar{A}_{n,1}$ combinations of Virasoro generators. However, the module of interest for the study of the loop model is rather the quotient module $\bAStTL{\!\!0}{\q^{\pm 2}}$. In this module we consider in particular the lowest-energy link state of lattice momentum $\Pscaled=0$, which in the continuum limit will correspond to the identity state $\mathbf{1}$ with conformal weights $(h,\bar{h})=(0,0)$. We act on this state with the Koo--Saleur generator $\KSgen_{-1} $. The norm of the resulting state $\KSgen_{-1} \mathbf{1} $ defined through the positive-definite scalar product is shown in Table \ref{Wbar_norm}. (The norm of $\bar{\KSgen}_{-1} \mathbf{1} $ yields the same values %result \lawrence{"leads to a similar discussion"?} 
by symmetry.)
%\linnea{Lawrence, can you modify this table to only include the desired values of x? By the way, I'm trying to figure out if left-adjusting the columns inside might look better to line up the digits, the drawback is that the middle column will be shifted to the left... what do you think? I've kept 8 decimals, please verify if this is consistent with your tables. I can easily change it.} \lawrence{1. Done. 2. I'm not attached to any particular formatting; feel free to experiment around and see which one you like best. 3. I've carried over the default setting of Mathematica, which is 6 significant figures. Those which appear on the surface to have fewer than 6 sig figs should be padded with zeros (e.g. 0.053899 to 0.0538990), which are automatically removed in the Mathematica interface.}

Within this module there is no state to project on that we expect to give a nonzero contribution in the limit $N\to\infty$, the only state with the proper conformal weights having been excluded by the quotient. Projecting on the lowest-energy state still remaining in the sector of the appropriate lattice momentum we therefore expect the result to approach zero at $N \to \infty$ (Table \ref{Wbar_proj}).\footnote{In these tables, the peculiar value $x = \pi /\sec^{-1}(2 \sqrt{2})-1$ corresponds to $Q = 1/2$, which is further studied in Appendix \ref{app:singlets}.}

%\hubert{Give results for $L_{-1}|0\rangle$. Other results? }

\begin{table}[H]
\centering
%\begin{tabular}{c|ccccc}
%$N$ & $x=\frac{\pi }{3}$ & $\frac{\pi }{2}$ & $\frac{\pi }{\sec ^{-1}(2 \sqrt{2})}-1$ & $\e$ & $\pi$ \\
\begin{tabular}{c|ll>{\hspace*{8pt}}lll}
\multicolumn{1}{c}{$N$}  \vline  & \multicolumn{1}{c}{$x= \frac{\pi}{3}$}  & \multicolumn{1}{c}{$\frac{\pi}{2}$}  & \multicolumn{1}{c}{$\frac{\pi }{\sec^{-1}(2 \sqrt{2})}-1$}& \multicolumn{1}{c}{$\e$} & \multicolumn{1}{c}{$\pi$} \\
 \hline
%6 & $0$ & $0$ & $0$ & $0$ & $0$ \\
8 & $0.00105459$ & $0.0134764$ & $0.0140696$ & $0.0319876$ & $0.0360179$ \\
10 & $0.00151863$ & $0.0183461$ & $0.0191326$ & $0.0430288$ & $0.0484952$ \\
12 & $0.0018035$ & $0.0212243$ & $0.0221219$ & $0.0495035$ & $0.0558428$ \\
14 & $0.00200139$ & $0.0231978$ & $0.0241704$ & $0.053899$ & $0.0608373$ \\
16 & $0.00215397$ & $0.0247167$ & $0.0257464$ & $0.0572352$ & $0.0646218$ \\
18 & $0.00228117$ & $0.0259884$ & $0.0270657$ & $0.0599884$ & $0.0677341$ \\
20 & $0.00239306$ & $0.0271153$ & $0.0282345$ & $0.0623992$ & $0.0704482$ \\
22 & $0.00249505$ & $0.0281508$ & $0.0293087$ & $0.0645961$ & $0.0729122$ \\
24 & $0.00259016$ & $0.0291244$ & $0.0303188$ & $0.0666511$ & $0.0752099$
\end{tabular}
\caption{%$\|L_{-1}\ket{0}\|_2$.
The value of $\|\KSgen_{-1}  \mathbf{1}  \|_2$ for a given length $N$ and parameter $x$. $\mathbf{1}$ is the field in the $j = 0$ sector with conformal weights $(h_{1,1}, h_{1,1}) = (0,0)$. 
}\label{Wbar_norm}
\end{table}

\begin{table}[H]
\centering
\begin{tabular}{c|ll>{\hspace*{8pt}}lll}
\multicolumn{1}{c}{$N$}  \vline  & \multicolumn{1}{c}{$x= \frac{\pi}{3}$}  & \multicolumn{1}{c}{$\frac{\pi}{2}$}  & \multicolumn{1}{c}{$\frac{\pi }{\sec^{-1}(2 \sqrt{2})}-1$}& \multicolumn{1}{c}{$\e$} & \multicolumn{1}{c}{$\pi$} \\
\hline
8  &  0.00105459 & 0.0134764 & 0.0140696 & 0.0319876 & 0.0360179 \\
10  &  0.00154453 & 0.0201482 & 0.0209567 & 0.0434477 & 0.0485887 \\
12  &  0.00140952 & 0.0207429 & 0.0216739 & 0.0447805 & 0.0501347 \\
14  &  0.00121929 & 0.0192739 & 0.0202699 & 0.0427614 & 0.0480059 \\
16  &  0.00103467 & 0.0170396 & 0.0180351 & 0.0394988 & 0.044548 \\
18  &  0.000875168 & 0.0147437 & 0.0156912 & 0.0359407 & 0.0407787 \\
20  &  0.000742847 & 0.0126649 & 0.0135396 & 0.0325055 & 0.0371365 \\
22  &  0.00063449 & 0.0108785 & 0.0116714 & 0.0293585 & 0.0337921 \\
24  &  0.000545883 & 0.0093767 & 0.0100887 & 0.0265463 & 0.0307936 \\
\hline
extrapolation & 0.0000850643 & 0.00442133 & 0.00495163 & 0.000157526 & $-0.000258504$
\end{tabular}
\caption{%$\| \Pi^1 L_{-1}\ket{0}\|_2$ \linnea{excluding N=6 since they are all zero.}\lawrence{do we also want to remove that in the first table?}\linnea{might as well. I've removed it. We can mention somewhere that we did this.}
%
%The value of 
$\|\Pi^{(1)}\KSgen_{-1}  \mathbf{1}  \|_2$ 
with the same conventions as in Table \ref{Wbar_norm}.
%for a given length $N$ and parameter $x$.
% $\mathbf{1}$ is the field in the $j = 0$ sector with conformal weights $(h_{1,1}, h_{1,1}) = (0,0)$. 
%
$\Pi^{(1)}$ is a projection to the state of lowest energy within the $j = 0$, $\Pscaled = 1$ sector. This is the state within this sector that has conformal weights $(h_{1,-1}, h_{1,1}) = (1,0)$. 
The extrapolation is obtained by fitting the last five data points to a curve of the form $c_0 + c_1/N + c_2/N^2 + c_3/N^3 + c_4/N^4$.
}\label{Wbar_proj}
\end{table}

\subsection{Numerical results for $\AStTL{j}{1}$}\label{loop_lines}

In this section we numerically illustrate the equations $A_{1,2}{\Phi}_{1,2}=\bar{A}_{1,2}\bar{\Phi}_{1,2}$ and ${L}_{-1}{\Phi}_{1,1}=\bar{L}_{-1}\bar{{\Phi}}_{1,1}$ from section \ref{loop_Wj}. The general strategy is discussed above.

At $j=2$ we find $\Phi_{1,2} , \bar{\Phi}_{1,2} $ in the sectors of $\Pscaled= N/2-2$ and  $\Pscaled= N/2+2$, respectively. Thus, the descendant states $\mathcal{A}_{1,2} {\Phi}_{1,2} $ and $\bar{\mathcal{A}}_{1,2} \bar{\Phi}_{1,2} $ both belong to the sector of $\Pscaled = N/2$. We show first in Table  \ref{normAX} the norm $\|  \mathcal{A}_{1,2} {\Phi}_{1,2}  \|_2$ (which by symmetry equals $\| \bar{\mathcal{A}}_{1,2} \bar{\Phi}_{1,2}  \|_2$). 
%We shall use this norm for comparison when considering whether $\| \mathcal{A}_{1,2} {\Phi}_{1,2}  -  \bar{\mathcal{A}}_{1,2} \bar{\Phi}_{1,2}  \|_{\underline 2}$ is indeed tending to zero as $N\to\infty$. 
The ratio\footnote{While we numerically do observe scaling-weak convergence of $\mathcal{A}_{1,2}  \Phi_{1,2}  - \bar{\mathcal{A}}_{1,2} \bar{\Phi}_{1,2}$ to zero in the sense of definition \eqref{generic_swc}, here we report the values of $\|\mathcal{A}_{1,2}  \Phi_{1,2}  - \bar{\mathcal{A}}_{1,2} \bar{\Phi}_{1,2}\|_{\underline 2}/\|\mathcal{A}_{1,2}  \Phi_{1,2}  \|_2$ and $\|\Pi^{(4)}(\mathcal{A}_{1,2}  \Phi_{1,2}  - \bar{\mathcal{A}}_{1,2} \bar{\Phi}_{1,2})\|_{\underline 2}/\|\Pi^{(4)}\mathcal{A}_{1,2}  \Phi_{1,2}  \|_2$ to give a measure of relative deviation from zero. The decay of the latter quantity to zero implies the scaling-weak convergence of $\mathcal{A}_{1,2}  \Phi_{1,2}  - \bar{\mathcal{A}}_{1,2} \bar{\Phi}_{1,2}$ so long as the norm $\|\Pi^{(4)}\mathcal{A}_{1,2}  \Phi_{1,2}  \|_2$ does not grow too quickly. That this is the case can be seen in Table \ref{normPiAX}.} $\| \mathcal{A}_{1,2} {\Phi}_{1,2}  -  \bar{\mathcal{A}}_{1,2} \bar{\Phi}_{1,2}  \|_{\underline 2} / \|  \mathcal{A}_{1,2} {\Phi}_{1,2}  \|_2  $ is shown in Table \ref{level2}. We then repeat the same measurements using the projector $\Pi^{(4)}$ onto the four lowest-energy states in the sector of $\Pscaled = N/2$, which we have identified as containing all fields up to conformal weights $(h_{1,2}+2,h_{1,-2}) = (h_{1,-2},h_{1,-2})$. The results are shown in Table \ref{normPiAX} for $\| \Pi^{(4)} \mathcal{A}_{1,2} {\Phi}_{1,2}  \|_2 $ and in Table \ref{level2proj} for the ratio $\| \Pi^{(4)}( \mathcal{A}_{1,2} {\Phi}_{1,2}  -  \bar{\mathcal{A}}_{1,2} \bar{\Phi}_{1,2} ) \|_{\underline 2} / \| \Pi^{(4)} \mathcal{A}_{1,2} {\Phi}_{1,2}  \|_2  $.

Similarly, at $j=1$ we find $\Phi_{1,1} , \bar{\Phi}_{1,1} $ in the sectors of $\Pscaled= N/2-1$,  $\Pscaled= N/2+1$, and the descendants in the sector of $\Pscaled=N/2$. The norm $\|  \KSgen_{-1} {\Phi}_{1,1}  \|_2 = \| \bar{\KSgen}_{-1} \bar{\Phi}_{1,1}  \|_2$ is given in Table \ref{normLX}, and the ratio $\|\KSgen_{-1} \Phi_{1,1}  - \bar{\KSgen}_{-1} \bar{\Phi}_{1,1} \|_{\underline 2}/\|\KSgen_{-1} \Phi_{1,1}  \|_2$ is given in Table \ref{level1}. The same measurements are repeated with the projector $\Pi^{(2)}$ onto the two lowest-energy states in the sector of $\Pscaled=N/2$, which we have identified as containing all fields up to conformal weights $(h_{1,1}+1,h_{1,-1}) = (h_{1,-1},h_{1,-1}) = (1,1)$. These results are shown in Table \ref{normPiLX} and Table \ref{level1proj}.

%\linnea{added summary/discussion of tables on July 12. may need rephrasing.}

Both at $j=2$ (Table \ref{level2proj}) and $j=1$ (Table \ref{level1proj}) we find that the results support the equations $A_{1,2}{\Phi}_{1,2}=\bar{A}_{1,2}\bar{\Phi}_{1,2}$ and ${L}_{-1}{\Phi}_{1,1}=\bar{L}_{-1}\bar{{\Phi}}_{1,1}$ from section \ref{loop_Wj} when we use projectors $\Pi^{(d)}$, with $d=4$ and $d=2$, respectively. Here we have used the lowest rank such that the states with the relevant conformal weights are included among the states we project on. As discussed earlier and illustrated in Figures \ref{x2vProj},\ref{x1vProj} we expect that the result in the limit $N \to \infty$ will remain the same for higher rank projectors. However, we do not expect the result to remain the same when no projector is applied. Indeed, the values in Tables \ref{level2},\ref{level1} do not tend to zero as $N$ increases. The numerical proximity of $\|  \mathcal{A}_{1,2} {\Phi}_{1,2}  \|_2$ to $\| \Pi^{(4)} \mathcal{A}_{1,2} {\Phi}_{1,2}  \|_2$ and of $\|\KSgen_{-1}  \Phi_{1,1}  \|_2$ to $\|\Pi^{(2)}\KSgen_{-1}  \Phi_{1,1}  \|_2$ strongly indicates that the lack of convergence can be attributed to parasitic couplings to higher states, however small these couplings may be.%\lawrence{I came up with this point in a delirium after hours of editing, I will make it sound better soon}

%%%%%%%%%%%%%%%%%%%% j=2 %%%%%%%%%%%%%%%%%%%%%%%%%%

\begin{table}[H]
\centering
%\begin{tabular}{c|ccccc}
%$N$ & $x=\frac{\pi }{3}$ & $\frac{\pi }{2}$ & $\frac{\pi }{\sec ^{-1}(2 \sqrt{2})}-1$ & $\e$ & $\pi$ \\
\begin{tabular}{c|ll>{\hspace*{8pt}}lll}
\multicolumn{1}{c}{$N$}  \vline  & \multicolumn{1}{c}{$x= \frac{\pi}{3}$}  & \multicolumn{1}{c}{$\frac{\pi}{2}$}  & \multicolumn{1}{c}{$\frac{\pi }{\sec^{-1}(2 \sqrt{2})}-1$}& \multicolumn{1}{c}{$\e$} & \multicolumn{1}{c}{$\pi$} \\
 \hline
 10 & 0.724502 & 0.855299 & 0.859681 & 0.965547 & 0.985531 \\
 12 & 0.76937 & 0.879924 & 0.883733 & 0.977938 & 0.996234 \\
 14 & 0.794399 & 0.89486 & 0.898393 & 0.987034 & 1.00452 \\
 16 & 0.808303 & 0.903779 & 0.907193 & 0.993487 & 1.01062 \\
 18 & 0.81569 & 0.908984 & 0.912369 & 0.99819 & 1.01523 \\
 20 & 0.819132 & 0.91201 & 0.915425 & 1.00203 & 1.01916 \\
 22 & 0.820131 & 0.913885 & 0.917377 & 1.00577 & 1.02311
\end{tabular}
\caption{%$\|\bar AX\|_2$.
The value of $\| \mathcal{A}_{1,2}  \Phi_{1,2}  \|_2$ for a given length $N$ and parameter $x$.
%
%The value of $\| \mathcal{A}_{1,2}  \Phi_{1,2}  \|_2$ for a given length $N$ and parameter $x$. 
$\Phi_{1,2} $ is the field in the $j = 2$ sector with conformal weights $(h_{1,2}, h_{1,-2})$.
%
%
%The extrapolation is obtained by fitting the data to a curve of the form $c_0 + c_1/N + c_2/N^2 + c_3/N^3 + c_4/N^4$.
}
\label{normAX}
\end{table}

\begin{table}[H]
\centering
%\begin{tabular}{c|ccccc}
%$N$ & $x=\frac{\pi }{3}$ & $\frac{\pi }{2}$ & $\frac{\pi }{\sec ^{-1}(2 \sqrt{2})}-1$ & $\e$ & $\pi$ \\
\begin{tabular}{c|ll>{\hspace*{8pt}}lll}
\multicolumn{1}{c}{$N$}  \vline  & \multicolumn{1}{c}{$x= \frac{\pi}{3}$}  & \multicolumn{1}{c}{$\frac{\pi}{2}$}  & \multicolumn{1}{c}{$\frac{\pi }{\sec^{-1}(2 \sqrt{2})}-1$}& \multicolumn{1}{c}{$\e$} & \multicolumn{1}{c}{$\pi$} \\
 \hline
 $10$ & $0.211762$ & $0.451277$ & $0.458346$ & $0.621868$ & $0.653152$ \\
 $12$ & $0.151572$ & $0.359414$ & $0.365882$ & $0.523169$ & $0.555244$ \\
 $14$ & $0.115725$ & $0.304912$ & $0.310899$ & $0.459319$ & $0.490658$ \\
 $16$ & $0.09305$ & $0.275545$ & $0.281251$ & $0.42097$ & $0.450592$ \\
 $18$ & $0.078301$ & $0.264041$ & $0.269673$ & $0.402367$ & $0.429735$ \\
 $20$ & $0.0688143$ & $0.265678$ & $0.271421$ & $0.399548$ & $0.424477$ \\
 $22$ & $0.0631192$ & $0.277084$ & $0.283096$ & $0.409437$ & $0.432035$
\end{tabular}
\caption{%The value of $\|\bar A X - A \bar X\|_2/\|\bar A X\|_2$ for a given length $N$ and parameter $x$. Note that $\|\bar A X\|_2 = \|A\bar X\|_2$. $X$ is a field in the $j = 2$ sector with conformal weights $(h_{1,-2}, h_{1,2})$. Same for $\bar X$ but with $(h_{1,2}, h_{1,-2})$.
%
%The value of 
$\|\mathcal{A}_{1,2}  \Phi_{1,2}  - \bar{\mathcal{A}}_{1,2} \bar{\Phi}_{1,2}\|_{\underline 2}/\|\mathcal{A}_{1,2}  \Phi_{1,2}  \|_2$ 
with the same conventions as in Table \ref{normAX}. 
%
%for a given length $N$ and parameter $x$. 
$\Phi_{1,2} $ and $ \bar{\Phi}_{1,2}$ are fields in the $j = 2$ sector with conformal weights $(h_{1,2}, h_{1,-2})$ and  $(h_{1,-2}, h_{1,2})$.  
%The extrapolation is obtained by fitting the data to a curve of the form $c_0 + c_1/N + c_2/N^2 + c_3/N^3 + c_4/N^4$.
}
\label{level2}
\end{table}

\begin{table}[H]
\centering
%\begin{tabular}{c|ccccc}
%$N$ & $x=\frac{\pi }{3}$ & $\frac{\pi }{2}$ & $\frac{\pi }{\sec ^{-1}(2 \sqrt{2})}-1$ & $\e$ & $\pi$ \\
\begin{tabular}{c|ll>{\hspace*{8pt}}lll}
\multicolumn{1}{c}{$N$}  \vline  & \multicolumn{1}{c}{$x= \frac{\pi}{3}$}  & \multicolumn{1}{c}{$\frac{\pi}{2}$}  & \multicolumn{1}{c}{$\frac{\pi }{\sec^{-1}(2 \sqrt{2})}-1$}& \multicolumn{1}{c}{$\e$} & \multicolumn{1}{c}{$\pi$} \\
 \hline
 10 & 0.724473 & 0.853535 & 0.857809 & 0.959148 & 0.977917 \\
 12 & 0.769136 & 0.874667 & 0.87823 & 0.964825 & 0.981331 \\
 14 & 0.793913 & 0.886208 & 0.889353 & 0.967146 & 0.982324 \\
 16 & 0.807582 & 0.891969 & 0.894866 & 0.96719 & 0.981519 \\
 18 & 0.814765 & 0.894091 & 0.89683 & 0.965628 & 0.979399 \\
 20 & 0.81803 & 0.89391 & 0.896546 & 0.962969 & 0.976361 \\
 22 & 0.818871 & 0.892283 & 0.894845 & 0.959593 & 0.972719
\end{tabular}
\caption{%$\|\Pi^{(4)}\bar AX\|_2$.
%
%The value of 
$\|\Pi^{(4)} \mathcal{A}_{1,2}  \Phi_{1,2}  \|_2$ 
with the same conventions as in Table \ref{normAX}.
%
%for a given length $N$ and parameter $x$. 
%$\Phi_{1,2} $ is the field in the $j = 2$ sector with conformal weights $(h_{1,2}, h_{1,-2})$.
%
$\Pi^{(4)}$ is a projection to the four states of lowest energy within the $j = 2$, $\Pscaled = N/2$ sector. These are the states within this sector that have conformal weights up to $(h_{1,-2}, h_{1,-2})$. }
%
%The extrapolation is obtained by fitting the data to a curve of the form $c_0 + c_1/N + c_2/N^2 + c_3/N^3 + c_4/N^4$.

\label{normPiAX}
\end{table}

\begin{table}[H]
\centering
%\begin{tabular}{c|ccccc}
%$N$ & $x=\frac{\pi }{3}$ & $\frac{\pi }{2}$ & $\frac{\pi }{\sec ^{-1}(2 \sqrt{2})}-1$ & $\e$ & $\pi$ \\
\begin{tabular}{c|>{\hspace*{8pt}}l>{\hspace*{8pt}}l>{\hspace*{8pt}}l>{\hspace*{8pt}}ll}
\multicolumn{1}{c}{$N$}  \vline  & \multicolumn{1}{c}{$x= \frac{\pi}{3}$}  & \multicolumn{1}{c}{$\frac{\pi}{2}$}  & \multicolumn{1}{c}{$\frac{\pi }{\sec^{-1}(2 \sqrt{2})}-1$}& \multicolumn{1}{c}{$\e$} & \multicolumn{1}{c}{$\pi$} \\
 \hline
 $10$ & $0.210763$ & $0.439128$ & $0.445749$ & $0.600649$ & $0.63123$ \\
 $12$ & $0.148284$ & $0.332095$ & $0.337881$ & $0.482952$ & $0.514155$ \\
 $14$ & $0.110152$ & $0.257777$ & $0.26275$ & $0.39576$ & $0.426568$ \\
 $16$ & $0.0852125$ & $0.204972$ & $0.209224$ & $0.329727$ & $0.359515$ \\
 $18$ & $0.0679876$ & $0.16651$ & $0.170153$ & $0.278915$ & $0.307392$ \\
 $20$ & $0.0555742$ & $0.13779$ & $0.14093$ & $0.239167$ & $0.26624$ \\
 $22$ & $0.0463202$ & $0.115847$ & $0.118574$ & $0.207563$ & $0.233243$ \\
 \hline
extrapolation & $\hspace*{-8pt}-0.0015914$ & $\hspace*{-8pt}-0.000137804$ & $0.0000985782$ & $\hspace*{-8pt}-0.000155346$ & $0.00040305$
\end{tabular}
\caption{%The value of $\|\Pi^{(4)}(\bar A X - A \bar X)\|_2/\|\Pi^{(4)}\bar A X\|_2$ for a given length $N$ and parameter $x$. $X$ is a field in the $j = 2$ sector with conformal weights $(h_{1,-2}, h_{1,2})$. Same for $\bar X$ but with $(h_{1,2}, h_{1,-2})$. $\Pi^{(4)}$ is a projection to the lowest four states in the $j = 2$, $\Pscaled = N/2$ sector. The extrapolation is obtained by fitting the data to a curve of the form $c_0 + c_1/N + c_2/N^2 + c_3/N^3$.
%
%The value of 
$\|\Pi^{(4)}(\mathcal{A}_{1,2}  \Phi_{1,2}  - \bar{\mathcal{A}}_{1,2} \bar{\Phi}_{1,2})\|_{\underline 2}/\|\Pi^{(4)}\mathcal{A}_{1,2}  \Phi_{1,2}  \|_2$
with the same conventions and the same projector as in Table \ref{normPiAX}.
% for a given length $N$ and parameter $x$. 
$\Phi_{1,2} $ and $ \bar{\Phi}_{1,2}$ are fields in the $j = 2$ sector with conformal weights $(h_{1,2}, h_{1,-2})$ and  $(h_{1,-2}, h_{1,2})$. 
%
%$\Pi^{(4)}$ is a projection to the four states of lowest energy within the $j = 2$, $\Pscaled = N/2$ sector. These are the states within this sector that have conformal weights $(h_{1,-2}, h_{1,-2})$. 
%
%The extrapolation is obtained by fitting the data to a curve of the form $c_0 + c_1/N + c_2/N^2 + c_3/N^3 + c_4/N^4$.
Extrapolation as in Table \ref{Wbar_proj}.
}
\label{level2proj}
\end{table}

%%%%%%%%%%%%%%% j=1 %%%%%%%%%%%%%%%%%%%%

\begin{table}[H]
\centering
%\begin{tabular}{c|ccccc}
%$N$ & $x=\frac{\pi }{3}$ & $\frac{\pi }{2}$ & $\frac{\pi }{\sec ^{-1}(2 \sqrt{2})}-1$ & $\e$ & $\pi$ \\
\begin{tabular}{c|ll>{\hspace*{8pt}}lll}
\multicolumn{1}{c}{$N$}  \vline  & \multicolumn{1}{c}{$x= \frac{\pi}{3}$}  & \multicolumn{1}{c}{$\frac{\pi}{2}$}  & \multicolumn{1}{c}{$\frac{\pi }{\sec^{-1}(2 \sqrt{2})}-1$}& \multicolumn{1}{c}{$\e$} & \multicolumn{1}{c}{$\pi$} \\
 \hline
 8 & 0.325797 & 0.350787 & 0.351607 & 0.372103 & 0.376185 \\
 10 & 0.329276 & 0.352661 & 0.353516 & 0.376516 & 0.381388 \\
 12 & 0.328903 & 0.351523 & 0.35239 & 0.3768 & 0.382216 \\
 14 & 0.327108 & 0.349302 & 0.350173 & 0.375408 & 0.381203 \\
 16 & 0.324801 & 0.346738 & 0.347609 & 0.37335 & 0.379417 \\
 18 & 0.32236 & 0.34414 & 0.34501 & 0.371078 & 0.377346 \\
 20 & 0.319948 & 0.341637 & 0.342507 & 0.368803 & 0.375224 \\
 22 & 0.317637 & 0.339281 & 0.340151 & 0.366621 & 0.373165
\end{tabular}
\caption{%$\|L_{-1}X\|_2$.
The value of 
$\|\KSgen_{-1}  \Phi_{1,1}  \|_2$ 
for a given length $N$ and parameter $x$.
%for a given length $N$ and parameter $x$. 
$\Phi_{1,1} $ is the field in the $j = 1$ sector with conformal weights $(h_{1,1}, h_{1,-1}) = (0,1)$.  
%
%The extrapolation is obtained by fitting the data to a curve of the form $c_0 + c_1/N + c_2/N^2 + c_3/N^3 + c_4/N^4$.
}
\label{normLX}
\end{table}

\begin{table}[H]
\centering
%\begin{tabular}{c|ccccc}
%$N$ & $x=\frac{\pi }{3}$ & $\frac{\pi }{2}$ & $\frac{\pi }{\sec ^{-1}(2 \sqrt{2})}-1$ & $\e$ & $\pi$ \\
\begin{tabular}{c|ll>{\hspace*{8pt}}lll}
\multicolumn{1}{c}{$N$}  \vline  & \multicolumn{1}{c}{$x= \frac{\pi}{3}$}  & \multicolumn{1}{c}{$\frac{\pi}{2}$}  & \multicolumn{1}{c}{$\frac{\pi }{\sec^{-1}(2 \sqrt{2})}-1$}& \multicolumn{1}{c}{$\e$} & \multicolumn{1}{c}{$\pi$} \\
 \hline
 $8$ & $0.0106586$ & $0.104747$ & $0.108207$ & $0.18703$ & $0.199715$ \\
 $10$ & $0.0114921$ & $0.118893$ & $0.12311$ & $0.224566$ & $0.241811$ \\
 $12$ & $0.0117599$ & $0.124101$ & $0.128674$ & $0.243359$ & $0.263743$ \\
 $14$ & $0.0119545$ & $0.127043$ & $0.131823$ & $0.255388$ & $0.278124$ \\
 $16$ & $0.0121676$ & $0.129476$ & $0.134403$ & $0.264557$ & $0.289152$ \\
 $18$ & $0.0124147$ & $0.131931$ & $0.136979$ & $0.272425$ & $0.298551$ \\
 $20$ & $0.012693$ & $0.134557$ & $0.139716$ & $0.279679$ & $0.307105$ \\
 $22$ & $0.012996$ & $0.13737$ & $0.142638$ & $0.286636$ & $0.315194$
\end{tabular}
\caption{%The value of $\|\bar{\KSgen}_{-1} X - \KSgen_{-1} \bar{X}\|_2/\|\bar{\KSgen}_{-1} X\|_2$ for a given length $N$ and parameter $x$. $X$ is a field in the $j = 1$ sector with conformal weights $(h_{1,-1}, h_{1,1}) = (1,0)$. Same for $\bar X$ but with $(0,1)$.
%
%The value of 
$\|(\KSgen_{-1}  \Phi_{1,1}  - \bar{\KSgen}_{-1} \bar{\Phi}_{1,1})\|_{\underline 2}/\|\KSgen_{-1}  \Phi_{1,1}  \|_2$ 
with the same conventions as in Table \ref{normLX}.
%for a given length $N$ and parameter $x$. 
$\Phi_{1,1} $ and $ \bar{\Phi}_{1,1}$ are fields in the $j = 1$ sector with conformal weights $(h_{1,1}, h_{1,-1}) = (0,1)$ and  $(h_{1,-1}, h_{1,1}) = (1,0)$.  %The extrapolation is obtained by fitting the data to a curve of the form $c_0 + c_1/N + c_2/N^2 + c_3/N^3 + c_4/N^4$.
}
\label{level1}
\end{table}

\begin{table}[H]
\centering
%\begin{tabular}{c|ccccc}
%$N$ & $x=\frac{\pi }{3}$ & $\frac{\pi }{2}$ & $\frac{\pi }{\sec ^{-1}(2 \sqrt{2})}-1$ & $\e$ & $\pi$ \\
\begin{tabular}{c|ll>{\hspace*{8pt}}lll}
\multicolumn{1}{c}{$N$}  \vline  & \multicolumn{1}{c}{$x= \frac{\pi}{3}$}  & \multicolumn{1}{c}{$\frac{\pi}{2}$}  & \multicolumn{1}{c}{$\frac{\pi }{\sec^{-1}(2 \sqrt{2})}-1$}& \multicolumn{1}{c}{$\e$} & \multicolumn{1}{c}{$\pi$} \\
 \hline
 8 & 0.325786 & 0.349822 & 0.350572 & 0.368581 & 0.37208 \\
 10 & 0.329244 & 0.350622 & 0.35134 & 0.36961 & 0.373372 \\
 12 & 0.328845 & 0.348905 & 0.349594 & 0.367684 & 0.371562 \\
 14 & 0.327027 & 0.34636 & 0.347031 & 0.364836 & 0.368753 \\
 16 & 0.324704 & 0.343605 & 0.344262 & 0.361773 & 0.365685 \\
 18 & 0.322251 & 0.340885 & 0.341533 & 0.358769 & 0.362653 \\
 20 & 0.319832 & 0.338301 & 0.338943 & 0.355932 & 0.359775 \\
 22 & 0.317515 & 0.335886 & 0.336524 & 0.353297 & 0.357092
\end{tabular}
\caption{%$\|\Pi^{(2)}L_{-1}X\|_2$.
%The value of 
$\|\Pi^{(2)}\KSgen_{-1}  \Phi_{1,1}  \|_2$ 
with the same conventions as in Table \ref{normLX}. 
%for a given length $N$ and parameter $x$.
%
%$\Phi_{1,1} $ is the field in the $j = 1$ sector with conformal weights $(h_{1,1}, h_{1,-1})$. 
%
$\Pi^{(2)}$ is a projection to the two states of lowest energy within the $j = 1$, $\Pscaled = N/2$ sector. These are the states within this sector that have conformal weights up to $(h_{1,-1}, h_{1,-1}) = (1,1)$. 
%
%The extrapolation is obtained by fitting the data to a curve of the form $c_0 + c_1/N + c_2/N^2 + c_3/N^3 + c_4/N^4$.
}
\label{normPiLX}
\end{table}

\begin{table}[H]
\centering
%\begin{tabular}{c|ccccc}
%$N$ & $x=\frac{\pi }{3}$ & $\frac{\pi }{2}$ & $\frac{\pi }{\sec ^{-1}(2 \sqrt{2})}-1$ & $\e$ & $\pi$ \\
\begin{tabular}{c| >{\hspace*{8pt}}l>{\hspace*{8pt}}l>{\hspace*{6pt}}lll}
\multicolumn{1}{c}{$N$}  \vline  & \multicolumn{1}{c}{$x= \frac{\pi}{3}$}  & \multicolumn{1}{c}{$\frac{\pi}{2}$}  & \multicolumn{1}{c}{$\frac{\pi }{\sec^{-1}(2 \sqrt{2})}-1$}& \multicolumn{1}{c}{$\e$} & \multicolumn{1}{c}{$\pi$} \\
 \hline
 $8$ & $0.0000527492$ & $0.0058436$ & $0.00627048$ & $0.0215309$ & $0.0251946$ \\
 $10$ & $0.0000191654$ & $0.00297587$ & $0.00322369$ & $0.0133976$ & $0.0161156$ \\
 $12$ & $8.26633\times 10^{-6}$ & $0.00167699$ & $0.00183148$ & $0.00901364$ & $0.0111168$ \\
 $14$ & $4.03988\times 10^{-6}$ & $0.00102009$ & $0.00112185$ & $0.00641199$ & $0.00809058$ \\
 $16$ & $2.16718\times 10^{-6}$ & $0.0006584$ & $0.000728471$ & $0.00475643$ & $0.00612835$ \\
 $18$ & $1.25094\times 10^{-6}$ & $0.000445437$ & $0.000495467$ & $0.00364534$ & $0.004787994$ \\
 $20$ & $7.62288\times 10^{-7}$ & $0.000313082$ & $0.000349894$ & $0.00286778$ & $0.00383424$ \\
 $22$ & $4.87306\times 10^{-7}$ & $0.000227096$ & $0.000254879$ & $0.0023049$ & $0.00313292$ \\
 \hline
extrapolation & $4.38043\times 10^{-7}$ & $\hspace*{-8pt}-0.0000700002$ & $\hspace*{-8pt}-0.0000675678$ & $0.000161454$ & $0.0000896441$
\end{tabular}
\caption{%The value of 
$\|\Pi^{(2)}(\KSgen_{-1}  \Phi_{1,1}  - \bar{\KSgen}_{-1} \bar{\Phi}_{1,1})\|_{\underline 2}/\|\Pi^{(2)}\KSgen_{-1}  \Phi_{1,1}  \|_2$ 
with the same conventions and the same projector as Table \ref{normPiLX}.
%for a given length $N$ and parameter $x$. 
$\Phi_{1,1} $ and $ \bar{\Phi}_{1,1}$ are fields in the $j = 1$ sector with conformal weights $(h_{1,1}, h_{1,-1}) = (0,1)$ and  $(h_{1,-1}, h_{1,1}) = (1,0)$. 
%
%$\Pi^{(2)}$ is a projection to the two states of lowest energy within the $j = 1$, $\Pscaled = N/2$ sector. These are the states within this sector that have conformal weights $(h_{1,-1}, h_{1,-1})$. 
%
%The extrapolation is obtained by fitting the data to a curve of the form $c_0 + c_1/N + c_2/N^2 + c_3/N^3 + c_4/N^4$.
The extrapolation is obtained by fitting the last six data points to a curve of the form $c_0 + c_1/N + c_2/N^2 + c_3/N^3 + c_4/N^4$.
}
\label{level1proj}
\end{table}

\section{Observation of singlet states in the loop model}
\label{app:singlets}

%\linnea{subsection title -> title for now}

In this appendix we report transfer matrix computations that support some of our main results.

We consider the transfer matrix of the loop model on $N=2L$ sites, with $2j$ through-lines, in the geometry (see section~\ref{sec:loops-clusters} for details)
that corresponds to a Potts model on a square
lattice, with $L$ spins in each row and periodic boundary conditions. The corresponding loop model then lives on a tilted square lattice (the medial
lattice of the original, axially oriented square lattice). Our method of diagonalization has been explained in much detail in Appendix A of \cite{JS}.

We focus here on one well-chosen value, $Q = 1/2$, which can be considered representative for the case of generic values of $Q$.
For each size $L=5,6,\ldots,13$ we compute the first several hundred eigenvalues in each sector $\AStTL{j}{z^2}$ with $j=1,2,3$ and $z^{2j}=1$, extracting
the multiplicity, finite-size scaling dimension and lattice momentum of each eigenvalue. The multiplicities are always found to be either 1 or 2, and
we pay special attention to the singlets. The lattice momentum $\Pscaled$ can be identified only up to a sign, and it coincides with the conformal spin
modulo $L$, that is:
\begin{equation} \label{confspin}
 s \equiv h - \bar{h} = \Pscaled \mbox{ mod } L \,.
\end{equation}

One major difficulty in the study is that the $(i_L)$'th largest eigenvalue in the finite-size spectrum corresponds to the $(i_L)$'th lowest-lying scaling state
only for $L$ sufficiently large, and for all but the smallest few values of $i_L$ this simple situation is reached only when $L$ is much larger than the attainable system size. To nevertheless study the scaling states numerically, one therefore has to identify the sequences $(i_5,i_6,\ldots,i_{13})$
that correspond to any desired scaling field, using a lot of patience and a general methodology that is explained in Appendix A.5 of \cite{JS}.
Polynomial extrapolations of the finite-size scaling dimensions are then possible, most often using the data for all sizes (and only occasionally excluding
the first few sizes), leading to quite accurate estimates of the conformal scaling dimension %\lawrence{Since $x$ is taken, maybe $\Delta$?}
$\Delta \equiv h+\bar{h}$. Moreover, comparing
the values of $\Pscaled$ for several different $L$ in the sequence will permit us to lift the ``modulo $L$'' qualifier in \eqref{confspin} and determine $s$
(again up to a sign).

The values of $\Delta$ and $\pm s$ allow us to identify the corresponding scaling field, up to a few ambiguities.
To be precise, we are able to identify the corresponding primary field and the level of descendance on both the chiral and antichiral sides,
up to a possible overall exchange of chiral and antichiral components (recall that $s$ is determined only up to a sign). Our notation below takes this ambiguity
into account. For instance, a field which is descendant at level $(3,2)$ of a primary $\Phi$ will be denoted $L_{-3} \bar{L}_{-2} \Phi$, 
although it might in fact be any linear combination of the form
$\left( L_{-3} + \alpha L_{-2} L_{-1} + \beta L_{-1}^3 \right) \left( \bar{L}_{-2} + \gamma \bar{L}_{-1}^2 \right) \Phi$
for some unknown coefficients $\alpha,\beta,\gamma$---or indeed the same field with chiral and antichiral components being exchanged.

\subsection{$j=1$}

Results for the sector $\AStTL{1}{1}$ are shown in Table~\ref{Tab_exp_V11}. We have in fact identified the scaling fields for all lines with $i_{13} \le 60$
in this case, but to keep the table concise we show only the first 10 fields, along with several other fields that are either a primary or a singlet (or both).
The ranks $i_L$ of singlet fields are shown in red color, while those of the doublets are in black. Small numbers refer to finite-size levels for
which the lattice momentum $\Pscaled$ differs from the conformal spin $s$ by a non-trivial multiple of $L$, cf.~\eqref{confspin} (for more details, see Appendix A.5 of \cite{JS}).
The table shows all singlets with
$i_{12} \le 200$ (for the last seven lines the diagonalization for $L=13$ was numerically too demanding). The extrapolation of the scaling
dimension in shown to about the number of significant digits to which it agrees with the exact result.

\begin{table}
\begin{center}
\begin{tabular}{|c|ccccccccc|ll|l|}
 \cline{1-13}
  $s$
    & \multicolumn{9}{|c|}{$(i_5,i_6,\ldots,i_{13})$} & \multicolumn{2}{|c|}{$\Delta$} & Identification \\  \cline{2-12}
   & 5 & 6 & 7 & 8 & 9 & 10 & 11 & 12 & 13 & Numerics & Exact & of scaling field \\
  \hline
 0 & \textcolor{red}{1} & \textcolor{red}{1} & \textcolor{red}{1} & \textcolor{red}{1} & \textcolor{red}{1} & \textcolor{red}{1} & \textcolor{red}{1} & \textcolor{red}{1} & \textcolor{red}{1} & 0.187014 & 0.187027 & $\phi_{0,1} \otimes \phi_{0,-1} \equiv \psi_0$ \\
 1 & 2 & 2 & 2 & 2 & 2 & 2 & 2 & 2 & 2 & 1.000003 & 1 & $\phi_{1,1} \otimes \phi_{1,-1} \equiv \psi_1$ \\
 1 & 3 & 3 & 3 & 3 & 3 & 3 & 3 & 3 & 3 & 1.187040 & 1.187027 & $L_{-1} \psi_0$ \\
 2 & 4 & 4 & 4 & 4 & 4 & 4 & 4 & 4 & 4 & 1.9998 & 2 & $\bar{L}_{-1} \psi_1$ \\
 0 & \textcolor{red}{6} & \textcolor{red}{6} & \textcolor{red}{6} & \textcolor{red}{5} & \textcolor{red}{5} & \textcolor{red}{5} & \textcolor{red}{5} & \textcolor{red}{5} & \textcolor{red}{5} & 2.0016 & 2 & $L_{-1} \psi_1$ \\
 0 & \textcolor{red}{9} & \textcolor{red}{8} & \textcolor{red}{7} & \textcolor{red}{7} & \textcolor{red}{7} & \textcolor{red}{7} & \textcolor{red}{7} & \textcolor{red}{7} & \textcolor{red}{6} & 1.9985 & 2 & $L_{-1} \psi_1$ \\
 2 & 5 & 5 & 5 & 6 & 6 & 6 & 6 & 6 & 7 & 2.1882 & 2.1870 & $L_{-2} \psi_0$ \\
 0 & \textcolor{red}{7} & \textcolor{red}{9} & \textcolor{red}{8} & \textcolor{red}{8} & \textcolor{red}{8} & \textcolor{red}{8} & \textcolor{red}{8} & \textcolor{red}{8} & \textcolor{red}{8} & 2.18708 & 2.18703 & $L_{-1} \bar{L}_{-1} \psi_0$ \\
 2 & 8 & 10 & 10 & 9 & 9 & 9 & 9 & 9 & 9 & 2.18710 & 2.18703 & $L_{-2} \psi_0$ \\
 3 & {\tiny 5} & 7 & 9 & 10 & 10 & 10 & 10 & 10 & 10 & 2.992 & 3 & $\bar{L}_{-2} \psi_1$ \\ \hline
 2 & 19 & 22 & 24 & 23 & 23 & 23 & 24 & 22 & 22 & 3.43895 & 3.43892 & $\phi_{2,1} \otimes \phi_{2,-1} \equiv \psi_2$ \\
 0 & \textcolor{red}{18} & \textcolor{red}{23} & \textcolor{red}{27} & \textcolor{red}{28} & \textcolor{red}{27} & \textcolor{red}{27} & \textcolor{red}{28} & \textcolor{red}{28} & \textcolor{red}{27} & 4.002 & 4 & $L_{-2} \bar{L}_{-1} \psi_1$ \\
 0 & \textcolor{red}{23} & \textcolor{red}{30} & \textcolor{red}{32} & \textcolor{red}{37} & \textcolor{red}{40} & \textcolor{red}{38} & \textcolor{red}{37} & \textcolor{red}{38} & \textcolor{red}{38} & 4.00016 & 4 & $L_{-2} \bar{L}_{-1} \psi_1$ \\
 0 & \textcolor{red}{20} & \textcolor{red}{24} & \textcolor{red}{29} & \textcolor{red}{34} & \textcolor{red}{36} & \textcolor{red}{34} & \textcolor{red}{35} & \textcolor{red}{37} & \textcolor{red}{39} & 4.1864 & 4.1870 & $L_{-2} \bar{L}_{-2} \psi_0$ \\
 0 & \textcolor{red}{33} & \textcolor{red}{45} & \textcolor{red}{47} & \textcolor{red}{47} & \textcolor{red}{49} & \textcolor{red}{52} & \textcolor{red}{49} & \textcolor{red}{46} & \textcolor{red}{46} & 4.18707 & 4.18703 & $L_{-2} \bar{L}_{-2} \psi_1$ \\
 0 & \textcolor{red}{\tiny 25} & \textcolor{red}{43} & \textcolor{red}{56} & \textcolor{red}{71} & \textcolor{red}{82} & \textcolor{red}{88} & \textcolor{red}{99} & \textcolor{red}{103} & & 6.03 & 6 & $L_{-3} \bar{L}_{-2} \psi_1$ \\
 0 & \textcolor{red}{51} & \textcolor{red}{78} & \textcolor{red}{95} & \textcolor{red}{111} & \textcolor{red}{114} & \textcolor{red}{119} & \textcolor{red}{118} & \textcolor{red}{115} & & 5.423 & 5.439 & $L_{-2} \psi_2$ \\
 0 & \textcolor{red}{53} & \textcolor{red}{79} & \textcolor{red}{100} & \textcolor{red}{116} & \textcolor{red}{120} & \textcolor{red}{123} & \textcolor{red}{121} & \textcolor{red}{117} & & 5.433 & 5.439 & $L_{-2} \psi_2$ \\
 0 & --- & --- & \textcolor{red}{57} & \textcolor{red}{76} & \textcolor{red}{94} & \textcolor{red}{112} & \textcolor{red}{117} & \textcolor{red}{120} & & 6.178 & 6.187 & $L_{-3} \bar{L}_{-3} \psi_0$ \\
 0 & --- & --- & \textcolor{red}{65} & \textcolor{red}{88} & \textcolor{red}{103} & \textcolor{red}{116} & \textcolor{red}{129} & \textcolor{red}{132} & & 5.989 & 6 & $L_{-3} \bar{L}_{-2} \psi_1$ \\
 0 & \textcolor{red}{55} & \textcolor{red}{86} & \textcolor{red}{115} & \textcolor{red}{133} & \textcolor{red}{144} & \textcolor{red}{145} & \textcolor{red}{151} & \textcolor{red}{155} & & 5.994 & 6 & $L_{-3} \bar{L}_{-2} \psi_1$ \\
 0 & \textcolor{red}{52} & \textcolor{red}{84} & \textcolor{red}{119} & \textcolor{red}{144} & \textcolor{red}{166} & \textcolor{red}{174} & \textcolor{red}{175} & \textcolor{red}{182} & & 6.32 & 6.19 & $L_{-3} \bar{L}_{-3} \psi_0$ \\
\hline
\end{tabular}
\end{center}
\caption{Conformal spectrum in the sector $\AStTL{1}{1}$ for $Q=1/2$.}
  \label{Tab_exp_V11}
\end{table}

The primaries
\begin{equation}
 \Phi_{e,j} = \phi_{e,j} \otimes \bar{\phi}_{e,-j} \,, \qquad
 \bar{\Phi}_{e,j} = \phi_{e,-j} \otimes \bar{\phi}_{e,j}
\end{equation}
with $j=1$ can be seen in the table for $e=0,1,2$, corresponding to the lines with $i_{13} = 1$, $2$, $22$. The latter two are doublets, while the first one is
a singlet, as $\Phi_{0,1} = \bar{\Phi}_{0,1}$ because of the identification $\phi_{r,s} = \phi_{-r,-s}$. For the same reason, we find several of the
spinless descendants of $\Phi_{0,1}$ to be singlets (e.g., $i_{13} = 8, 39$ and $i_{12} = 120, 182$).

A more remarkable finding is the singlet nature of the pair of lines with $i_{13} = 5, 6$. If we had been dealing with a product of Verma modules, these
would have formed a degenerate doublet. Instead we see here a manifestation of the duality
\begin{equation}
 L_{-1} \Phi_{1,1} = \bar{L}_{-1} \bar{\Phi}_{1,1}
\end{equation}
and so the two singlets should instead be identified with the top and bottom fields in a Jordan-cell representation of the type \eqref{diamondPhi12},
with $(e,j) = (1,1)$.

Much lower in the spectrum, we similarly remark the singlet nature of the lines with $i_{12} = 115, 117$. They show the duality
\begin{equation}
 A_{2,1} \Phi_{2,1} = \bar{A}_{2,1} \bar{\Phi}_{2,1}
\end{equation}
and give evidence for a Jordan cell with $(e,j) = (2,1)$.

\subsection{$j=2$}

In the same way, we show results for the sectors $\AStTL{2}{z^2}$ in Table~\ref{Tab_exp_V2}. 
Notice that the results for all permissible cases, $z^{2j}=1$, are shown in the same table and the corresponding scaling levels
are marked by the additional label $k=0, 1$ for $z^2 = \e^{2 \pi \i k / j}$.

\begin{table}
\begin{center}
\begin{tabular}{|c|ccccccccc|ll|l|}
\cline{1-13}
  $k, s$
    & \multicolumn{9}{|c|}{$(i_5,i_6,\ldots,i_{13})$} & \multicolumn{2}{|c|}{$\Delta$} & Identification \\  \cline{2-12}
   & 5 & 6 & 7 & 8 & 9 & 10 & 11 & 12 & 13 & Numerics & Exact & of scaling field \\
  \hline
 0, 0 & \textcolor{red}{1} & \textcolor{red}{1} & \textcolor{red}{1} & \textcolor{red}{1} & \textcolor{red}{1} & \textcolor{red}{1} & \textcolor{red}{1} & \textcolor{red}{1} & \textcolor{red}{1} & 1.1095698 & 1.1095673 & $\phi_{0,2} \times \phi_{0,-2} \equiv \phi_0$ \\
 1, 1 & 2 & 2 & 2 & 2 & 2 & 2 & 2 & 2 & 2 & 1.312824 & 1.312810 & $\phi_{1/2,2} \times \phi_{1/2,-2} \equiv \phi_{1/2}$ \\
 0, 2 & 3 & 3 & 3 & 3 & 3 & 3 & 3 & 3 & 3 & 1.92264 & 1.92254 & $\phi_{1,2} \times \phi_{1,-2} \equiv \phi_1$ \\
 0, 1 & 5 & 4 & 4 & 4 & 4 & 4 & 4 & 4 & 4 & 2.1099 & 2.1096 & $\bar{L}_{-1} \phi_{0}$ \\
 1, 2 & 4 & 5 & 5 & 5 & 5 & 5 & 5 & 5 & 5 & 2.31304 & 2.31281 & $\bar{L}_{-1} \phi_{1/2}$ \\
 1, 0 & 7 & 7 & 6 & 6 & 6 & 6 & 6 & 6 & 6 & 2.31297 & 2.31281 & $L_{-1} \phi_{1/2}$ \\
 0, 3 & {\tiny 6} & 8 & 8 & 7 & 7 & 7 & 7 & 7 & 7 & 2.9230 & 2.9225 & $\bar{L}_{-1} \phi_{1}$ \\
 0, 2 & 6 & 9 & 9 & 9 & 9 & 8 & 8 & 8 & 8 & 3.1075 & 3.1096 & $\bar{L}_{-2} \phi_{0}$ \\
 1, 3 & {\tiny 8} &10 & 11 & 11 & 10 & 10 & 10 & 9 & 9 & 2.9393 & 2.9388 & $\phi_{3/2,2} \times \phi_{3/2,-2} \equiv \phi_{3/2}$ \\
 0, 1 & 13 & 13 & 13 & 13 & 11 & 11 & 11 & 10 & 10 & 2.9229 & 2.9225 & $L_{-1} \phi_{1}$ \\ \hline
 0, 0 & \textcolor{red}{9} & \textcolor{red}{11} & \textcolor{red}{12} & \textcolor{red}{15} & \textcolor{red}{12} & \textcolor{red}{12} & \textcolor{red}{12} & \textcolor{red}{12} & \textcolor{red}{12} & 3.1101 & 3.1096 & $L_{-1} \bar{L}_{-1} \phi_{0}$ \\
 0, 0 & \textcolor{red}{19} & \textcolor{red}{24} & \textcolor{red}{28} & \textcolor{red}{30} & \textcolor{red}{28} & \textcolor{red}{29} & \textcolor{red}{27} & \textcolor{red}{25} & \textcolor{red}{24} & 3.9244 & 3.9225 & $L_{-2} \phi_1$ \\
 0, 0 & 21 & 27 & 31 & 32 & 32 & 33 & 31 & 30 & 25 & 3.9231 & 3.9225 & $L_{-2} \phi_1$ \\
 0, 0 & \textcolor{red}{31} & \textcolor{red}{40} & \textcolor{red}{40} & \textcolor{red}{42} & \textcolor{red}{41} & \textcolor{red}{40} & \textcolor{red}{36} & \textcolor{red}{37} & \textcolor{red}{35} & 3.9202 & 3.9225 & $L_{-2} \phi_1$ \\
 0, 4 & {\tiny 23} & {\tiny 34} & {\tiny 37} & 43 & 47 & 48 & 45 & 46 & & 4.356 & 4.361 & $\phi_{2,2} \times \phi_{2,-2} \equiv \phi_{2}$ \\
\hline
\end{tabular}
\end{center}
\caption{Conformal spectrum in the sector $\AStTL{2}{z^2}$ for $Q=1/2$. The label $k$ corresponds to $z^2 = (-1)^k$.}
  \label{Tab_exp_V2}
\end{table}

The primary $\Phi_{0,2}$ on the line with $i_{13}=1$, and its descendant at level $(1,1)$ on the line with $i_{13}=12$, are both singlets due to self-duality.
But more importantly, we find a pair of singlets on the lines with $i_{12} = 24, 35$. They show the duality
\begin{equation}
 A_{1,2} \Phi_{1,2} = \bar{A}_{1,2} \bar{\Phi}_{1,2} \,,
\end{equation}
providing evidence of a Jordan cell with $(e,j) = (1,2)$. By contrast, the remaining two states with conformal weights $(h_{1,2}+2,h_{1,2}+2)$
can be identified as the doublet on the line with $i_{13}=25$. They correspond to a level-$(2,0)$ descendant of $\Phi_{1,2}$ and a level-$(0,2)$ descendant
of $\bar{\Phi}_{1,2}$, with coefficients that are unknown but different from those of the operators $A_{1,2}$ and $\bar{A}_{1,2}$, respectively.

\subsection{$j=3$}

Finally, the results for the sectors $\AStTL{3}{z^2}$ are given in Table~\ref{Tab_exp_V3}. The labels $k=0,1,2$ refer to the
cases $z^2=1$, $z^2=\e^{2\pi\i/ 3}$ and $z^2 = \e^{4\pi\i/ 3}$, respectively.

We observe here a pair of singlets on the lines with $i_{11} = 111, 169$. They show the duality
\begin{equation}
 A_{3,1} \Phi_{3,1} = \bar{A}_{3,1} \bar{\Phi}_{3,1}
\end{equation}
and give evidence for a Jordan cell with $(e,j) = (3,1)$.

\medskip

Summarizing, we have identified pairs of singlets to give evidence of the structure \eqref{diamondPhi12} and its extension to the general
case of $\Phi_{e,j} = \phi_{e,j} \otimes \bar{\phi}_{e,-j}$ and $\bar{\Phi}_{e,j} = \phi_{e,-j} \otimes \bar{\phi}_{e,j}$ for the pairs
$(e,j) = (1,1)$, $(2,1)$, $(1,2)$ and $(1,3)$.

\begin{table}
\begin{center}
\begin{tabular}{|c|cccccccc|ll|l|}
\cline{1-12}
  $k, s$
    & \multicolumn{8}{|c|}{$(i_5,i_6,\ldots,i_{12})$} & \multicolumn{2}{|c|}{$\Delta$} & Identification \\  \cline{2-12}
   & 5 & 6 & 7 & 8 & 9 & 10 & 11 & 12  & Numerics & Exact & of scaling field \\
  \hline
 0, 0 & \textcolor{red}{1} & \textcolor{red}{1} & \textcolor{red}{1} & \textcolor{red}{1} & \textcolor{red}{1} & \textcolor{red}{1} & \textcolor{red}{1} & \textcolor{red}{1}  & 2.64732 & 2.64713 & $\phi_{0,3} \times \phi_{0,-3} \equiv \zeta_0$ \\
 1, 1 & 2 & 2 & 2 & 2 & 2 & 2 & 2 & 2  & 2.73773 & 2.73746 & $\phi_{1/3,3} \times \phi_{1/3,-3} \equiv \zeta_{1/3}$ \\
 2, 2 & 3 & 3 & 3 & 3 & 3 & 3 & 3 & 3  & 3.00867 & 3.00846 & $\phi_{2/3,3} \times \phi_{2/3,-3} \equiv \zeta_{2/3}$ \\
 0, 3 & 5 & 5 & 4 & 4 & 4 & 4 & 4 & 4  & 3.463 & 3.46011 & $\phi_{1,3} \times \phi_{1,-3} \equiv \zeta_1$ \\
 0, 0 & \textcolor{red}{10} & \textcolor{red}{13} & \textcolor{red}{16} & \textcolor{red}{17} & \textcolor{red}{19} & \textcolor{red}{17} & \textcolor{red}{17} & \textcolor{red}{17}  & 4.637 & 4.647 & $L_{-1} \bar{L}_{-1} \zeta_0$ \\
 2, 0 & --- & --- & \textcolor{red}{36} & \textcolor{red}{48} & \textcolor{red}{61} & \textcolor{red}{74} & \textcolor{red}{84} & \textcolor{red}{84}  & 6.640 & 6.647 & $L_{-2} \bar{L}_{-2} \zeta_{0}$ \\
% $V_{3?0}$ & --- & \textcolor{red}{1} & \textcolor{red}{16} & \textcolor{red}{48} & \textcolor{red}{105} & & & & & 11.9 & ?? & ?? \\
 0, 0 & \textcolor{red}{31} & \textcolor{red}{55} & \textcolor{red}{79} & \textcolor{red}{100} & \textcolor{red}{107} & \textcolor{red}{109} & \textcolor{red}{111} & \textcolor{red}{108}  & 6.454 & 6.460 & $L_{-3} \zeta_1$ \\
% $V_{3?0}$ & --- & --- & \textcolor{red}{1} & \textcolor{red}{17} & \textcolor{red}{61} & \textcolor{red}{140} & & & & 13.7 & ? & ? \\
 0, 0 & \textcolor{red}{39} & \textcolor{red}{66} & \textcolor{red}{98} & \textcolor{red}{122} & \textcolor{red}{140} & \textcolor{red}{149} & \textcolor{red}{144} & &  6.6478 & 6.6471 & $L_{-2} \bar{L}_{-2} \zeta_0$ \\
 0, 0 & \textcolor{red}{43} & \textcolor{red}{85} & \textcolor{red}{126} & \textcolor{red}{155} & \textcolor{red}{163} & \textcolor{red}{175} & \textcolor{red}{169} & &  6.468 & 6.460 & $L_{-3} \zeta_1$ \\
 0, 0 & \textcolor{red}{39} & \textcolor{red}{91} & \textcolor{red}{153} & \textcolor{red}{216} & \textcolor{red}{280} & \textcolor{red}{327} & \textcolor{red}{367} & &  8.70 & 8.64 & $L_{-3} \bar{L}_{-3} \zeta_0$ \\
 0, 0 & \textcolor{red}{31} & \textcolor{red}{91} & \textcolor{red}{169} & \textcolor{red}{251} & \textcolor{red}{332} & \textcolor{red}{402} & \textcolor{red}{456} & &  8.85 & 8.64 & $L_{-3} \bar{L}_{-3} \zeta_0$ \\
% $V_{3?0}$ & {\tiny 42} & {\tiny 111} & \textcolor{red}{215} & \textcolor{red}{324} & \textcolor{red}{414} & \textcolor{red}{473} & & & & 8.57 & ? & ? \\
% $V_{3?0}$ & \textcolor{red}{62} & \textcolor{red}{175} & \textcolor{red}{335} & \textcolor{red}{483} & & & & & & 9.15 & ? & ? \\

\hline
\end{tabular}
\end{center}
\caption{Conformal spectrum in the sector $\AStTL{3}{z^2}$ for $Q=1/2$. The label $k$ %\lawrence{just $k$ (no ``${}=0$'')?} 
corresponds to $z^2 = (\e^{2\pi\i/3})^k$.
%\jesper{Computations for the last two columns are still ongoing. Adjust number of significant digits later.}
}
  \label{Tab_exp_V3}
\end{table}

\FloatBarrier

\section{Parity and the structure of the modules}

In this Appendix, added post publication, we correct our previous reasoning in which we concluded that the number of singlets or doublets observed will directly yield insight into the structure of the continuum limit Virasoro modules. The error in this reasoning is clear when comparing to \cite{GSJS}, where we perform a similar investigation in the XXZ spin chain. We there have the same spectrum of the Hamiltonian and must thus have the same number of singlets and doublets, while the numerical results from using the Koo-Saleur generators indicate that the structures of the continuum limit Virasoro modules are not the same.  

To distinguish between the two types of modules that appear for loop models and the XXZ spin chain, we must instead think more carefully about the symmetry under parity, under which chiral and anti-chiral are mapped to each other. On the lattice this corresponds to a mapping of site $j$ to $-j$, which by \eqref{generators} maps the Koo-Saleur generator $\KSgen_{n}$ to $\bar{\KSgen}_n$. The parity operation is idempotent, so we can distinguish the states by its eigenvalues, $P  = \pm 1$.
 Let us consider the two states $\Phi_{1,2} = \phi_{1,2}\otimes \bar{\phi}_{1,-2}$ and $\bar{\Phi}_{1,2}=\phi_{1,-2}\otimes \bar{\phi}_{1,2}$, which are mapped to each other under parity. Different situations can occur for their descendants $L_{-1}^2 \Phi_{1,2},\; L_{-2} \Phi_{1,2} ,\; \bar{L}_{-1}^2 \bar{\Phi}_{1,2} $ and $ \bar{L}_{-2} \bar{\Phi}_{1,2}$.  
On the lattice, the four scaling states that have the correct lattice momenta and energies to be identified with these descendants form two singlets and one doublet. The states $v_1,v_2$ in the doublet are mapped to each other by parity and can thus be combined into one $P=1$ state $v_1+v_2$ and one $P=-1$ state $v_1 - v_2$. We now wish to distinguish between the types of modules using the parity of the singlets.

If the four descendants are independent we can form four linear combinations that are eigenstates of the parity operator:
\begin{equation}
\begin{split}
L_{-1}^2 \Phi_{1,2} & \pm \bar{L}_{-1}^2 \bar{\Phi}_{1,2} \\
L_{-2} \Phi_{1,2} & \pm \bar{L}_{-2} \bar{\Phi}_{1,2}
\end{split}
\end{equation}
Out of these two have parity $P=1$ and two have $P=-1$. %There are four scaling states on the lattice that correspond to some to us unknown linear combinations of these descendants. What we do know is that in this scenario the scaling states must also be sorted into two states with $P = 1$, two with $P = -1$. Note that they may all have different energies, or appear as a doublet and two singlets, or two doublets. In this case we know that there is one doublet and two singlets, the singlets must thus have opposite parities $P=1,P=-1$. 

Now consider instead the states depicted in the diagram of \eqref{diamondPhi12}, reproduced here for convenience:
\begin{equation} 
\begin{tikzpicture}[>=stealth]
\coordinate (PSI) at (0, 1.5) {};
\coordinate (XY) at (-3.5, 0) {};
\coordinate (YX) at (3.5, 0) {};
\coordinate (AXY) at (0, -1.5) {};
\node(axy) at (AXY) {$A_{1,2}\Phi_{1,2}=\bar{A}_{1,2}\bar{\Phi}_{1,2}$};
\node(yx) at (YX) {$\bar{\Phi}_{1,2}=\phi_{1,-2}\otimes \bar{\phi}_{1,2}$};
\node(xy) at (XY) {$\Phi_{1,2}=\phi_{1,2}\otimes \bar{\phi}_{1,-2}$};
\node(psi) at (PSI) {$\Psi_{1,2}$};
\draw[->] (psi) -- (yx);
\draw[->] (psi) -- (xy);
\draw[->] (xy) -- (axy);
\draw[->] (yx) -- (axy);
\draw[->] (psi) -- (axy);
\node(L0) at ([xshift=6ex,yshift=-2ex]$(psi)!0.5!(axy)$) {\small$(L_0-h_{-1,2})$};
\node(A+) at ([yshift=3ex]$(psi)!0.5!(xy)$) {\small$A^{\dagger}$};
\node(A+bar) at ([yshift=3ex]$(psi)!0.5!(yx)$) {\small$\bar{A}^{\dagger}$};
\node(A) at ([yshift=-2ex]$(xy)!0.4!(axy)$) {\small$A$};
\node(Abar) at ([yshift=-2ex]$(yx)!0.4!(axy)$) {\small$\bar{A}$};
\end{tikzpicture}
\end{equation}
The four descendants are no longer independent. The bottom field $A_{1,2}\Phi_{1,2}=\bar{A}_{1,2}\bar{\Phi}_{1,2}$ is clearly invariant under parity. Meanwhile the top field satisfies $A A^\dag \Psi = \bar{A}\bar{A}^\dag \Psi$ and therefore also has $P=1$. These two fields should both appear as singlets, while the doublet would correspond to the two linear combinations that can be formed with what remains (not shown in the diagram). %Once again, the scaling states must be sorted by parity in the same way as the continuum fields.

The above argument has been exposed for the continuum formulation. In order to validate our whole approach of inferring properties of the continuum theory from lattice discretizations, a similar scenario had better hold on the lattice as well. We therefore return to the finite-size numerics to seek the verdict. When acting with $P: j \to -j$ on the two singlets, we find that the results depend on the representation. In the loop model, both singlets have $P=1$ corresponding to the situation in the diagram above, while in the XXZ spin chain we find that one has $P=1$, one $P=-1$, so that we rather have the parities expected from four linearly independent descendants. The two lattice discretizations thus indeed confirm the general argument, and we find that only the loop model has the Jordan-cell structure \eqref{diamondPhi12} with the dependence $A_{1,2} \Phi_{1,2} = \bar{A}_{1,2} \bar{\Phi}_{1,2}$, which is one of the main points of this paper.

\FloatBarrier

\end{document}